\documentclass[aps,prstab,reprint,groupedaddress,amsmath,amssymb,showpacs,showkeys,floatfix,longbibliography]{revtex4-1}

\usepackage{graphicx}
\usepackage{dcolumn}
\usepackage{bm}

\begin{document}

\title{Lattice optimization of the Novosibirsk fourth-generation light source SKIF
}

\author{ G.~Baranov }
  \affiliation{Budker Institute of Nuclear Physics SB RAS, Novosibirsk 630090, Russia}
\author{ A.~Bogomyagkov }
  \email{A.V.Bogomyagkov@inp.nsk.su}
  \affiliation{Budker Institute of Nuclear Physics SB RAS, Novosibirsk 630090, Russia}
\author{ E.~Levichev }
  \affiliation{Budker Institute of Nuclear Physics SB RAS, Novosibirsk 630090, Russia}
  \affiliation{Novosibirsk State Technical University, Novosibirsk 630073, Russia}
\author{ S.~Sinyatkin }
  \affiliation{Budker Institute of Nuclear Physics SB RAS, Novosibirsk 630090, Russia}

\date{\today}

\begin{abstract}
We discuss the choice of the magnetic lattice and parameter optimization of the fourth generation light source SKIF (Russian acronym for Siberian Circular Photon Source) under construction in Novosibirsk. The study compares several basic lattice cells to procure one with low emittance and large dynamic aperture. 
The result is a developed lattice of SKIF with 72~pm natural emittance (at zero beam current and absent betatron coupling) at 3~GeV beam energy and 476~m circumference. Only two families of sextupoles provide horizontal and vertical dynamic apertures of 12~mm and 3.5~mm respectively and energy acceptance more than 5\%. To check the potential of the found solution, we used slightly modified SKIF lattice to design 6~GeV light source and reached 33~pm natural emittance for 1075~m circumference with 40 straight sections.  Again, only two sextupole families ensure sufficient dynamic aperture (7~mm horizontal, 4~mm vertical) and energy acceptance more than 5\% for simple efficient injection and sufficient beam lifetime.
\end{abstract}


\maketitle

\section{Introduction}
A circular storage ring of the intense electron beams is a primary source of powerful electromagnetic radiation in the wide wave length range (from infra-red and ultra-violet to hard X-ray), used in different scientific disciplines: chemistry, biology, solid-state physics, geology etc. There are more than 50 working light sources in the world \cite{LightSource}.

The practical quantity describing the light source's efficiency is brightness defined at a particular photon wavelength as
the radiated flux per unit area of the source and per unit solid angle of emission. The main way to increase the brightness of the light source is to reduce the phase volume (emittance) of the electron beam, which results from the equilibrium between the radiation damping and quantum excitation of the betatron oscillations. 

Let the magnetic lattice of a storage ring be a sequence of identical bending magnets, with the bending angle $2\varphi$ each, then horizontal emittance is
\begin{equation}
\label{eq:TME-emittance}
\varepsilon_x=F \frac{C_q\gamma^2}{J_x}(2\varphi)^3,
\end{equation}
where $\gamma\gg1$ -- Lorentz factor, $J_x\approx 1\div 2$ -- damping number, $F$ is factor depending on type of magnetic lattice and its optimization,
\begin{equation*}
C_q=\frac{55}{32\sqrt{3}}\frac{\hbar}{mc}=3.84\cdot 10^{-13} \text{ m}.
\end{equation*}
For the magnet with a uniform magnetic field, factor $F$ is minimal when horizontal betatron function $\beta_x$ and dispersion function $\eta$ reach the certain values at the magnet center, dependent on the magnet length $2L$ and bending angle $2\varphi$
\begin{align}
\beta_{x,min}&=\frac{L}{\sqrt{15}}, & \eta_{min}&=\frac{L\varphi}{6}.
\end{align}
This configuration received the name TME (Theoretical Minimum Emittance) \cite{Teng:1984cz,Teng:1985gm,Lee:1991st,Antoniou:2013uva,Cai:2018bvb} and factor $F$ is
\begin{equation}
\label{eq:TME-F}
F_{TME}=\frac{1}{12\sqrt{15}}.
\end{equation}
The caveat is that it is practically impossible to achieve conditions for the minimal emittance simultaneously with an acceptable length of the cell (so that the whole ring is not too long and expensive), feasible quadrupole and sextupole strengths, and sufficient transverse and longitudinal dynamic apertures. Therefore, the realistic lattice is just an approximation of TME. The simplest optically periodical cell, built around the dipole in the center, requires two quadrupole doublets on the sides of the magnet. Exactly this type of cell is fundamental for all light sources of the 4th generation.

Minimization of $F$ unavoidably stiffens focusing, increases natural chromaticity, correction of which requires strong sextupoles. Aside from the technical difficulties of this high field magnets creation, there is a more fundamental problem -- nonlinear motion limits the stability area (dynamic aperture). The decrease of the dynamic aperture is the main difficulty in the process of designing a storage ring with small emittance.

Further, we show that the described cell, consisting of dipole in the center and two doublets of quadrupoles, allows, besides TME, other optical solutions with different optical functions, emittances, chromaticities, sextupoles strengths and dynamical apertures. As the result of the study we choose one solution as the fundamental for magnetic lattice of the 4th generation synchrotron radiation (SR) light source SKIF, being constructed in Novosibirsk. The initial requirements of the facility are 3~GeV beam energy, circumference $\Pi<500$~m, horizontal emittance $\varepsilon_x<100$~pm.

\section{The choice of the basic cell}
\begin{figure}[!htb]
\centering
\includegraphics*[width=.45\textwidth,trim=50 35 65 20, clip]{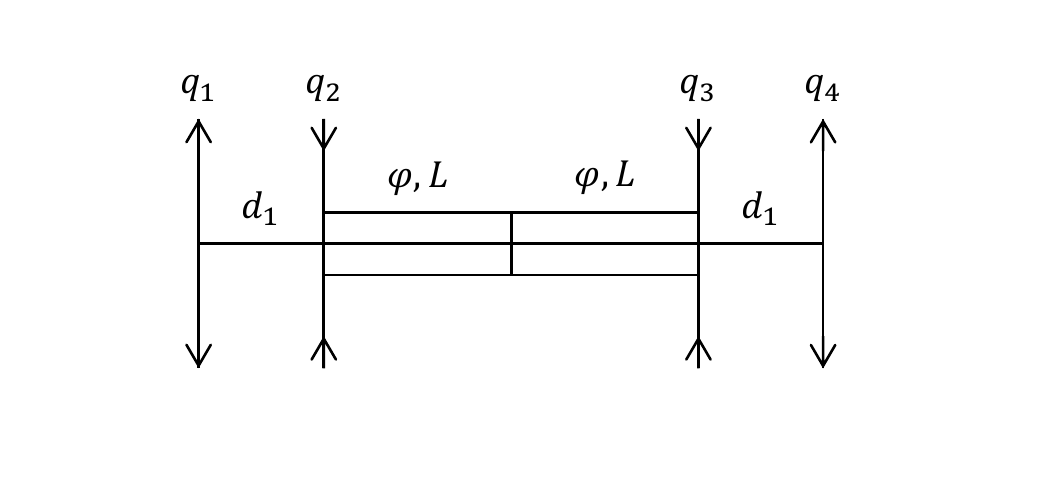}
\caption{Low emittance cell layout.}
\label{fig:TME-layout}
\end{figure}

Consider the simplified cell layout on FIG.~\ref{fig:TME-layout}. 
The dipole in the center has a length of $2L$ and a bending angle $\varphi_c=2\varphi$. On both sides of the dipole are thin quadrupoles $q_{1,2}$ with integrated strengths $p_{1,2}=p_{4,3}=(B'l)_{1,2}/B\rho$, and sextupoles, superimposed with quadrupoles, with strengths $s_{1,2}=s_{4,3}=(B''l)_{1,2}/B\rho$, where $B\rho$ is magnetic rigidity, and $d_1$ is the length of the drift between the quadrupoles. The total length of the cell is $L_c=2(L+d_1)$.

\begin{figure}[htb]
\centering
\includegraphics*[width=.48\columnwidth,angle=0,trim=100 60 150 15,clip]{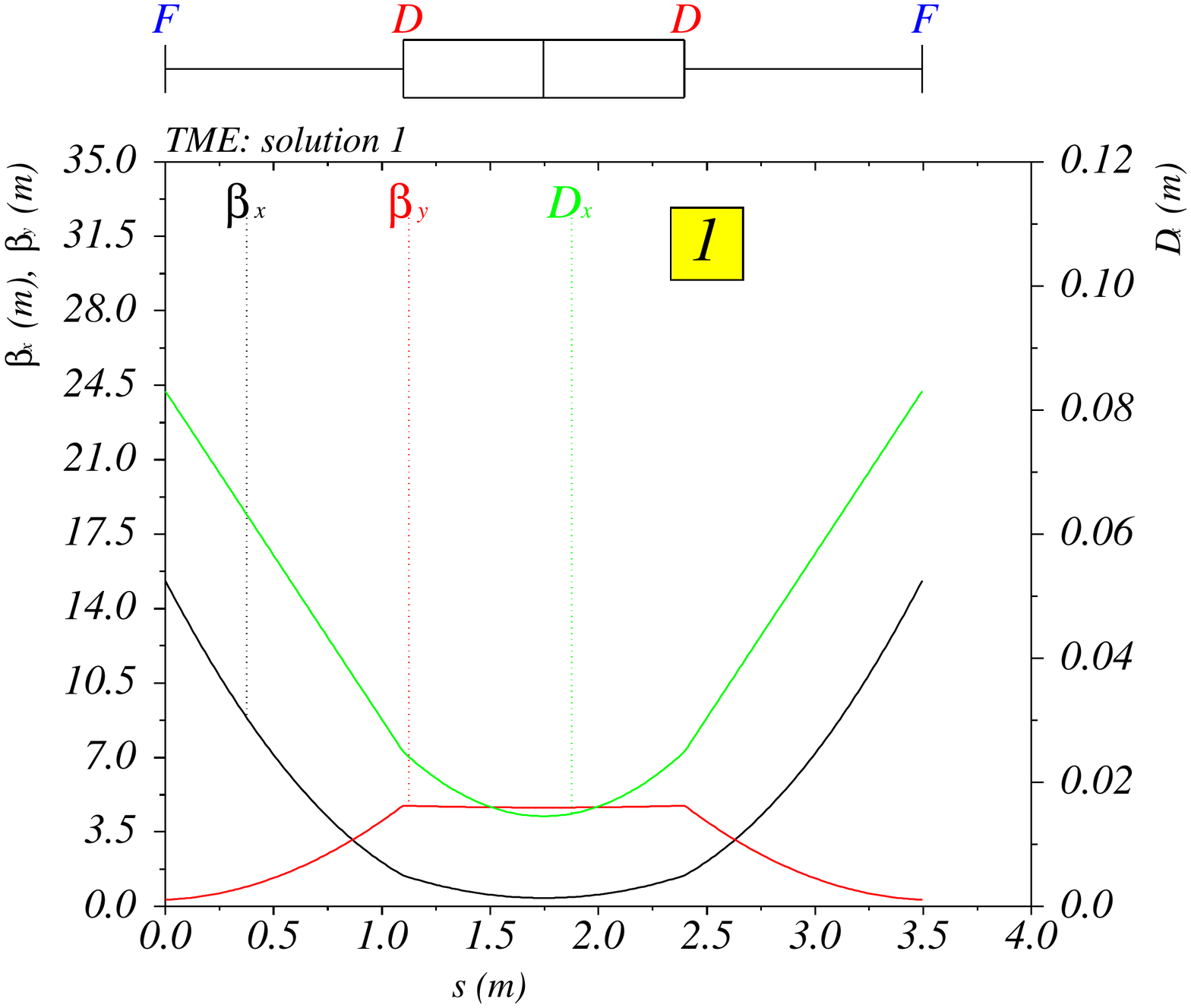}
\includegraphics*[width=.48\columnwidth,angle=0,trim=100 60 150 15,clip]{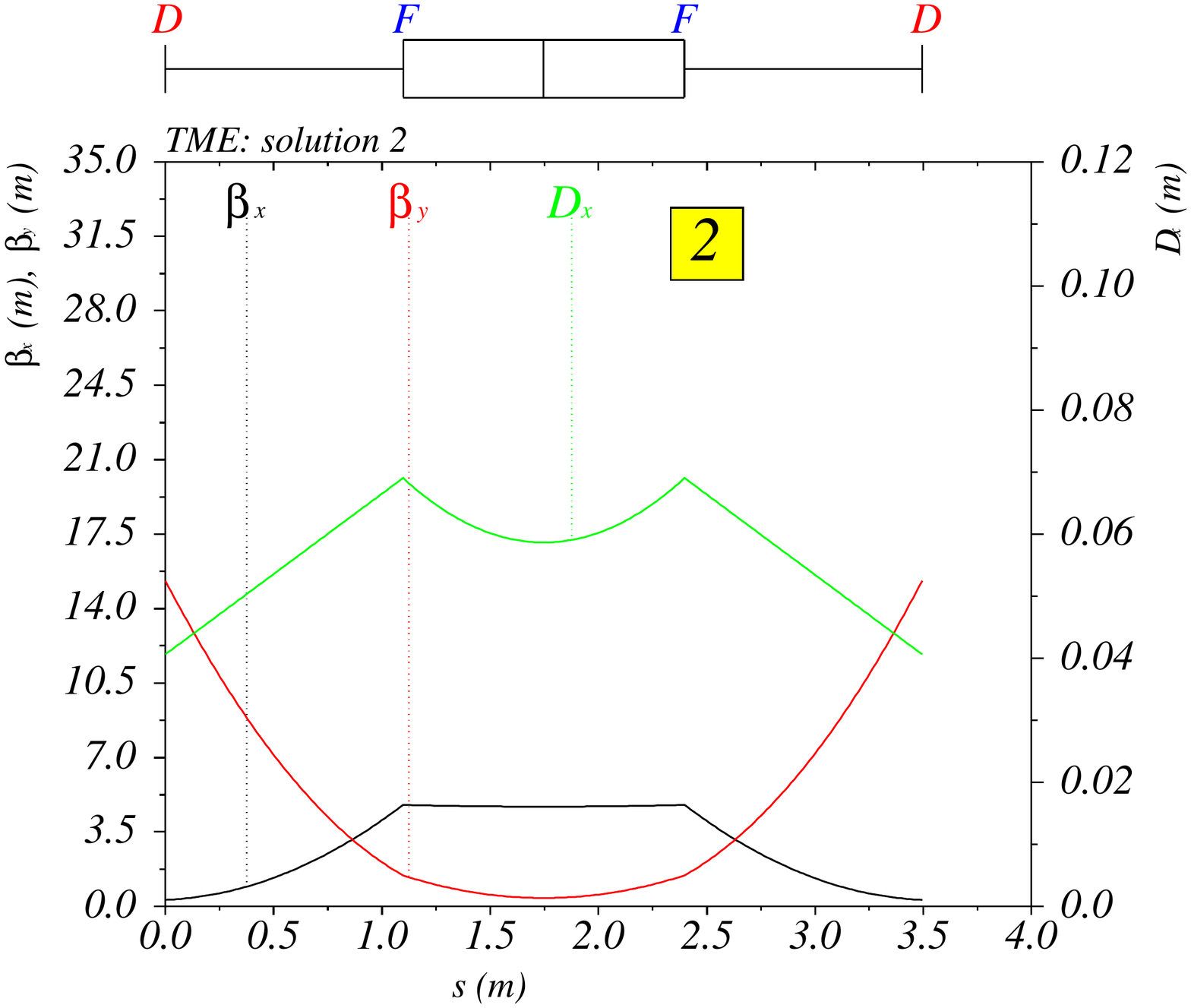}
\includegraphics*[width=.48\columnwidth,angle=0,trim=100 60 150 15,clip]{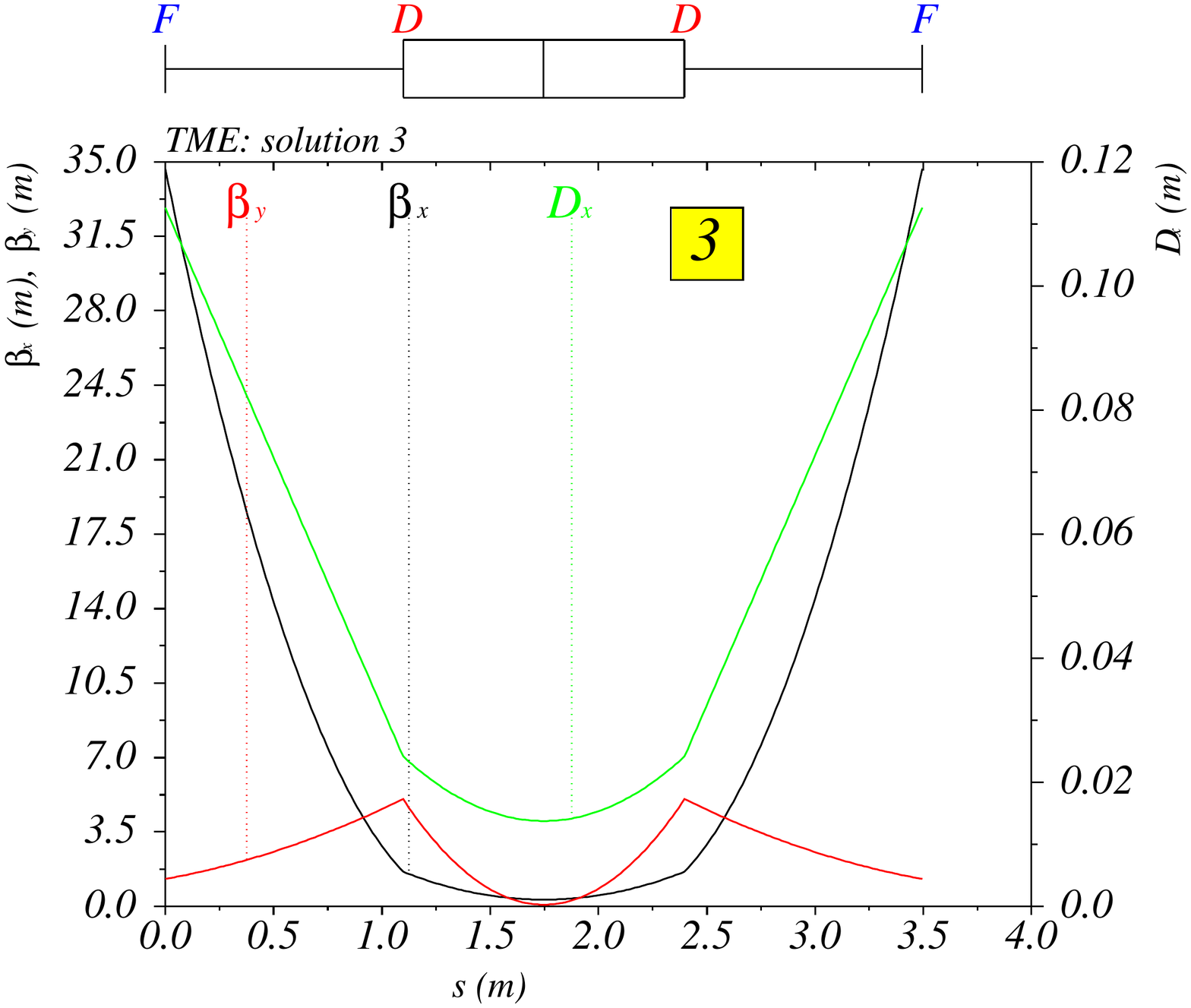}
\includegraphics*[width=.48\columnwidth,angle=0,trim=100 60 150 15,clip]{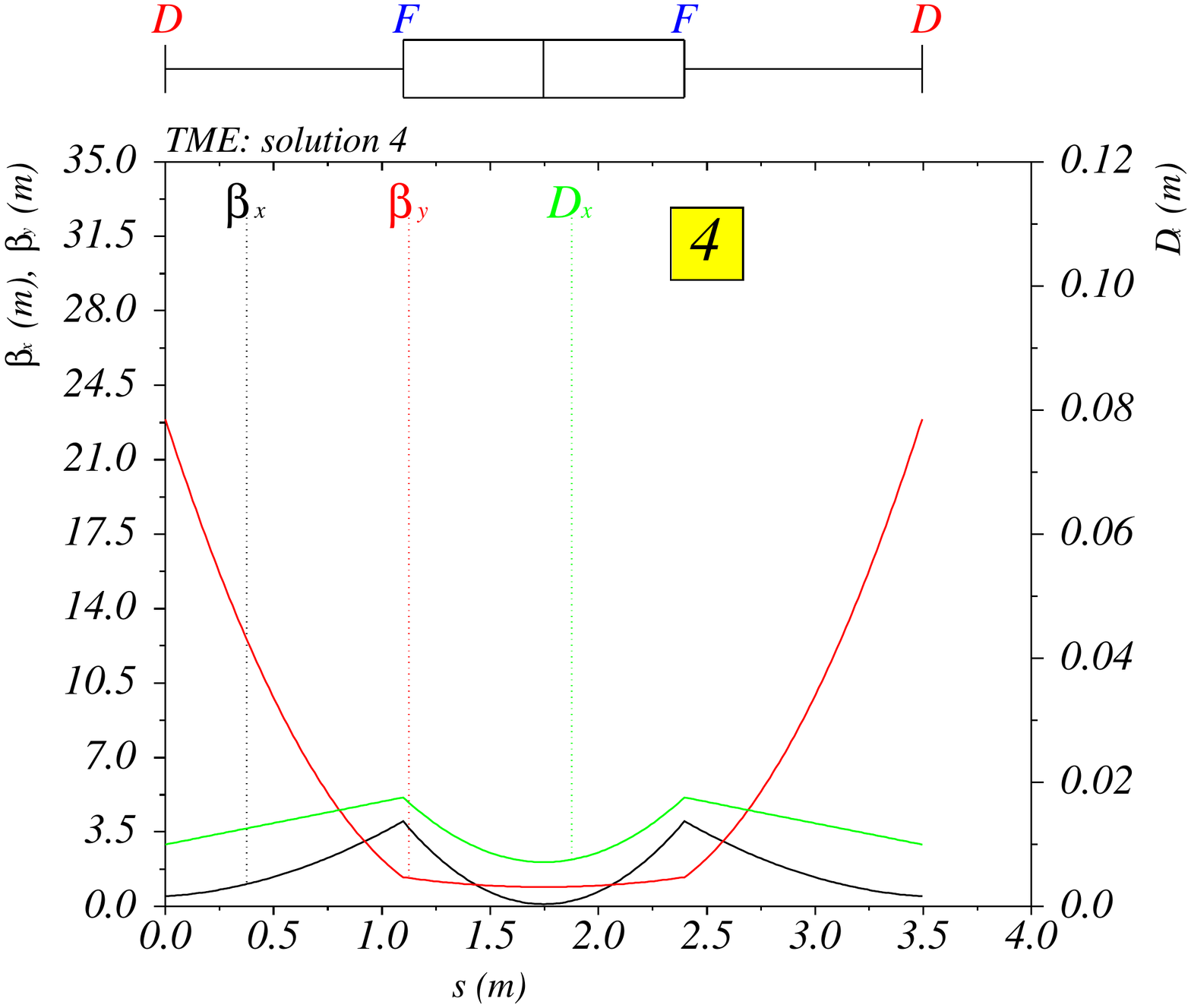}
\caption{Four periodic optical solutions for the cell from FIG.~\ref{fig:TME-layout}. F and D stand for focusing and defocusing quadrupole, respectively. Solution 4 provides minimum emittance (TME).}
\label{fig:optics-TME-1}
\end{figure}

Periodic solutions were found with the computer algebra system WOLFRAM MATHEMATICA \cite{Mathematica} and verified with MAD-X \cite{MADX}. Analytical calculations agree with numerical. FIG.~\ref{fig:optics-TME-1} shows the optical functions of the four solutions.
The graphs demonstrate that solution 2 dispersion function at the center of the dipole is too large; therefore, it will not provide low emittance.

Since our goal is to find a basic cell for light source SKIF, besides the emittance we are interested in the quadrupoles and sextupoles strengths, which should be small for the magnet fabrication and for obtaining a large dynamic aperture.

For estimations, let us use the value of SKIF's horizontal emittance $\varepsilon_x=75$~pm. At the beam energy $E=3$~GeV and $J_x=1$, according to \eqref{eq:TME-F} and \eqref{eq:TME-emittance}, the bending angle of the cell is $\varphi_c=2\varphi=0.0641=2\pi/98$. From the circumference of the light source ($\approx500$~m) we designate $30\%$ for the straight and matching sections ($\approx150$~m), then the total length occupied by basic cells is $\approx350$~m, and the length of one cell is $L_c\approx 3.5$~m. In the cell we choose $d_1\approx 1$~m, then the dipole half-length is $L\approx 0.75$~m.  We will investigate the influence of these lengths on emittance and other accelerator parameters later.

\begin{figure}[htb]
\centering
\includegraphics*[width=.48\columnwidth,angle=0,trim=0 0 0 0,clip]{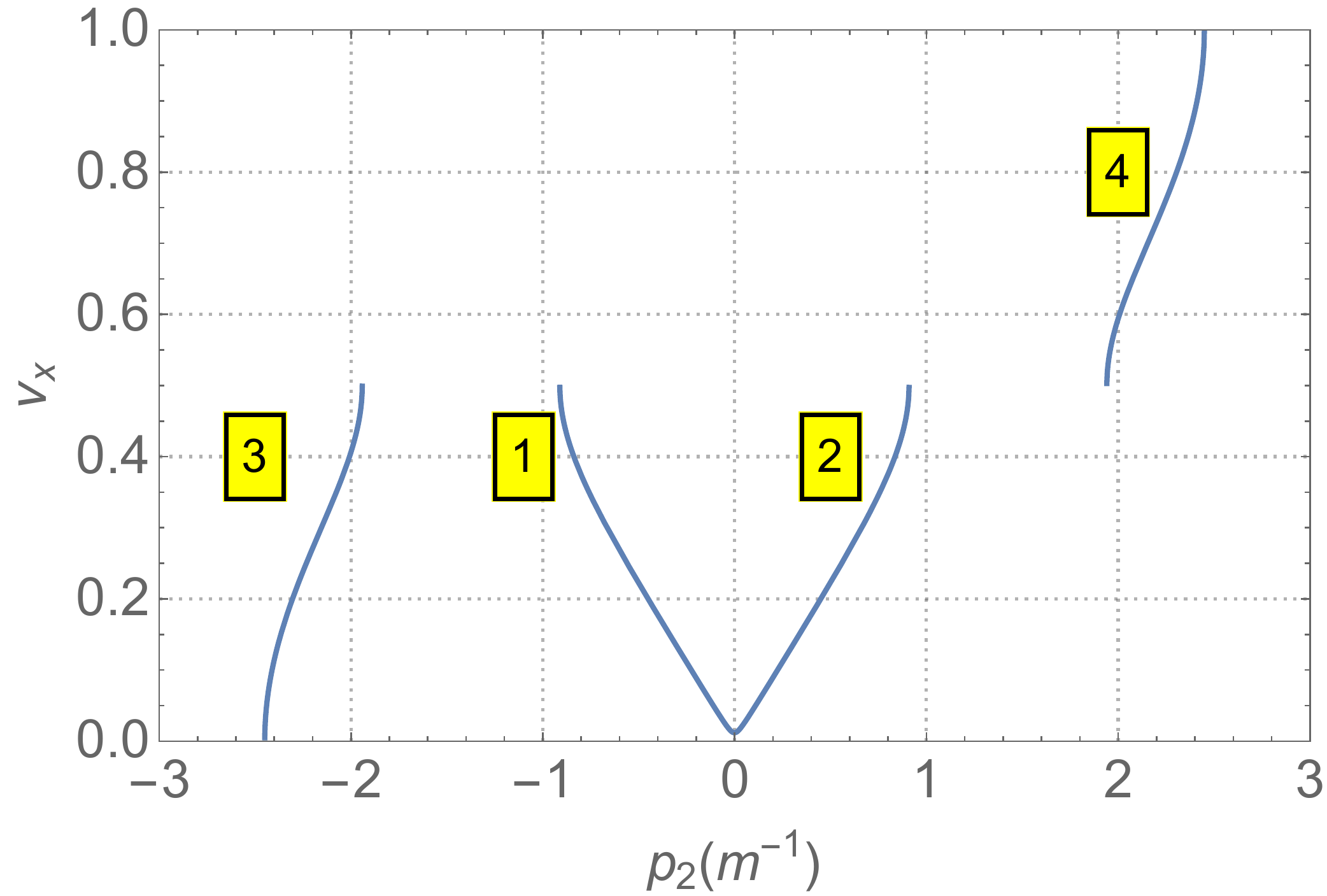}
\includegraphics*[width=.48\columnwidth,angle=0,trim=0 0 0 0,clip]{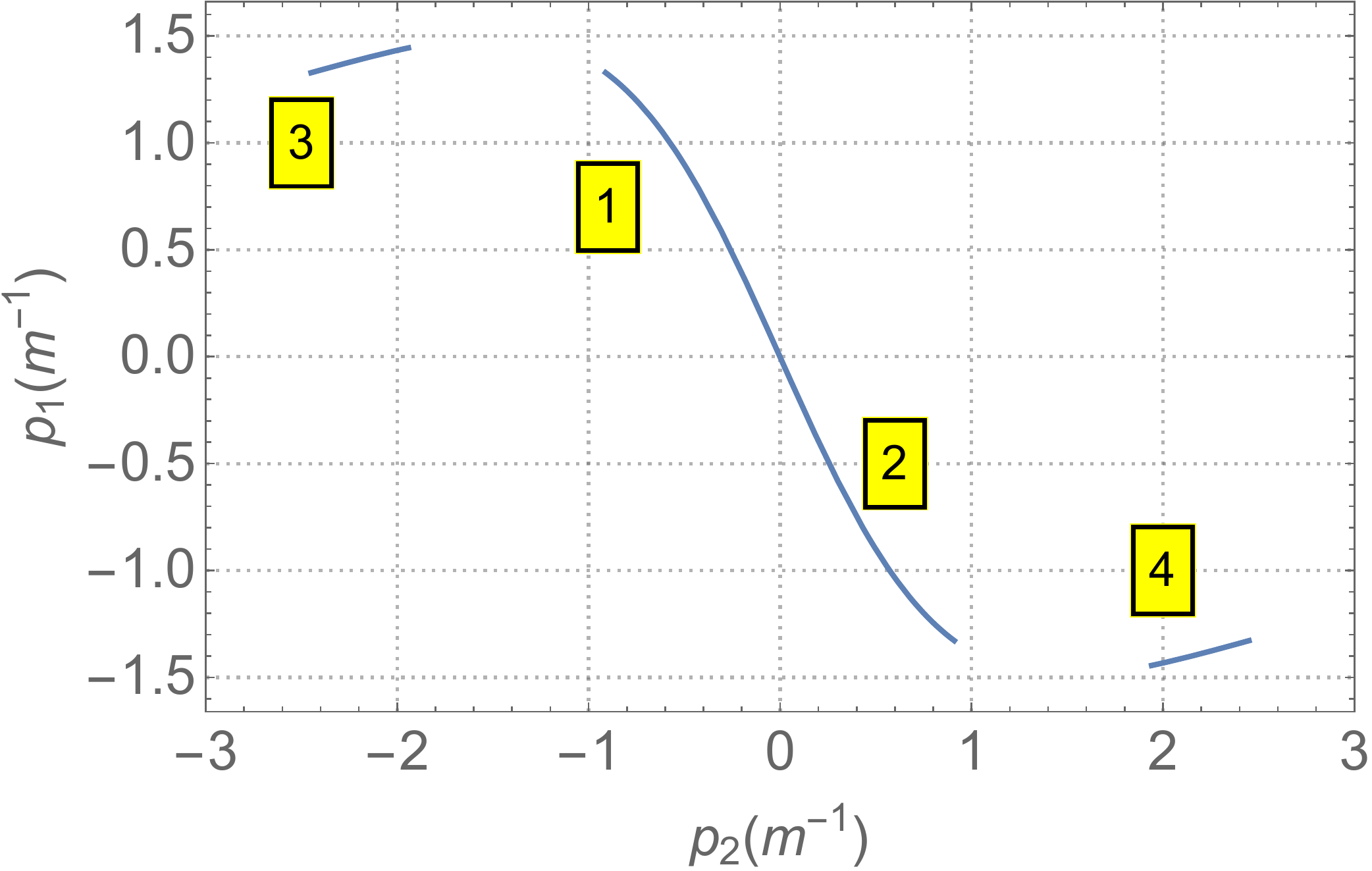}
\caption{Stability area of the cell in the form of tune $\nu_x$ (left) and integrated strength $p_1$ of the first quadrupole (right) on integrated strengths $p_2$ of the second quadrupole.}
\label{fig:stability-TME-1}
\end{figure}

Calculating stability area of the cell from FIG.~\ref{fig:TME-layout}, we found dependance of the cell tune (FIG.~\ref{fig:stability-TME-1} left) and the first quadrupole strength (FIG.~\ref{fig:stability-TME-1} right) on the integrated strength $p_2$  of the second quadrupole (right).
One can clearly distinguish the four stability areas corresponding to the four optical solutions. Moreover, the first three solutions have $0\leq\nu_x\leq0.5$, and the fourth, the TME, has  $0.5\leq\nu_x\leq1$. Curiously, solutions 3 and 4 (TME) possess equally strong quadrupoles strengths, only of opposite signs, $\left|p_2\right|>2$~m$^{-1}$, $\left|p_1\right|\approx1.5$~m$^{-1}$, regardless of the fourth solution tune (the measure of the lattice rigidity) being greater than the third's. Given the quadrupole length $l=0.2$~m and beam energy $E=3$~GeV, then $p_2=2$~m$^{-1}$ corresponds to $B'=100$~T/m gradient, which is rather large value even for small apertures of $\varnothing25\div30$~mm, common for the 4th generation light sources. For solutions 1 and 2,  $\left|p_2\right|<1$~m$^{-1}$ and $\left|p_1\right|<1.5$~m$^{-1}$ and the gradients are moderate.

\begin{figure}[htb]
\centering
\includegraphics*[width=.48\columnwidth,angle=0,trim=0 0 0 0,clip]{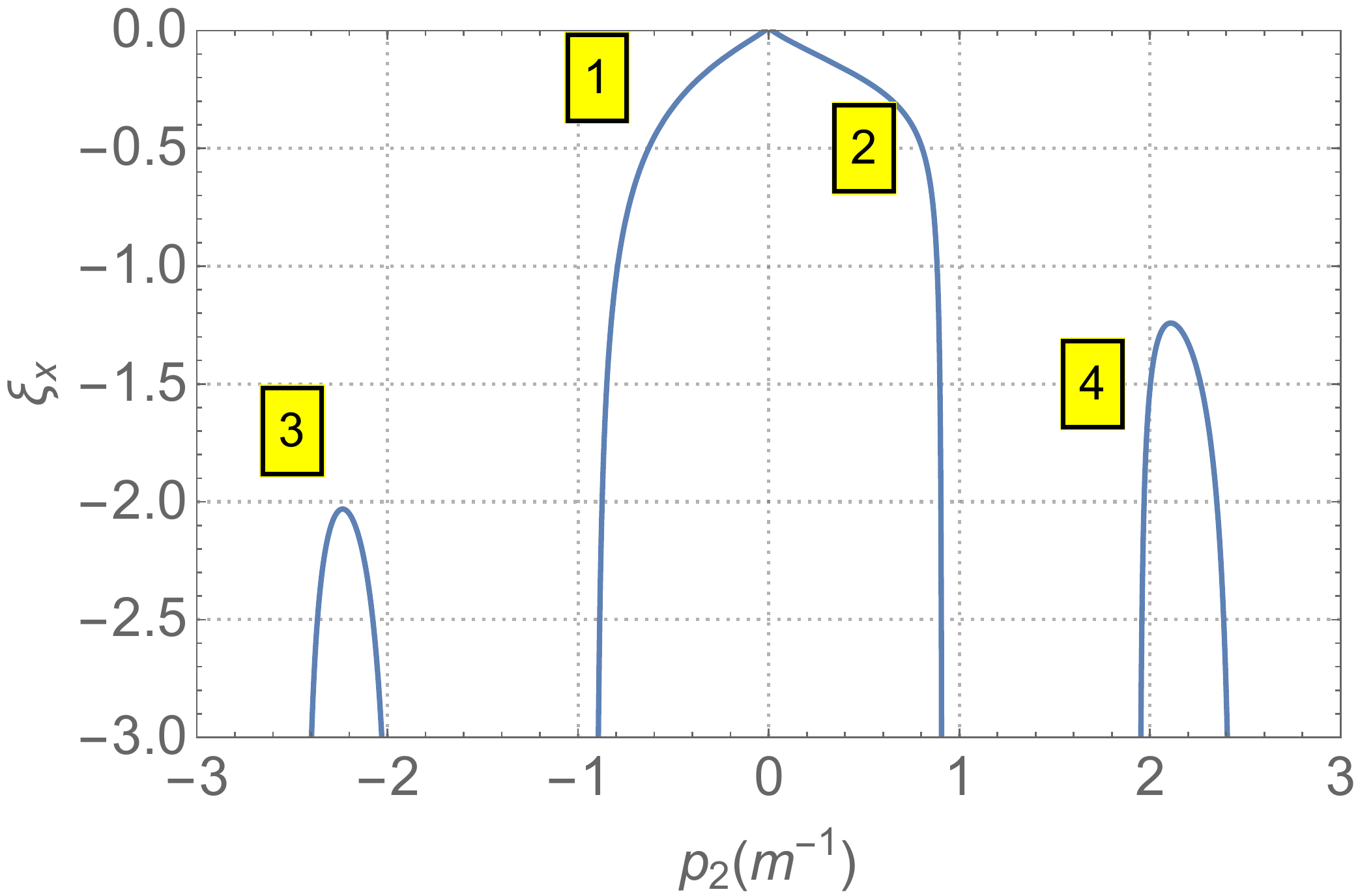}
\includegraphics*[width=.48\columnwidth,angle=0,trim=0 0 0 0,clip]{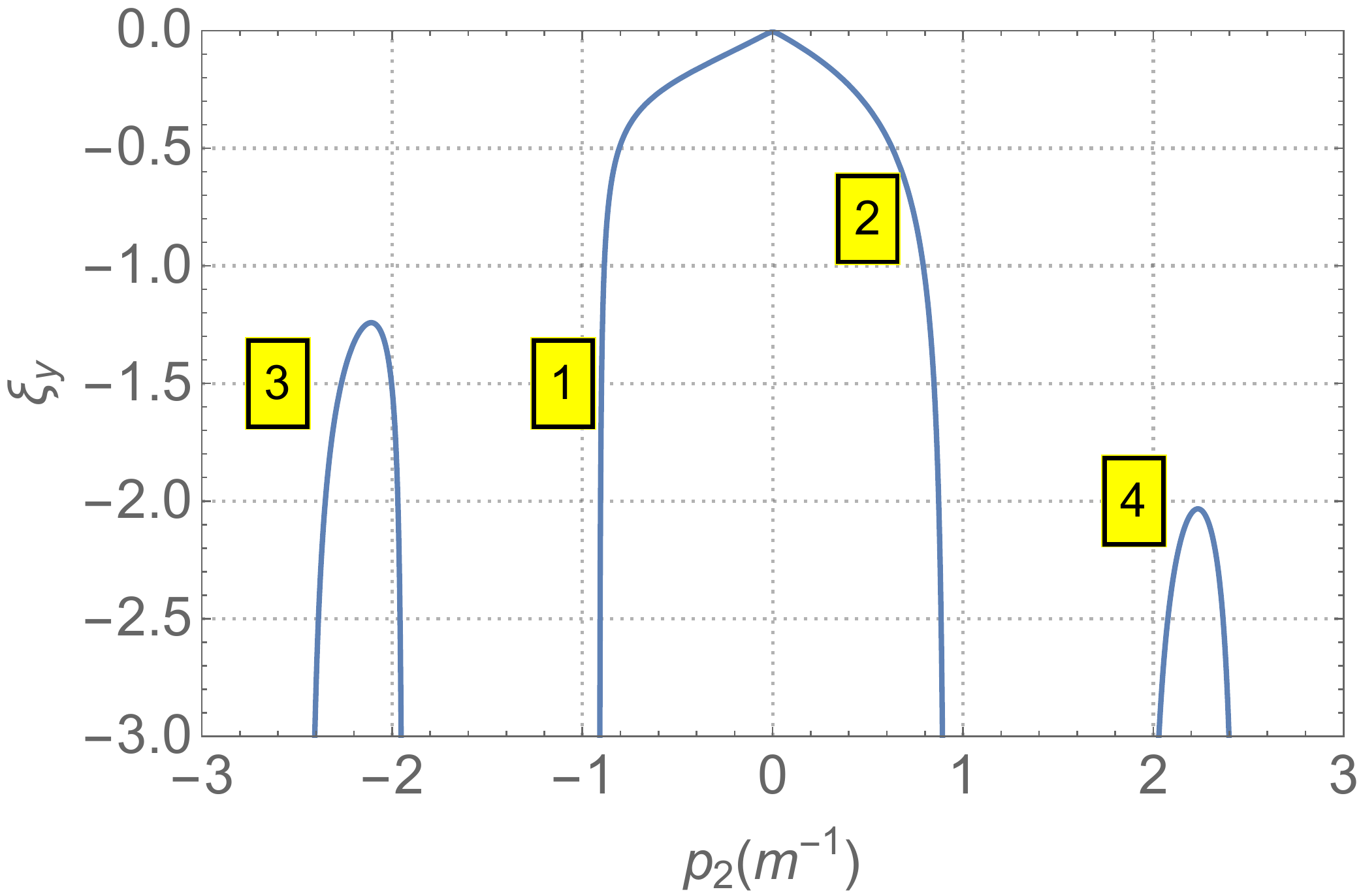}
\caption{Horizontal and vertical chromaticity versus second quadrupole strength.}
\label{fig:chromaticity-TME}
\end{figure}

\begin{figure}[htb]
\centering
\includegraphics*[width=.48\columnwidth,angle=0,trim=0 0 0 0,clip]{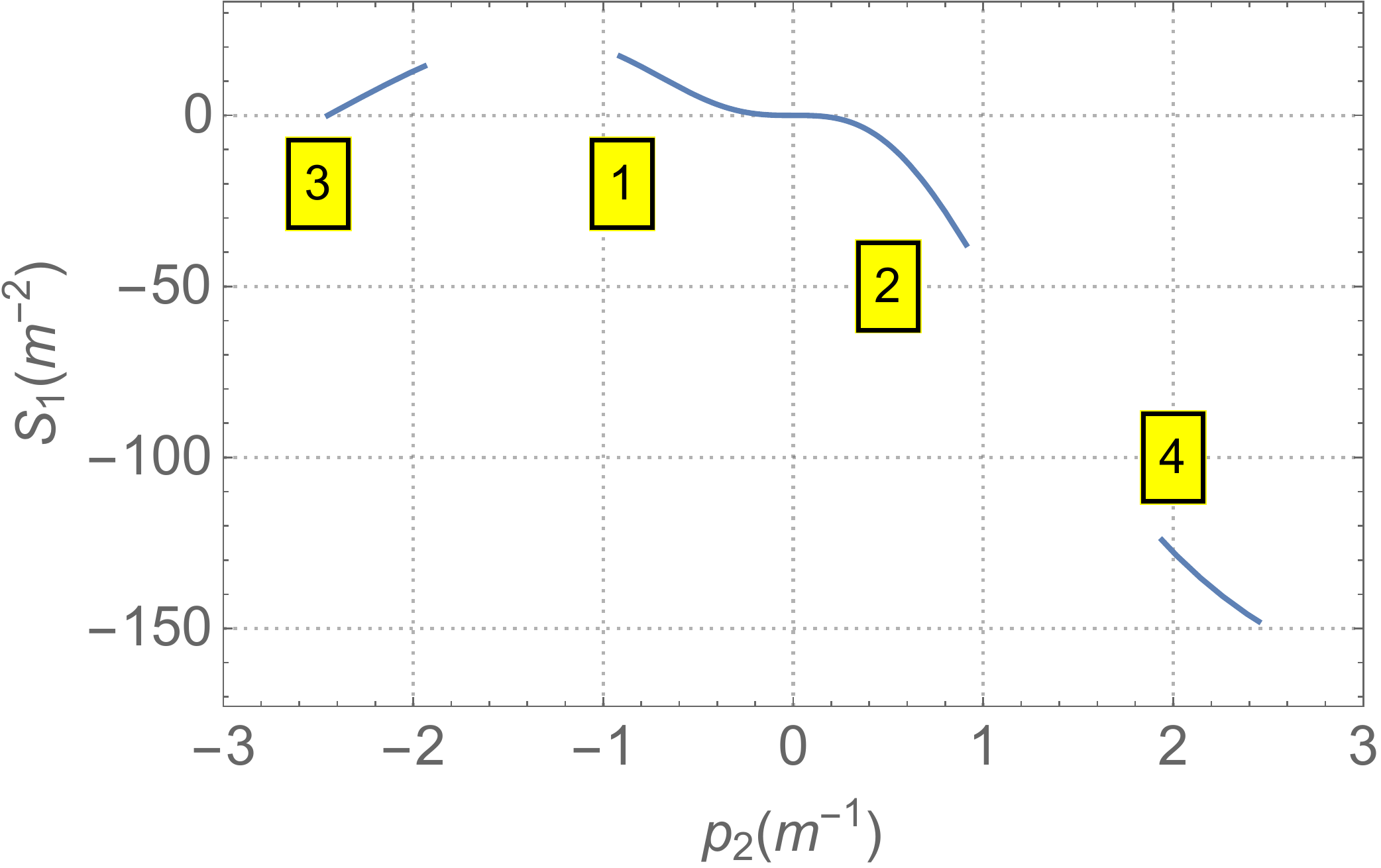}
\includegraphics*[width=.48\columnwidth,angle=0,trim=0 0 0 0,clip]{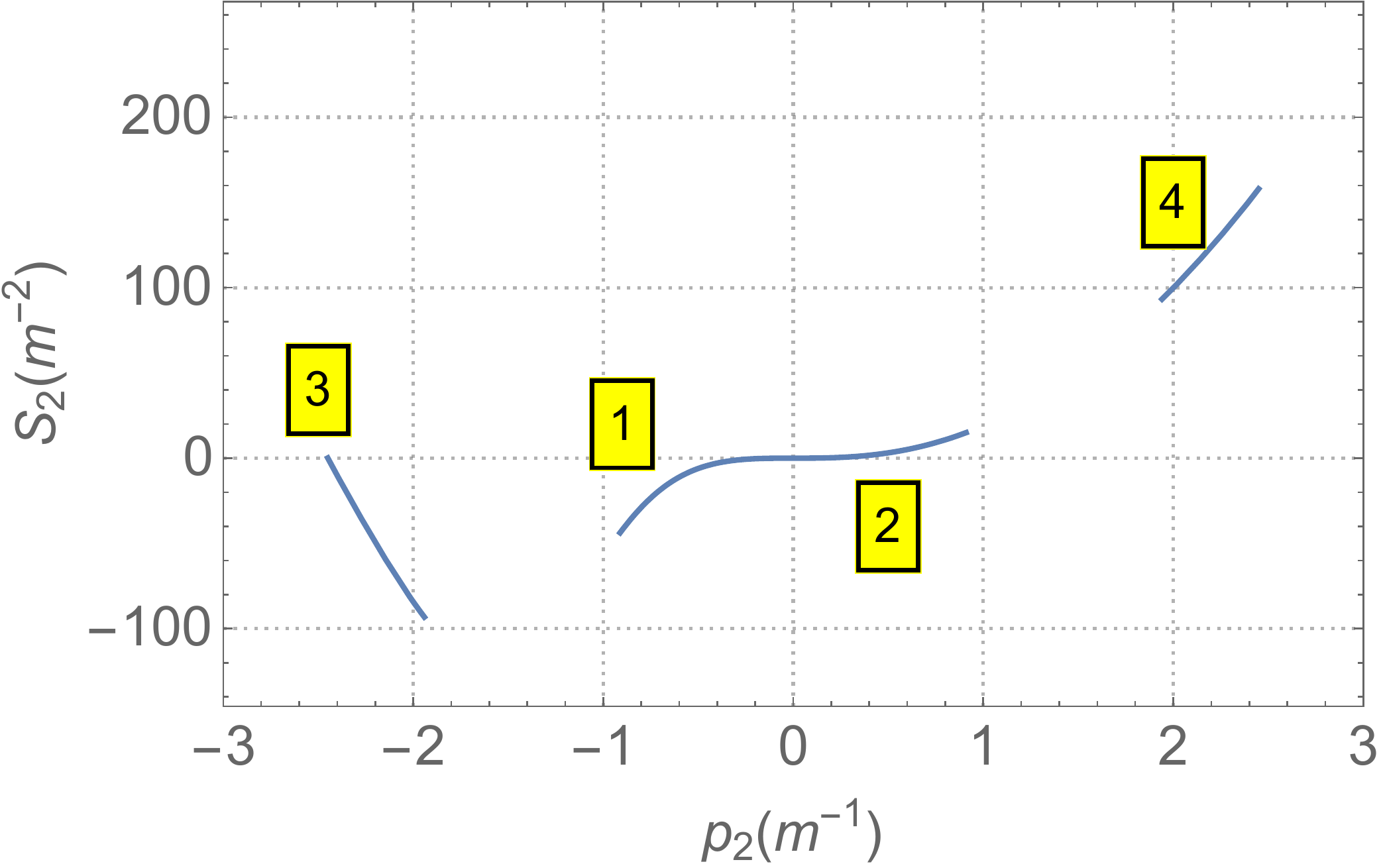}
\caption{The sextupole strengths of the four solutions versus the second quadrupole strength.}
\label{fig:sextupoles-TME}
\end{figure}

FIG.~\ref{fig:chromaticity-TME} presents the natural chromaticity of all solutions. Solutions 1 and 2 possess $\left|\xi_{x,y}\right|<1$, while solutions 3 and 4 have either horizontal or vertical chromaticity's absolute value exceeding 2, which results in strong sextupoles compensating this chromaticity. It is confirmed on FIG.~\ref{fig:sextupoles-TME} showing the integrated sextupole strengths $s_{1,2}$ as a function of $(p_2)$.
The 4th solution's sextupoles are significantly greater than of the 1st and the 2nd solutions. For typical value $s_{1,2}=100$~m$^{-2}$ of the fourth solution with sextupole length $l=0.2$~m, the second gradient is $B''=5000$~T/m$^2$, which is a large and not feasible value.

\begin{figure}[htb]
\centering
\includegraphics*[width=.48\columnwidth,angle=0,trim=0 0 0 0,clip]{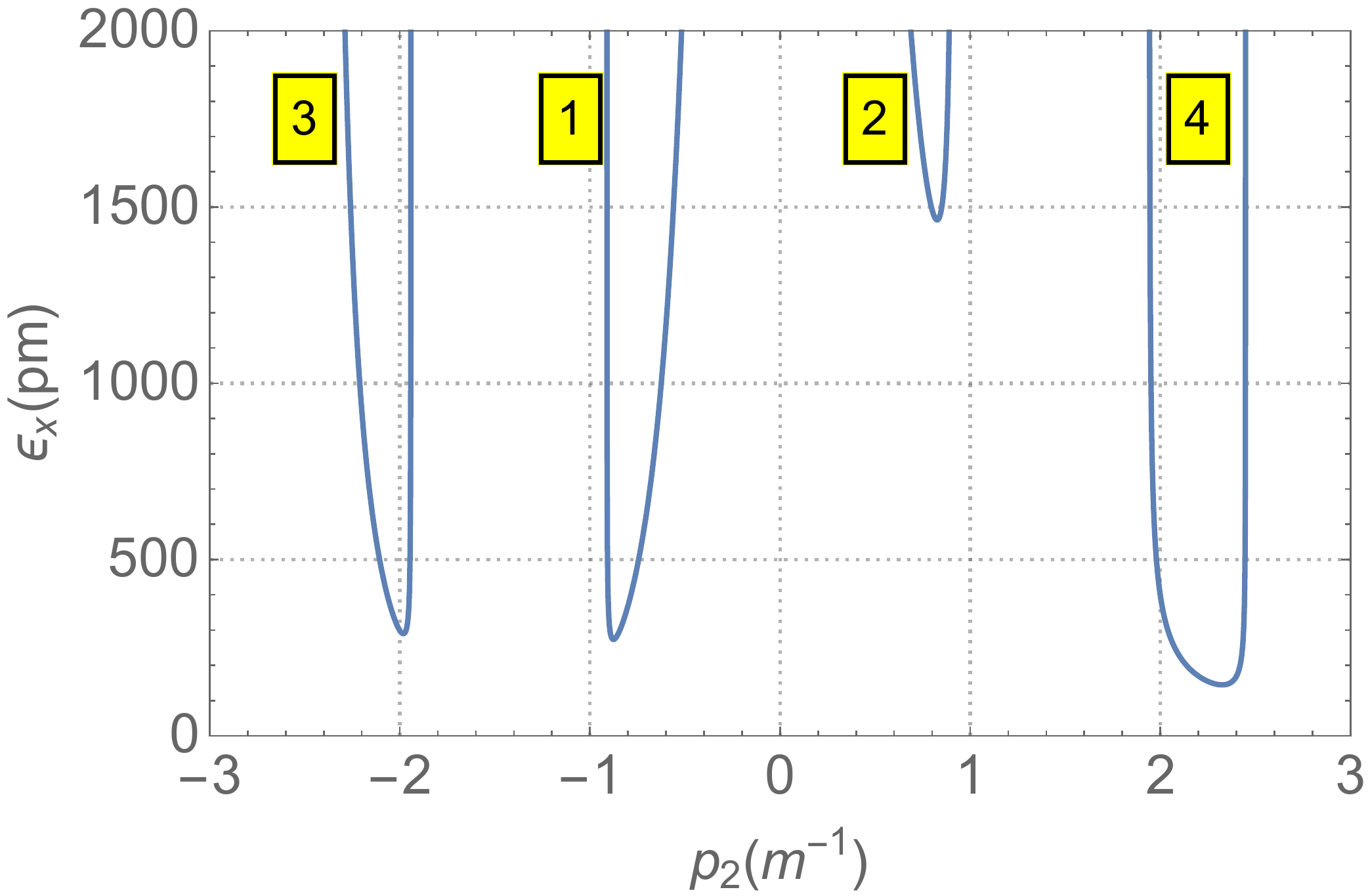}
\includegraphics*[width=.48\columnwidth,angle=0,trim=0 0 0 0,clip]{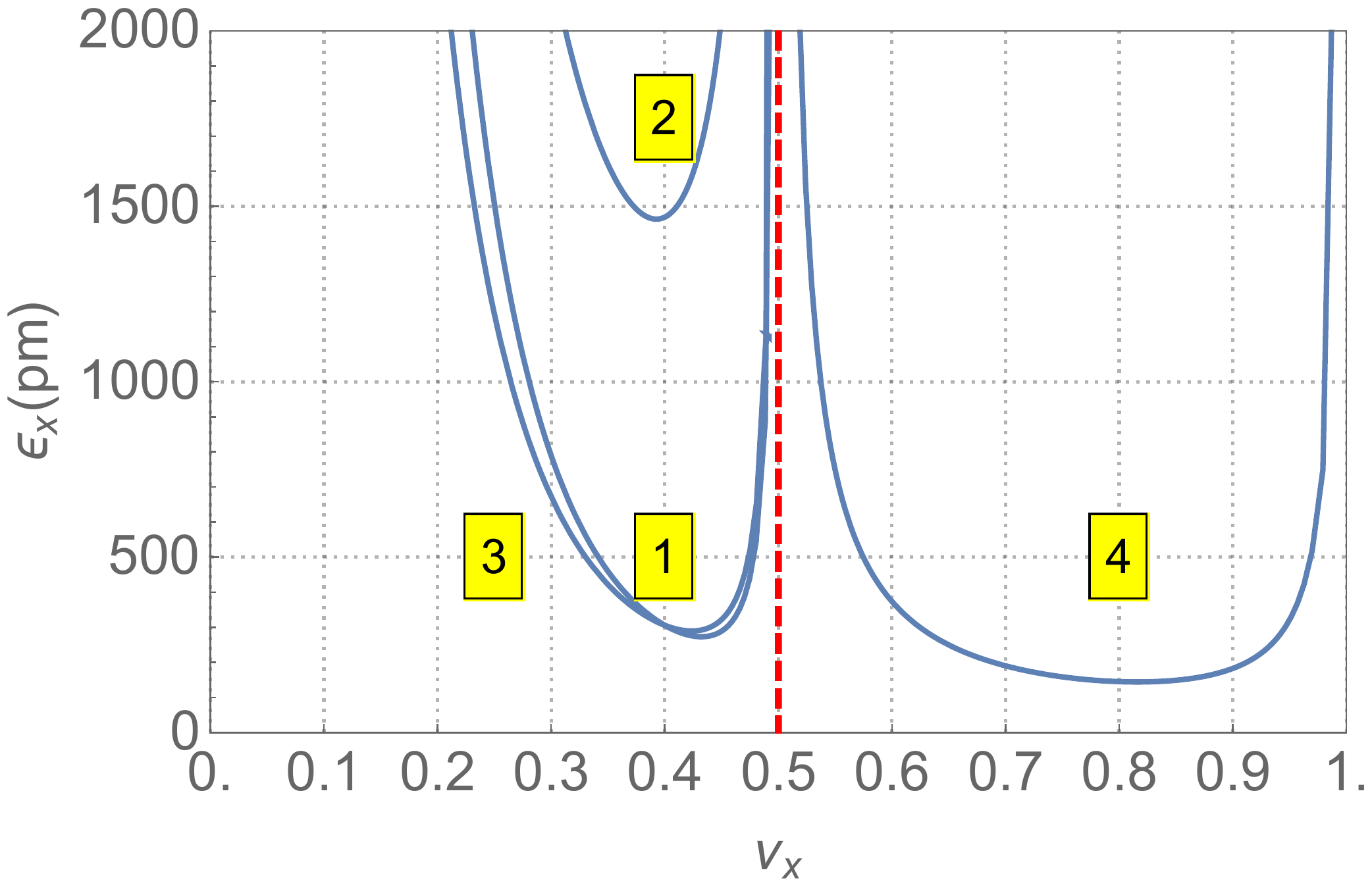}
\caption{Horizontal emittance as a function of the second quadrupole's strength (left) and as a function of tune (right).}
\label{fig:emittance-TME-1}
\end{figure}

Finally, FIG.~\ref{fig:emittance-TME-1} shows emittance of the four solutions as a function of the second quadrupole strength and the tune. Solution 2 indeed possesses too large emittance with respect to solution 4 (TME) and, therefore, unfit for the magnetic lattice of SKIF.
At the same time, the slightly different solutions 1 and 3, having two times larger emittance than TME, offer an advantage in lower sextupole and quadrupole strengths (solution 1), and therefore, in expected dynamic aperture. It is possible to lower their emittance by slightly reducing the bending angle (resulting in the perimeter increase). These changes are not significant due to the strong emittance dependance on $\varphi^3$. Thus, emittance reduction by factor two requires reduction of the bending angle less than 30\% ($\sqrt[3]{2}\approx 1.26$).

TABLE~\ref{tbl:TME-four-solutions} summarizes the main parameters of the four solutions. We chose the horizontal betatron tune in the vicinity of the emittance minimum (FIG.~\ref{fig:emittance-TME-1}). The vertical tune was calculated from the linear map with the condition of $\cos(2\pi\nu_x)=\cos(2\pi\nu_y)$.
\begin{table}[htb]
\caption{
\label{tbl:TME-four-solutions}
Main parameters of the four solutions
}
\begin{ruledtabular}
\begin{tabular}{lcccc}
Solution							& $1$				& $2$			& $3$			& $4$			\\
\colrule
Energy, $E$ [GeV]						& \multicolumn{4}{c} {3}										\\
$d_1$ [m]							& \multicolumn{4}{c} {1.0974}							 		\\
$L$ [m]							& \multicolumn{4}{c} {0.65}							 		\\
$L_c=2(L+d_1)$ [m]					& \multicolumn{4}{c} {3.4948}							 		\\
Circumference, $\Pi$ [m]					& \multicolumn{4}{c} {342.49}						 			\\
Number of cells, $N_c$					& \multicolumn{4}{c} {98}								 		\\
Cell's bending angle, $\varphi_c$				& \multicolumn{4}{c} { $\pi/49$}						 			\\
Dipole field, $B$ [T]						& \multicolumn{4}{c} {0.493}							 		\\
$\varepsilon_{TME}$ [pm]					& \multicolumn{4}{c} {75}								 		\\
$\nu_x/\nu_y$						& $0.4/0.4$			& $0.4/0.4$		& $0.4/0.6$		& $0.74/0.26$	\\
$\varepsilon_x$ [pm]					& $305.5$			& $1470.5$		& $305.6$		& $163.5$		\\
Momentum compaction 					&				&			&			&			\\
factor, $\alpha\cdot 10^{-4}$				& $3.3$			& $11.39$		& $3.16$		& $1.95 $		\\
Natural chromaticity, $\xi_x$				& $-1.36$			& $-0.61$		& $-3.43$		& $-1.36$		\\
Natural chromaticity, $\xi_y$				& $-0.61$			& $-1.36$		& $-1.46$		& $-2.04$		\\
Beta functions at the 					&				&			&			&			\\
dipole center, $\beta_{x0}$ [m]				& $0.39 $			& $4.69$		& $0.32$		& $0.11 $		\\
Dispersion function at the 					&				&			&			&			\\
dipole center, $\eta_{x0}$ [m]				& $0.014$			& $0.059$		& $0.014$		& $0.07$		\\
$p_1$ [m$^{-1}$]						& $1.27$			& $-1.27$		& $1.43$		& $-1.38$		\\
$p_2$ [m$^{-1}$]						& $-0.84$			& $0.84$		& $-2.01$		& $2.22$		\\
$s_1$ [m$^{-2}$]						& $15.35$			& $-31.39$		& $12.71$		& $-138.74$		\\
$s_2$ [m$^{-2}$]						& $-33.59$			& $12.14$		& $-83.04$		& $126.26$		\\
\end{tabular}
\end{ruledtabular}
\end{table}

TABLE~\ref{tbl:TME-four-solutions} shows that solution 2 emittance is about 10 times larger than solution 4, but emittances of solutions 1 and 3 are larger only by factor 2. Solution 4 has the lowest emittance, but sextupoles are too strong and are expected to provide a small dynamical aperture. The negative sextupole of solution 3 is significantly larger than of solution 1.

The magnetic lattice of the cell providing minimum emittance has been extensively studied \cite{Teng:1984cz,Teng:1985gm,Lee:1991st,Antoniou:2013uva,Jiao:2011zza,Cai:2018bvb,Riemann:2018kga}, including solution 1. In \cite{Jiao:2011zza} solution 1 was named modified TME (mTME), moreover, it was mentioned that betatron tune for mTME is $\nu_x < 0.5$, while for the true TME $\nu_x>0.5$. However, none of the studies scrutinized the solution 1 in regard of feasible quadrupoles' and sextupoles' strengths, and dynamic aperture.

Hence, we will abandon solutions 2 and 3 and will examine solution 1 (according to \cite{Jiao:2011zza} to be called mTME) as a candidate for the basic cell of SKIF, and solution 4 for comparison (to be called TME).

\section{Scrutiny of \lowercase{m}TME and TME}
\begin{figure}[htb]
\includegraphics*[width=.48\columnwidth,angle=0,trim=0 0 0 0,clip]{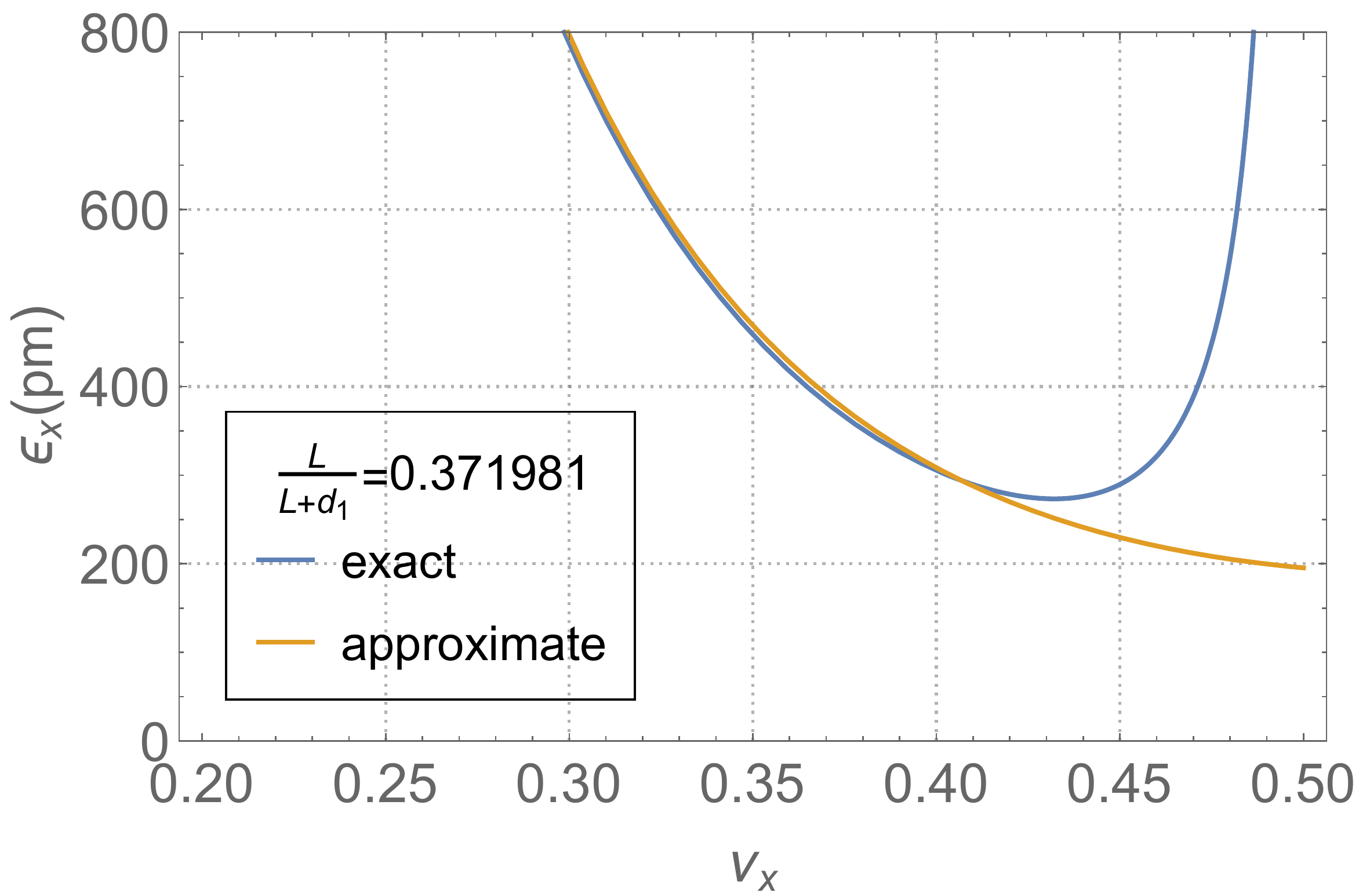}
\includegraphics*[width=.48\columnwidth,angle=0,trim=0 0 0 0,clip]{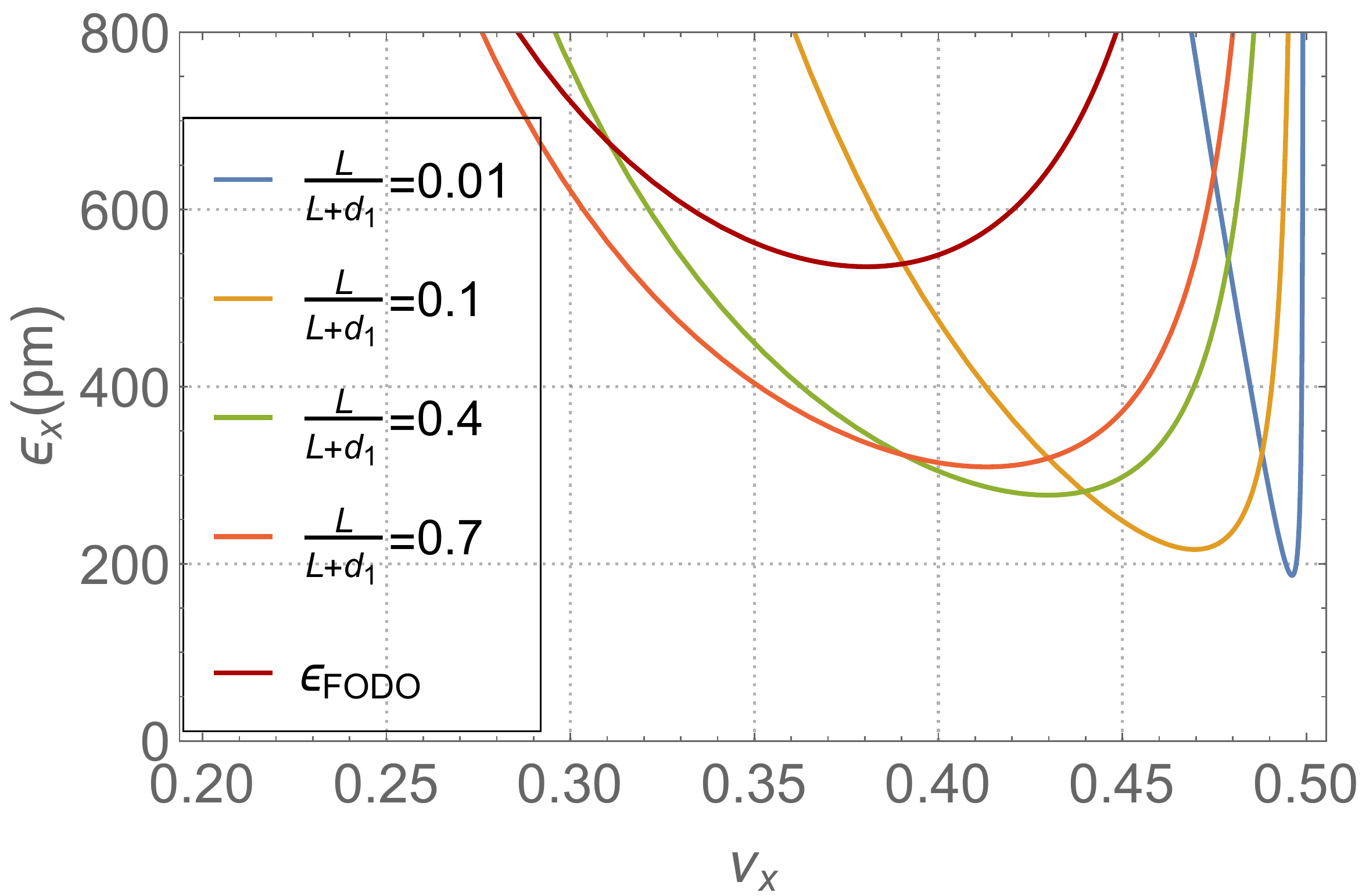}
\caption{Exact and approximate mTME emittance (left) as a function of tune. The mTME emittance for different ratios of the magnet and cell lengths (right). The red curve is FODO emittance. }
\label{fig:emittance-TME-14}
\end{figure}

Derivation of factor $F$ \eqref{eq:TME-F} does not use any assumptions about the magnetic cell. However, for the examined lattice (FIG.~\ref{fig:TME-layout}) factor $F$ depends on not only the betatron tune but also on lengths $L$ and $d_1$. Introducing the ratio of the dipole length to the cell length
\begin{equation}
u=\frac{2L}{2(L+d_1)},
\end{equation}
we derived the function $F_{mTME}(\nu_x,u)$ with the help of the computer algebra system WOLFRAM MATHEMATICA \cite{Mathematica}. Obtained expression agrees with computer simulation, however, it is cumbersome for analysis. The series expansion in $\mu_x=2\pi\nu_x$ with assumption $u<1$ is
\begin{equation}
\label{eq:F-factor-expanded}
\begin{split}
F&=\frac{8}{u\mu_x^3}+\frac{f_{-2}(u)}{\mu_x^2}+\frac{f_{-1}(u)}{\mu_x} +f_0(u) \\
&\qquad+\mu_x f_1(u) +\mu_x^2 f_2(u)+\mu_x^3 f_3(u)+\dots,
\end{split}
\end{equation}
where $f_n(u)$ are polynomials of u. This expression is also inconvenient for analysis because with the small number of terms, it poorly describes the exact solution, and with the large number of terms, it becomes cumbersome. The left graph of  FIG.~\ref{fig:emittance-TME-14} shows an example of exact and approximate (up to $\mu_x^3$ as in \eqref{eq:F-factor-expanded}) description of mTME emittance with $u\approx0.372$ as a function of horizontal tune. Note that with $\mu_x\to 0$ mTME emittance grows as $\mu_x^{-3}$. The right plot on FIG.~\ref{fig:emittance-TME-14} presents emittance of mTME cell for different ratios of the magnet and cell lengths; it also displays the emittance curve for FODO cell with identical cell length and bending angle. It is evident that for realistic values of $u$, emittance of mTME cell is significantly lower than one of FODO.

We did not find an analytical expression for optimum ratio of the dipole and drift lengths providing the minimum emittance of mTME cell and minimum value of $F_{mTME}$ similar to \eqref{eq:TME-F}. It is safe to assume from FIG.~\ref{fig:emittance-TME-14} that minimal emittance is reached for $u\to 0$ and $\nu_x\to 0.5$.

\begin{figure}[htb]
\includegraphics*[width=.48\columnwidth,angle=0,trim=0 0 0 0,clip]{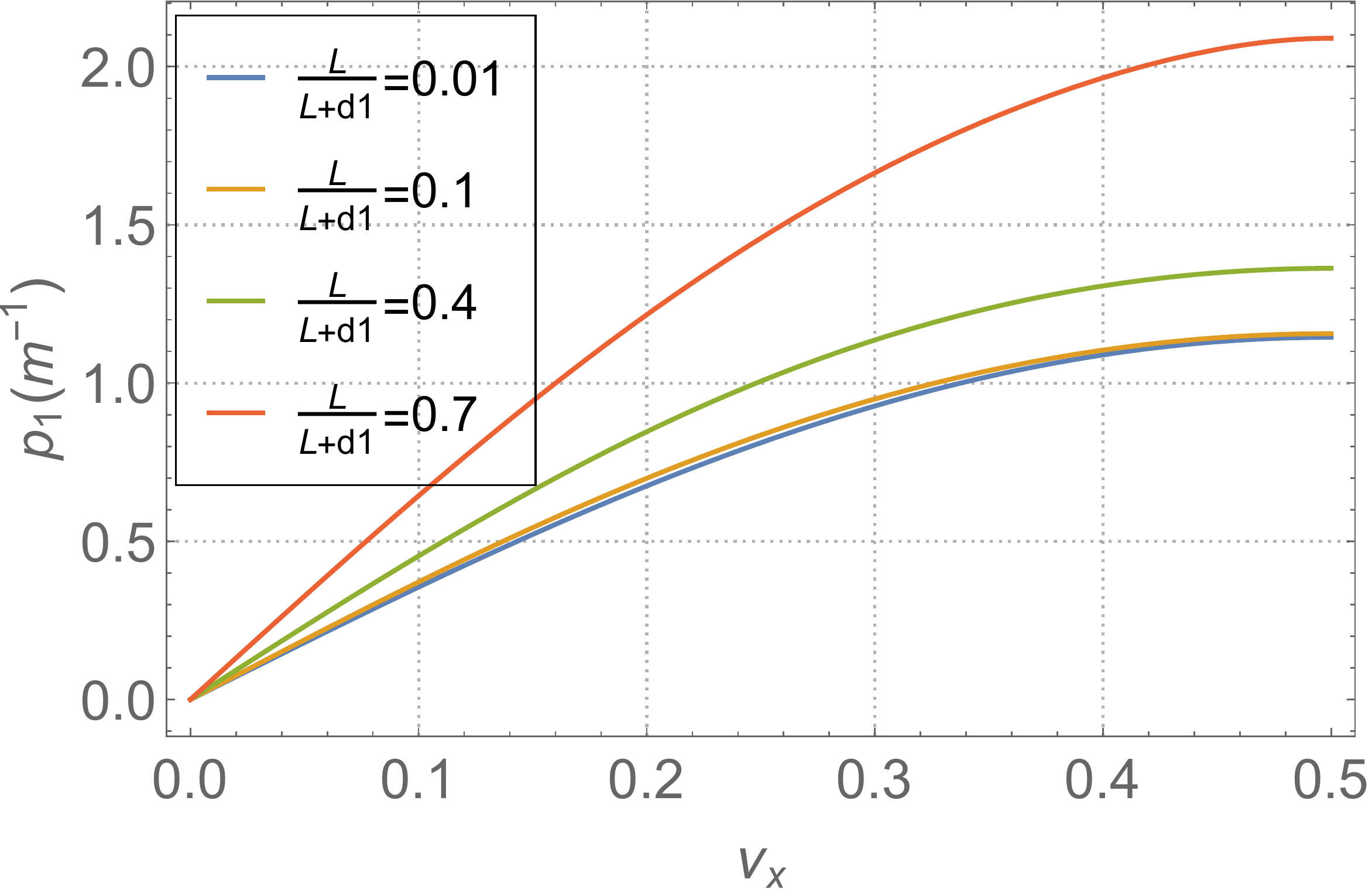}
\includegraphics*[width=.48\columnwidth,angle=0,trim=0 0 0 0,clip]{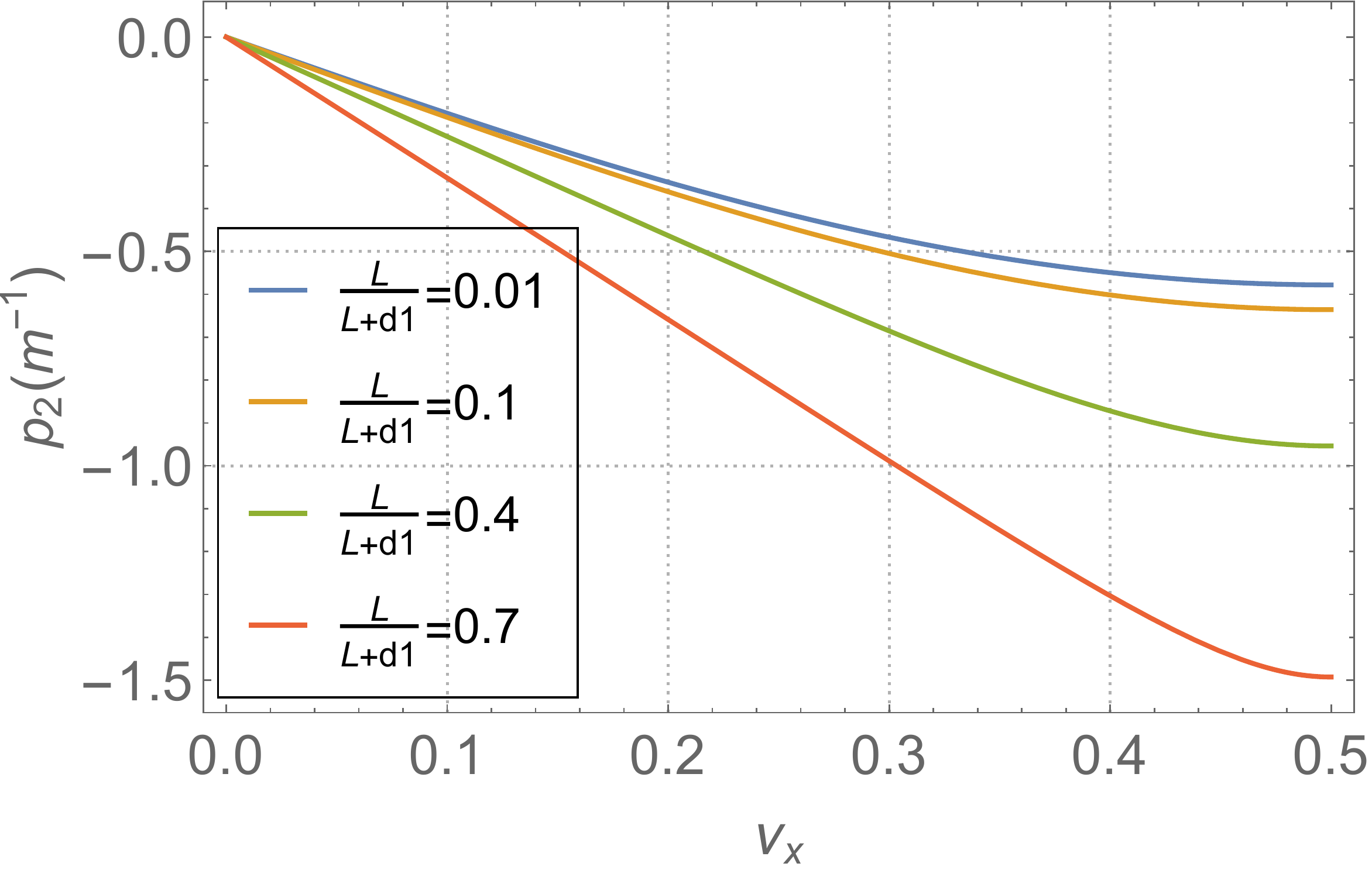}
\caption{Quadrupoles' strengths of mTME cell as a function of horizontal tune.}
\label{fig:quadrupoles-TME-14}
\end{figure}

\begin{figure}[htb]
\includegraphics*[width=.48\columnwidth,angle=0,trim=0 0 0 0,clip]{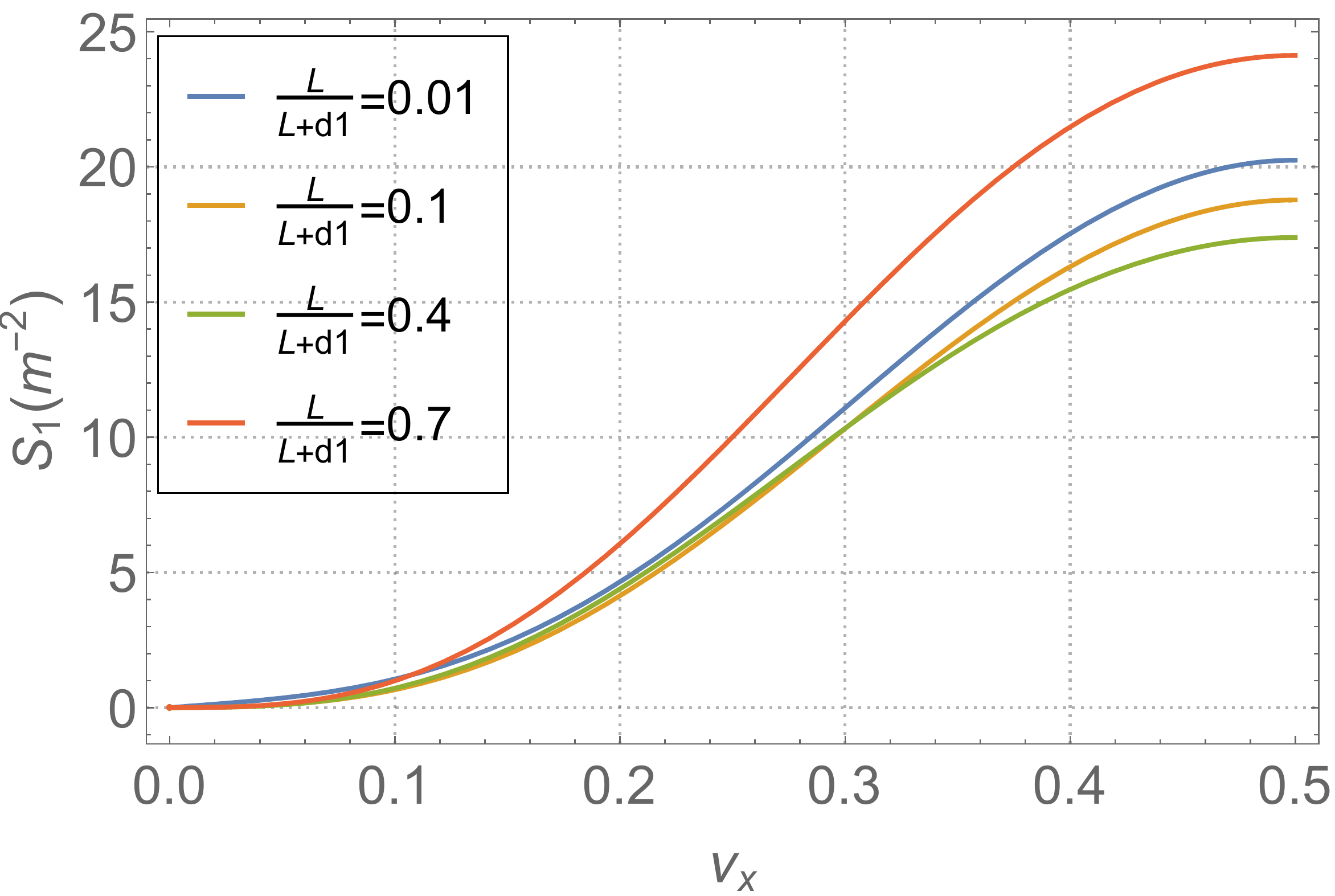}
\includegraphics*[width=.48\columnwidth,angle=0,trim=0 0 0 0,clip]{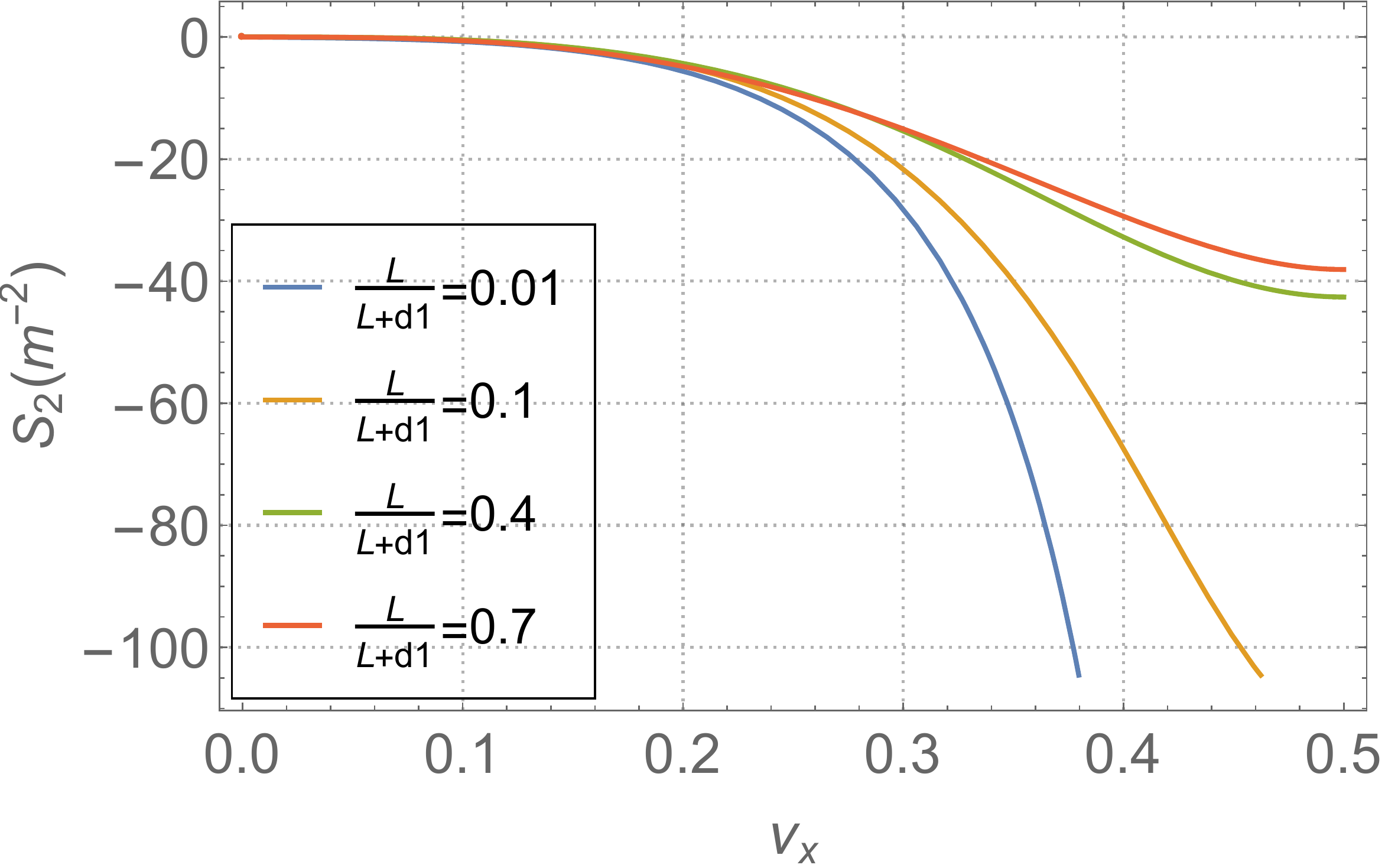}
\caption{Sextupoles' strengths of mTME cell as a function of horizontal tune.}
\label{fig:sextupoles-TME-14}
\end{figure}

However, the price of emittance minimization is the growth of quadrupoles' (FIG.~\ref{fig:quadrupoles-TME-14}) and sextupoles' strengths (FIG.~\ref{fig:sextupoles-TME-14}).

\begin{figure}[htb]
\centering
\includegraphics*[width=.7\columnwidth,angle=0,trim=0 0 0 0,clip]{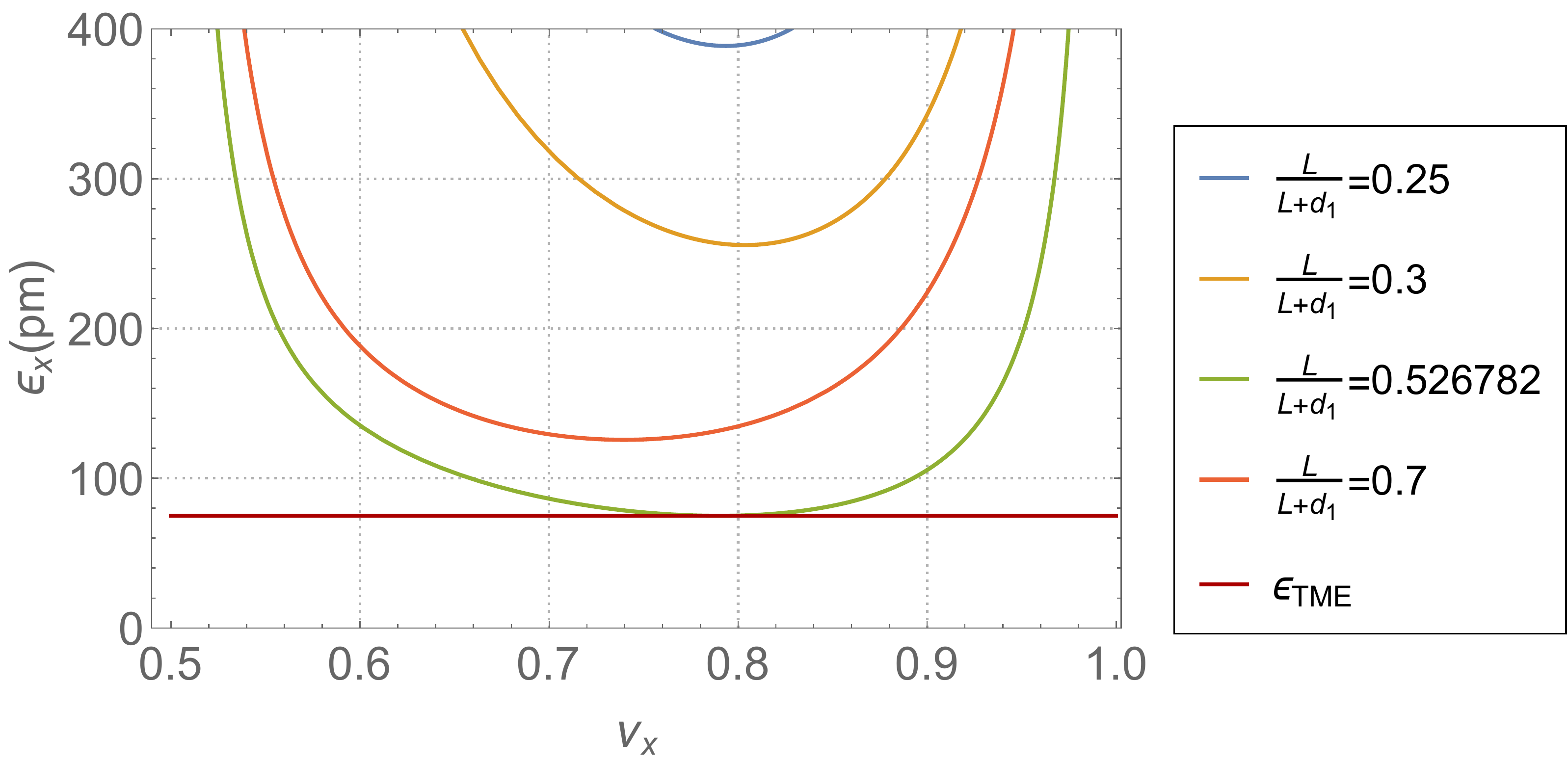}
\caption{TME cell emittance for different u. The red line indicates minimum emittance equal to 75~pm with parameters from TABLE~\ref{tbl:TME-four-solutions}.}
\label{fig:emittance-TME-sol4}
\end{figure}

For TME exists an analytical expression describing emittance dependance on $\nu_x$ and $u$ for the cell from FIG.~\ref{fig:TME-layout}, which also turns out to be cumbersome. However, the behavior of $F_{TME}(u)$ (FIG.~\ref{fig:emittance-TME-sol4}) differs from mTME.
The minimal value of emittance falls with the increase of $u$ (ratio of the dipole to cell lengths), whereas the position of the minimum in variable $\nu_x$ does not change. Numerical solution of the equation for minimum emittance gives expression \eqref{eq:TME-emittance} at $\nu_x=1-\arctan(\sqrt{3/5})\approx 0.79$ and $u\approx0.52678$. In other words, to achieve the TME emittance the dipole length must be approximately half the cell's length. Computer simulation of the cell with different ratios confirms the result.

The above conclusions are valid only for the thin-lens cell layout in FIG.~\ref{fig:TME-layout}, if one introduces a drift between quadrupole $q_2$ and dipole, or if one considers the realistic length of quadrupoles then numerical values will change.

In order to have a fair comparison of dynamic apertures of  mTME and TME cells we made two rings with approximately the same emittances and circumferences. In order to achieve that we lowered emittance of mTME (the first column of TABLE~\ref{tbl:TME-four-solutions}) to reach the value of TME (the fourth column of TABLE~\ref{tbl:TME-four-solutions}) by changing the bending angle from $\varphi_c=\pi/49$ to $\varphi_c=\pi/58$. Consequently, the number of cells increased from $98$ to $116$, and, with unchanged cell's length, it resulted in the total cells' length enlargement from 342.5~m to 405.4~m. Since we allowed circumference to enlarge for mTME cell, we treated TME cell in the same way, however, for emittance preservation, we decreased TME cell's phase advance, expecting that less rigid focusing would have relaxed sextupoles. At the end we compared the dynamic apertures of both rings by tracking.

\begin{figure}[htb]
\includegraphics*[width=.48\columnwidth,angle=0,trim=100 60 150 15,clip]{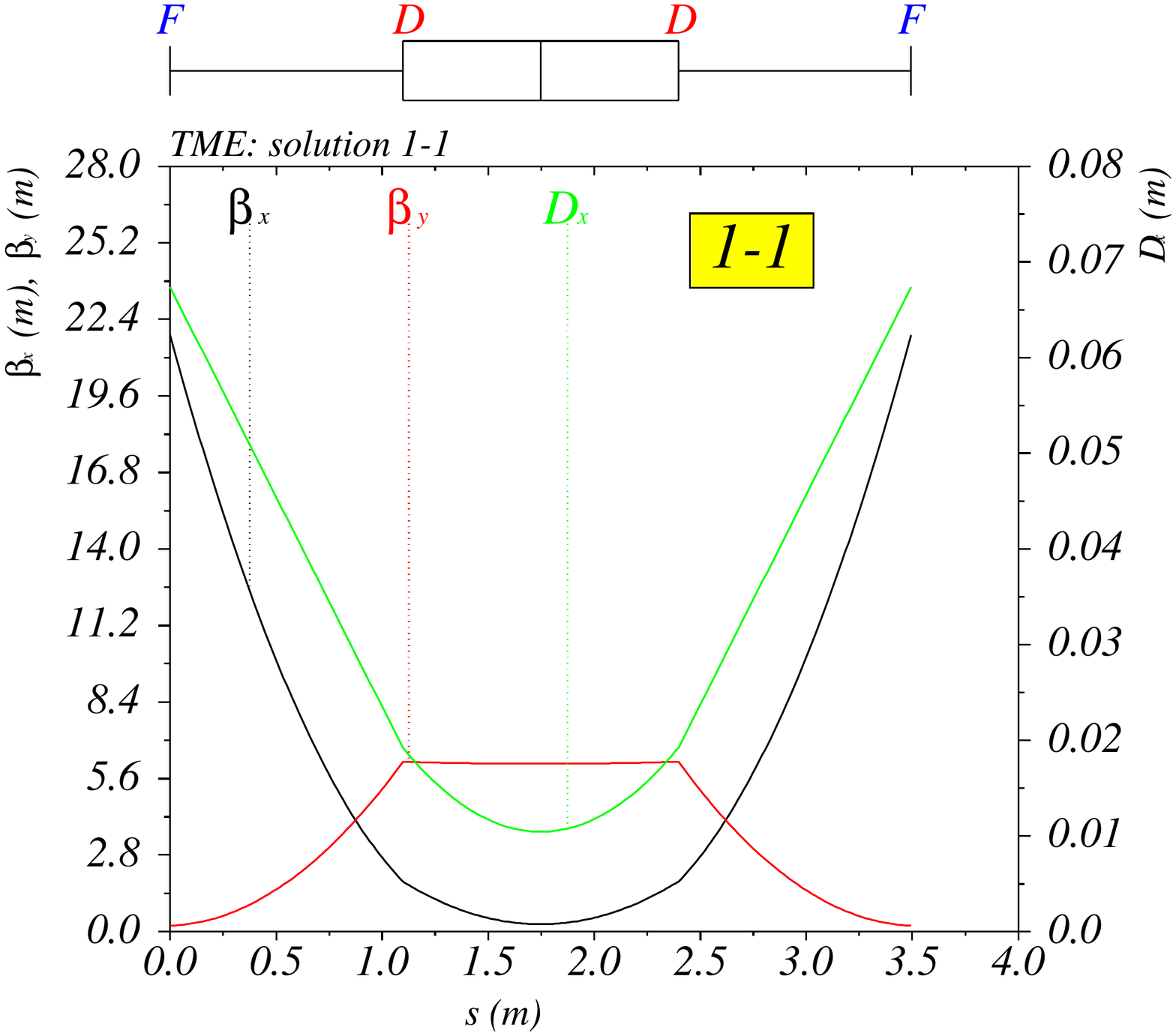}
\includegraphics*[width=.48\columnwidth,angle=0,trim=100 60 150 15,clip]{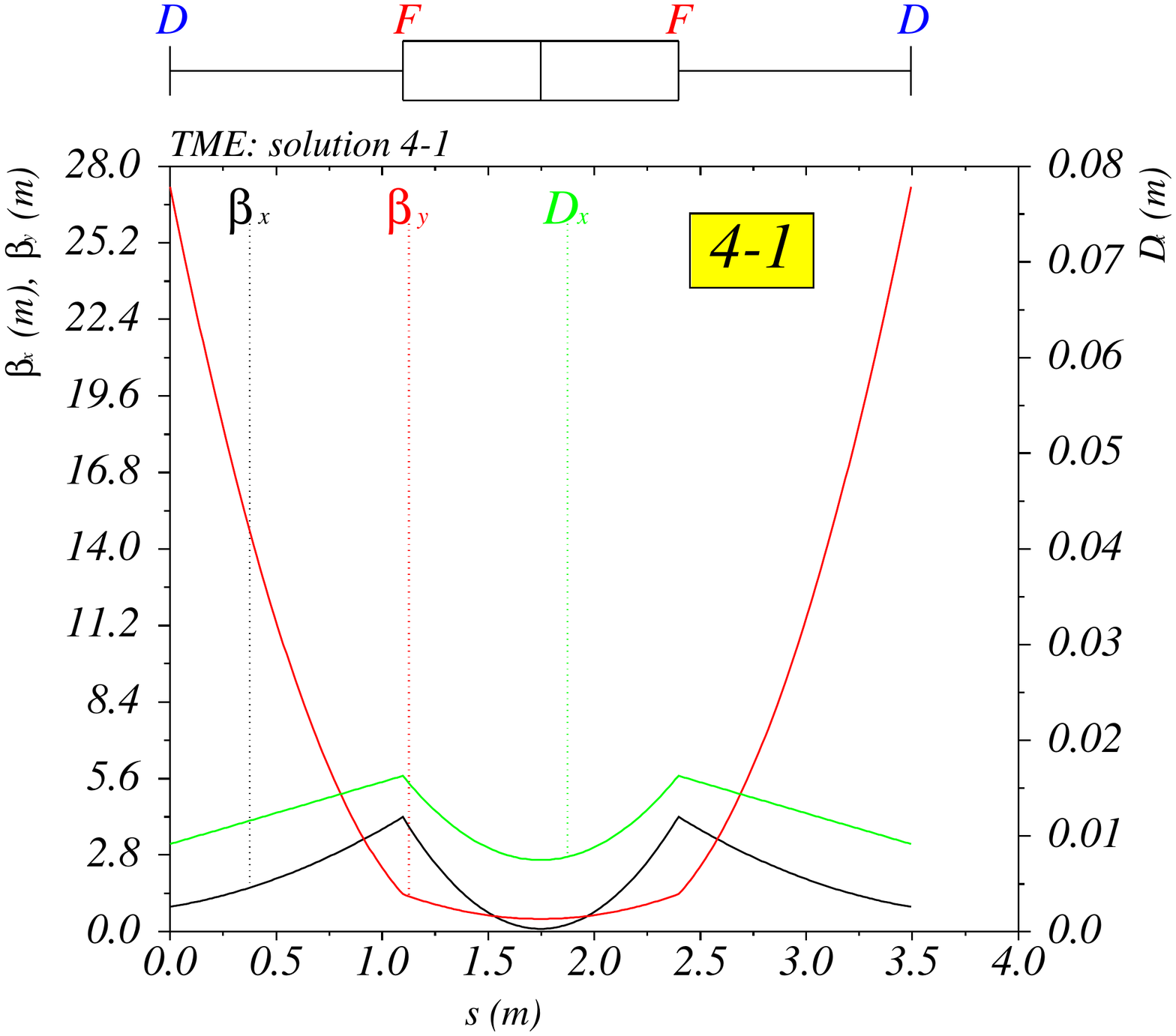}
\caption{Optical functions for model cells mTME  (left) and TME (right).}
\label{fig:optics-TME-solutions-1-4}
\end{figure}

\begin{figure}[htb]
\centering
\includegraphics*[width=.48\columnwidth,angle=0,trim=290 5 5 10,clip]{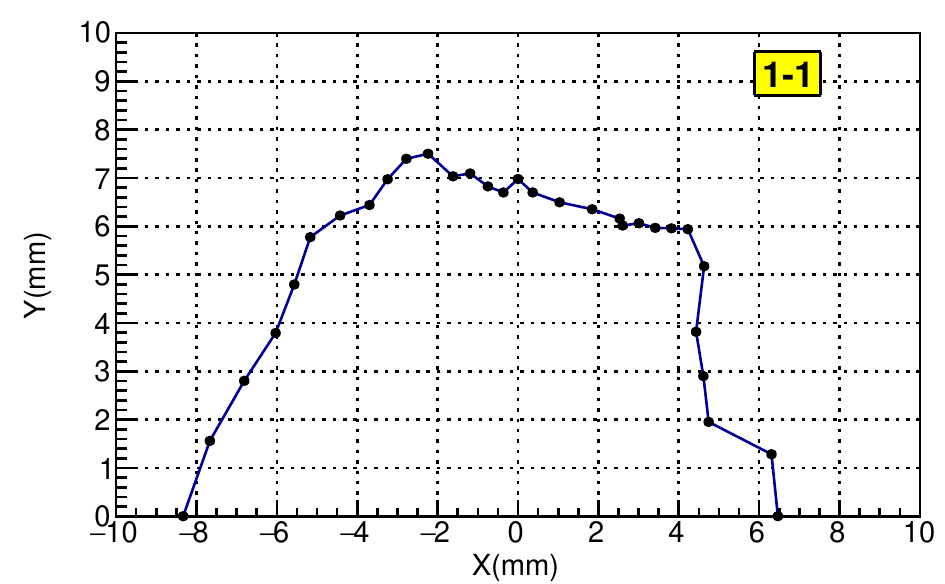}
\includegraphics*[width=.48\columnwidth,angle=0,trim=290 5 5 10,clip]{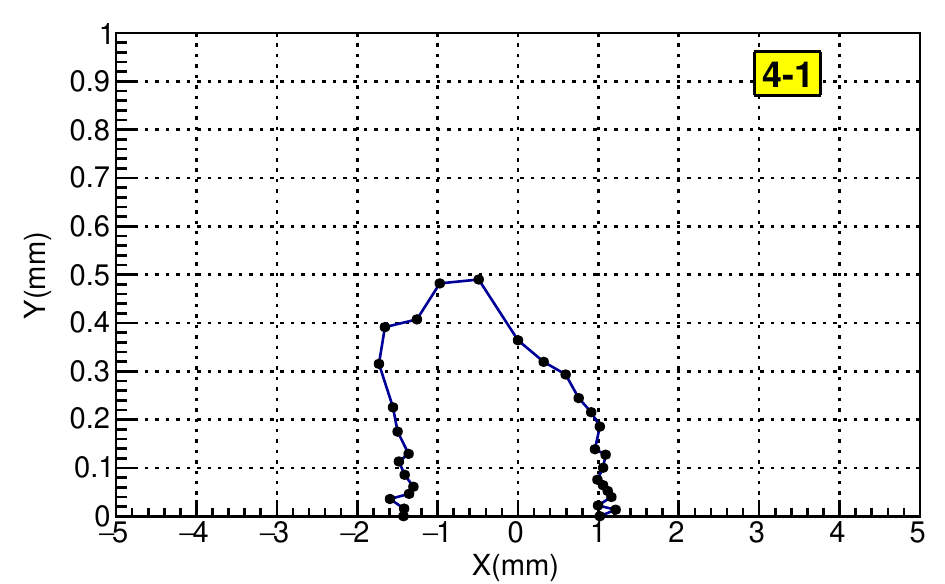}
\caption{Normalized to $\beta_x=15$~m, $\beta_y=5$~m dynamic apertures of mTME (left) and TME (right).}
\label{fig:da-norm-TME-solutions-1-4}
\end{figure}

The optical functions of the MAD-X models of described cells are shown on  FIG.~\ref{fig:optics-TME-solutions-1-4}.
TABLE~\ref{tbl:TME-solutions-1-4} presents main parameters of the cells, where energy, magnet and cell lengths are taken from TABLE~\ref{tbl:TME-four-solutions}. Now, energy, emittance and circumference of the two cells are identical; horizontal tune of TME cell is reduced, but sextupole strength is still several times larger than for mTME.
\begin{table}[htb]
\caption{
\label{tbl:TME-solutions-1-4} 
Main parameters of the model cells mTME and TME
}
\begin{ruledtabular}
\begin{tabular}{lcc}
Solution							& mTME 				& TME				\\
\colrule
Circumference, $\Pi$ [m]					& \multicolumn{2}{c} {$405.4$}					\\
Cell's number, $N_c$					& \multicolumn{2}{c} {$116$}					\\
Cell's bending angle, $\varphi_c$				& \multicolumn{2}{c} {$\pi/58\approx3.1^\circ$}		\\
$\nu_x/\nu_y$						& $0.43/0.43$			& $0.64/0.36$		\\
$\varepsilon_x$ [pm]					& $165$				& $161$			\\
Natural chromaticity, $\xi_x/\xi_y$			& $-2.0/-0.84$			& $-1.27/-2.63$		\\
$p_1/p_2$ [m$^{-1}$]					& $1.3/-0.87$			& $-1.42/2.06$		\\
$S_1/S_2$ [m$^{-2}$]					& $19.33/-45.45$			& $-155.14/126.62$	\\
\end{tabular}
\end{ruledtabular}
\end{table}
Dynamic aperture is found by usual definition as a set of initial conditions $\{x_0,x_0'=0,y_0,y_0'=0\}$ for particle surviving 1024 turns.
Since dynamic aperture depends on the values of the beta functions, we normalized both apertures to $\beta_x=15$~m, $\beta_y=5$~m (FIG.~\ref{fig:da-norm-TME-solutions-1-4}).
Noticing an unambiguous advantage of mTME cell in dynamic aperture we chose it as a unit cell of SR light source SKIF.

\section{Magnetic lattice of the light source SKIF}
\begin{figure}[htb]
\includegraphics*[width=.48\columnwidth,angle=0,trim=115 60 140 10,clip]{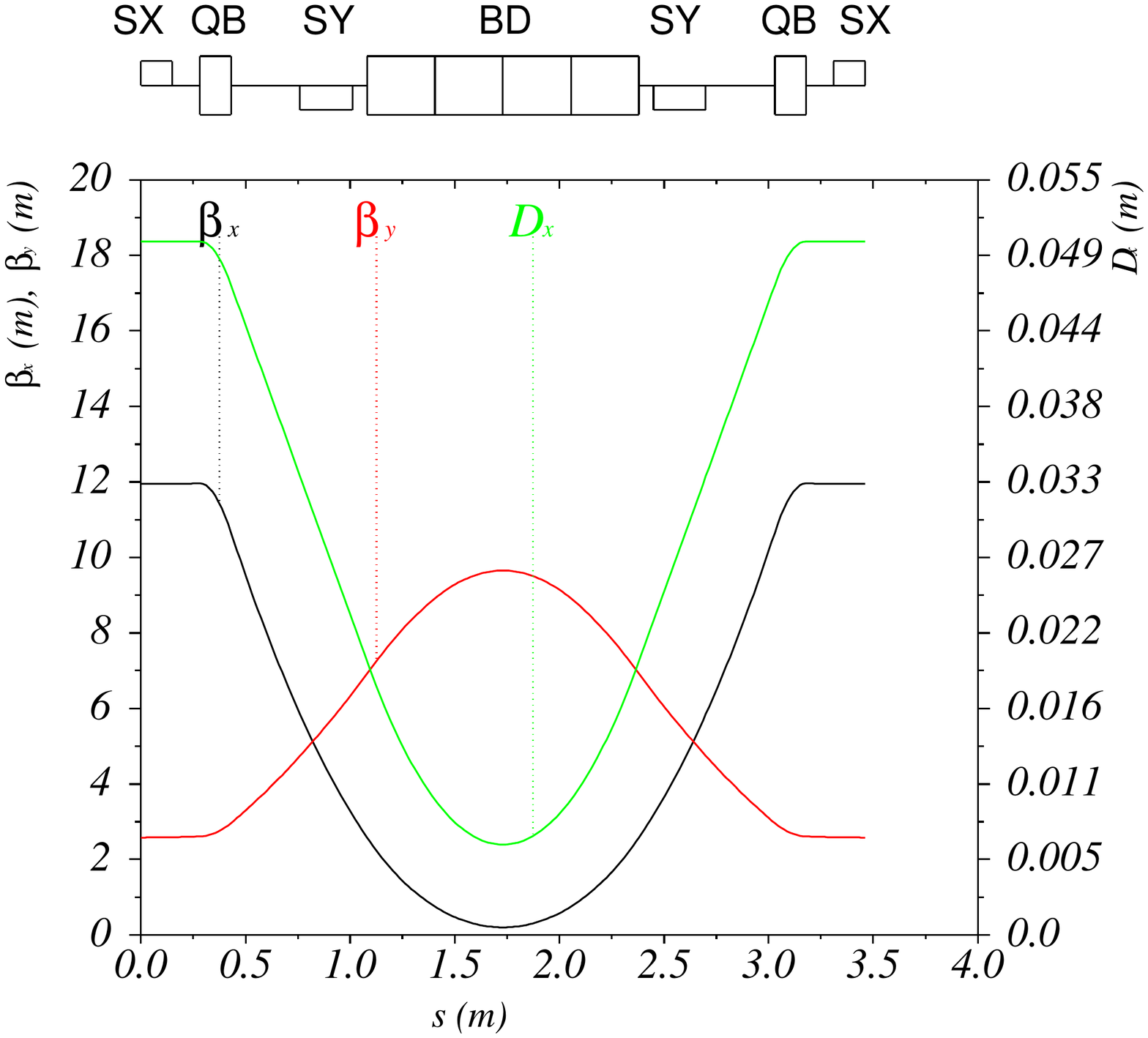}
\includegraphics*[width=.48\columnwidth,angle=0,trim=115 60 140 10,clip]{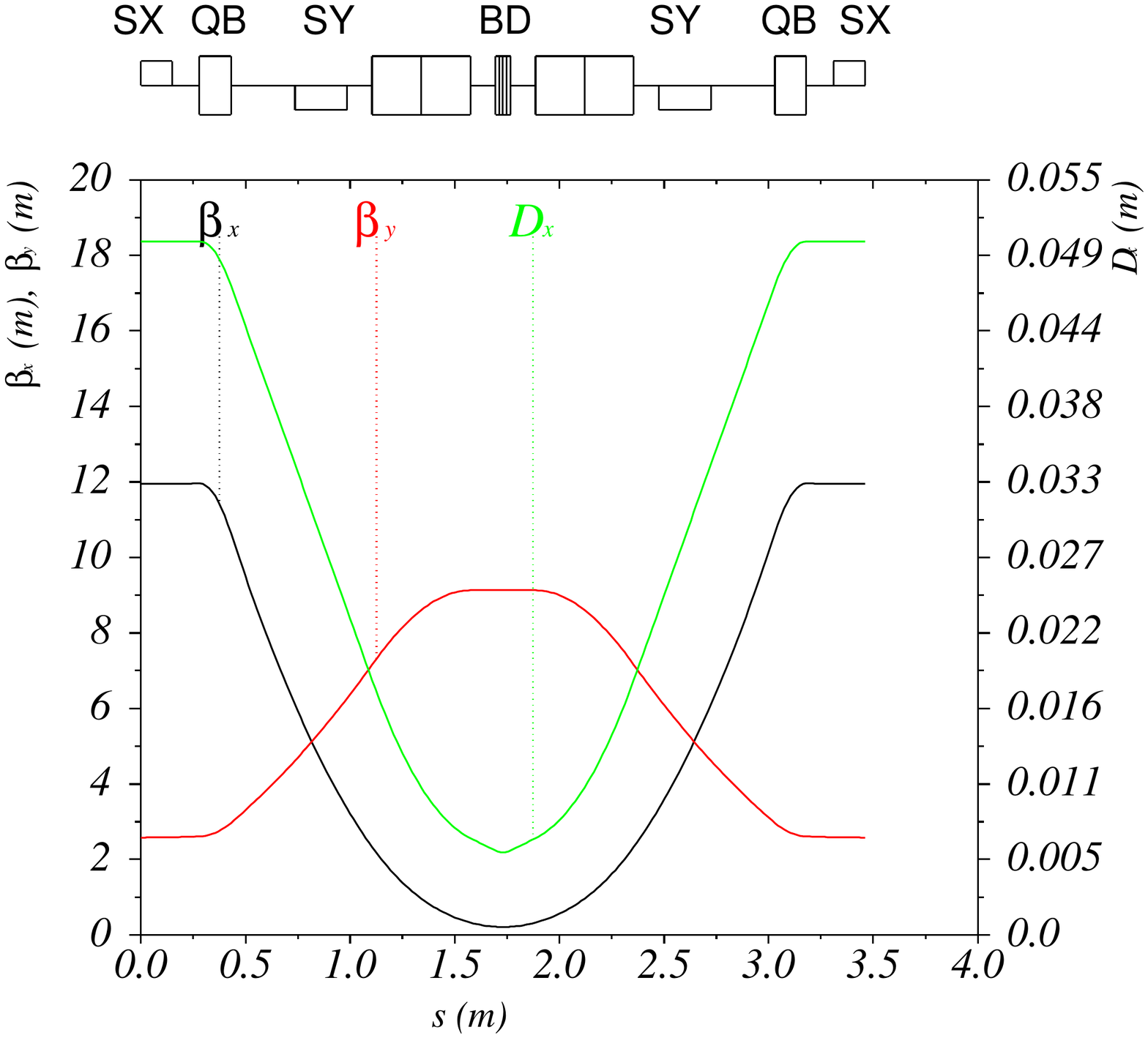}
\caption{Optical functions of SKIF base cell (left) and cell with high field dipole (right).}
\label{fig:SKIF-cell}
\end{figure}

The basic cell of the light source SKIF is mTME cell, but with some changes (left on FIG.~\ref{fig:SKIF-cell}):
\begin{itemize}
\item to minimize the cell's length, we transferred the gradient of the two quadrupoles $q_2$ into the bending magnet BD, which also changed damping number $J_x$;
\item to improve matching of the dispersion, we shifted the focusing quadrupoles horizontally, creating a combined function reverse bend QB \cite{Streun:2014gna}, which allowed to adjust damping number $J_x$;
\item we optimized strengths and relative position of the magnets to increase dynamic aperture, to minimize emittance and chromaticity, and adjust the horizontal damping number $J_x\approx 2$.
\end{itemize}

The cell possesses mirror symmetry relative to the center of BD dipole, which has a small negative gradient providing vertical focusing. Two quadrupoles QB focus the beam horizontally and being shifted in the horizontal plane, bend the beam in the opposite to the main dipole direction (reverse bend). The reverse bending helps with matching of the dispersion functions and emittance minimization. Two sextupoles SY and one sextupole SX (the cells start and end with SX) correct the chromaticity.

The small bending angle, necessary for emittance minimization, implies a weak magnetic field of the central dipole (0.55~T), disadvantageous for hard X-ray generation. To improve the situation, in the center cell of each super period, we replaced the weak BD dipole with the section of three dipoles consisting of two weak BD1 with small negative gradient and uniform BS dipole between them with the field of 2~T (right on FIG.~\ref{fig:SKIF-cell}). The remaining elements have slight differences from their counterparts of the regular cell. The length of the modified cell (HF -- high field) is identical to the regular one ($\approx3.5$~m, LF -- low field).  The distance between the elements in both cells is sufficient for the installation of vacuum, diagnostic and other equipment.

Now, we can compare the estimated earlier parameters of mTME (TABLE~\ref{tbl:TME-solutions-1-4}) with obtained parameters of the real cell (TABLE~\ref{tbl:SKIF-cell}).
\begin{table}[!htb]
\caption{
\label{tbl:SKIF-cell}
Parameters of SKIF LF and HF cells}
\begin{ruledtabular}
\begin{tabular}{lcc}
								& LF					& HF				\\
\colrule
Length, $L_c$ [m]						& \multicolumn{2}{c} {$3.4948$}					\\
Cells's bending angle, $\varphi_c$ [$^\circ$]		& $3.682$				& $3.642$			\\
$\nu_x/\nu_y$						& $0.446/0.124$			& $0.442/0.122$		\\
Momentum compaction					& 					& 				\\
 factor, $\alpha$						& $1.4\cdot 10^{-4}$		& $8.1\cdot 10^{-6}$	\\
$\varepsilon_x$ [pm]					& $70$				& $75$			\\
Energy spread, $\sigma_E/E$				& $8.73\cdot 10^{-4}$		& $1.3\cdot 10^{-3}$	\\
Energy loss per turn,					& 					& 				\\
 $U_0$ [keV]							& $4.72$				& $9.83$			\\
Damping numbers, $J_x/J_y$				& $2.24/0.76$			& $1.7/1.3$			\\
Natural chromaticity, $\xi_x/\xi_y$			& $-1.32/-0.36$			& $-1.27/-0.34$		\\
\end{tabular}
\end{ruledtabular}
\end{table}
To satisfy requirements on the number and length of straight sections, and the size of the facility, we increased the cell's bending angle up to $\varphi_c\approx 3.6^\circ$. Nevertheless, horizontal damping number $J_x\approx 2$ and reverse bends allowed achieving of $\varepsilon_x\approx 70$~pm. To relax vertical focusing and chromaticity, we decreased the vertical tune to $\nu_y\approx0.12$. Accordingly, the new chromaticity of SKIF cell became $\xi_{x,y}\approx-1.3/-0.4$. The integral normalized gradient in the LF cell is $(K_1L)_y\approx-1.3$~m$^{-1}$ (instead of $2p_2=-1.74$~m$^{-1}$), and $(K_1L)_x\approx0.8$~m$^{-1}$ (instead of $p_1\approx 1.3$~m$^{-1}$). The distinction is explained by different placement and realistic length of the quadrupoles. Estimation of sextupole integrated strengths $s_{1,2}\approx 20/-45$~m$^{-2}$ also differs from obtained $(K_2L)_{x,y}\approx 70/-59$~m$^{-2}$. This difference is because the model cell (on FIG.~\ref{fig:TME-layout}) has the sextupoles installed in the optimum places (where the corresponding beta reaches the maximum). In the real cell, sextupoles' position is a result of the dynamic aperture (transverse and longitudinal) optimization, where the phase advance between the sextupoles was varied, and, therefore, differs from the model's.

\begin{figure}[htb]
\includegraphics*[width=.95\columnwidth,angle=0,trim=115 60 140 10,clip]{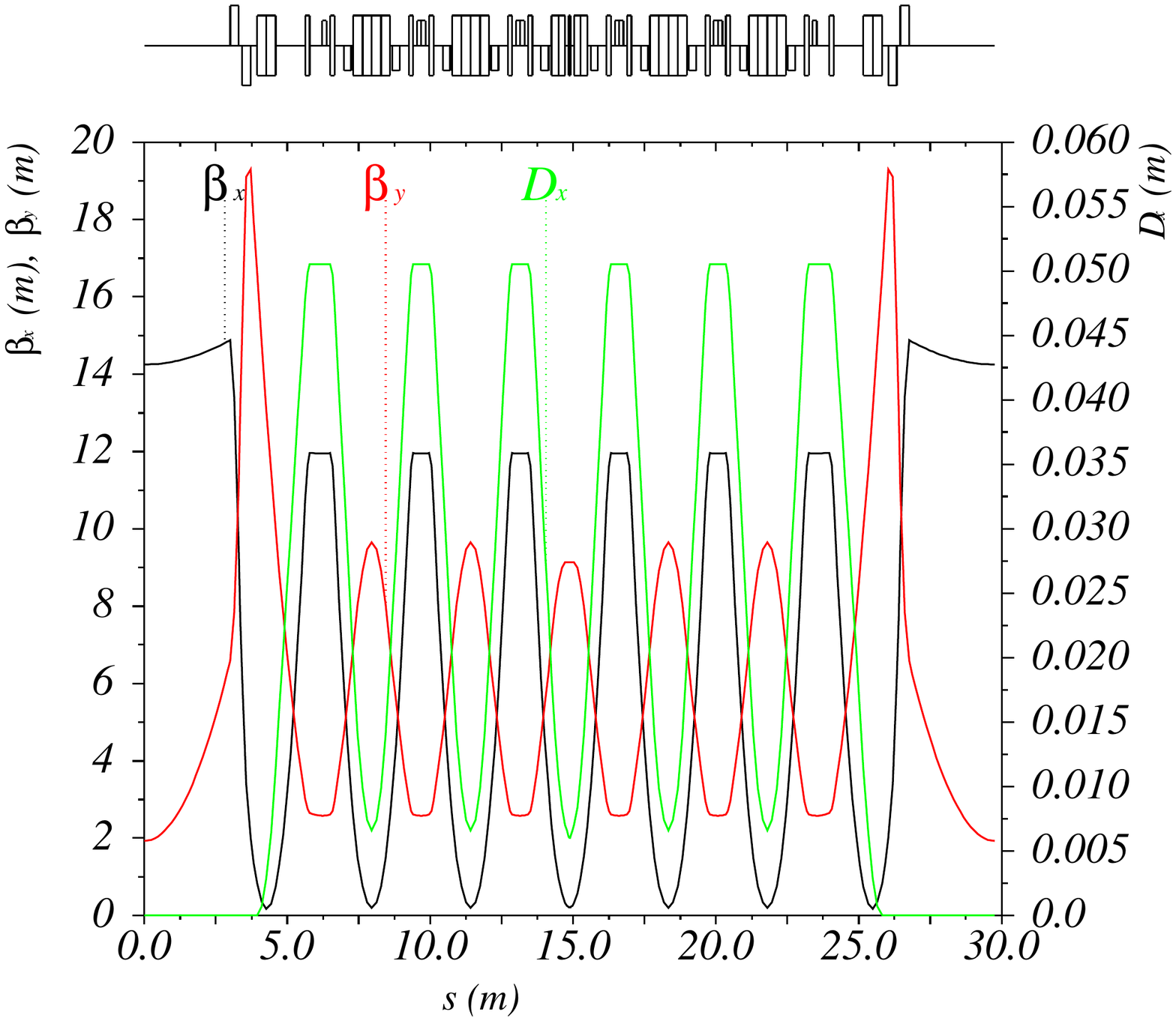}
\caption{Optical functions of SKIF super period, beginning and ending at the center of the straight section.}
\label{fig:SKIF-superperiod}
\end{figure}

SKIF's magnetic lattice consists of 16 super periods of 7BA type with five regular cells, described above, and two at the ends to cancel dispersion in the 6~m long straight section. Two doublets of quadrupoles provide sufficiently large $\beta_x$ (for horizontal injection optimization) and low $\beta_y$ (to minimize the influence of the vertical field insertion devices on beam dynamics). FIG.~\ref{fig:SKIF-superperiod} shows the whole lattice of the super period starting and ending in the middle of the straight section.
Magnetic lattice is rather simple and symmetrical, therefore, possesses a minimal number of structural resonances. Optical functions in the straight section (zero dispersion, high horizontal beta and low vertical) ensure traditional injection in the horizontal plane and are suitable for RF cavities and insertion devices -- undulators with a relatively low field. TABLE~\ref{tbl:SKIF-ring} covers the main parameters of the storage ring.
\begin{table}[!htb]
\caption{
\label{tbl:SKIF-ring}
Main parameters of SKIF lattice from FIG.~\ref{fig:SKIF-superperiod}
}
\begin{ruledtabular}
\begin{tabular}{lc}
Beam energy [GeV]				& $3$				\\
Symmetry					& $16$			\\
Circumference [m]				& $476.14$			\\
Revolution period [mks]			& $1.59$			\\
Horizontal emittance [pm]			& $72$			\\
Energy spread				& $1\cdot 10^{-3}$	\\
Energy loss per turn [keV]			& $535$			\\
Betatron tunes ($x/y$)			& $50.88/17.76$		\\
Momentum compaction factor		& $7.6\cdot 10^{-5}$	\\
Chromaticity ($x/y$)			& $-162/-58$		\\
RF harmonic					& $567$			\\
RF frequency [MHz]			& $357$			\\
RF voltage [MV]				& $0.77$			\\
Energy acceptance				& $2.6\%$			\\
Synchrotron tune				& $1.13\cdot 10^{-3}$	\\
Bunch length [mm]				& $5.5$			\\
Damping numbers ($x/y$)			& $1.91/1.09$		\\
Damping time ($x/y$) [ms]			& $9/16$			\\
\end{tabular}
\end{ruledtabular}
\end{table}

\begin{figure*}[htb]
\includegraphics*[width=.95\textwidth,angle=0,trim=0 0 0 0,clip]{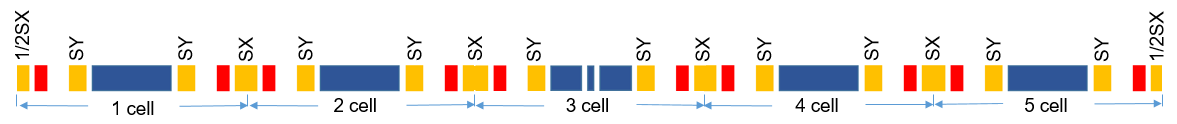}
\caption{Positioning of chromatic sextupoles.}
\label{fig:SKIF-sextupoles}
\end{figure*}

Two families of sextupoles in the regular cells compensate linear chromaticity of SKIF lattice (FIG.~\ref{fig:SKIF-sextupoles})

Since the cell begins and ends with horizontal sextupole, the five cell lattice has half-length sextupole at the ends. TABLE~\ref{tbl:SKIF-sextupoles} registers sextupole parameters.
\begin{table}[htb]
\caption{
\label{tbl:SKIF-sextupoles}
Main parameters of chromatic sextupoles
}
\begin{ruledtabular}
\begin{tabular}{lccc}
			& $l$, m		& $B''$,T/m$^2$		& $(K_2L)$, m$^{-2}$		\\
\colrule
SY			& $0.25$		& $-2370$					& $-59.25$		\\
SX			& $0.30$		& $ 2358$					& $ 70.74$		\\
\end{tabular}
\end{ruledtabular}
\end{table}

The size of the stable nonlinear motion (dynamic aperture) depends on the strength of betatron resonances and their position relative to the chosen tune point. Since it is impossible to evaluate theoretically the resonances' strengths (especially of the high orders) produced by sextupoles, the main way of dynamic aperture optimization is the numerical simulation of particle motion (tracking).

\begin{figure}[htb]
\includegraphics*[width=.48\columnwidth,angle=0,trim=25 360 450 15,clip]{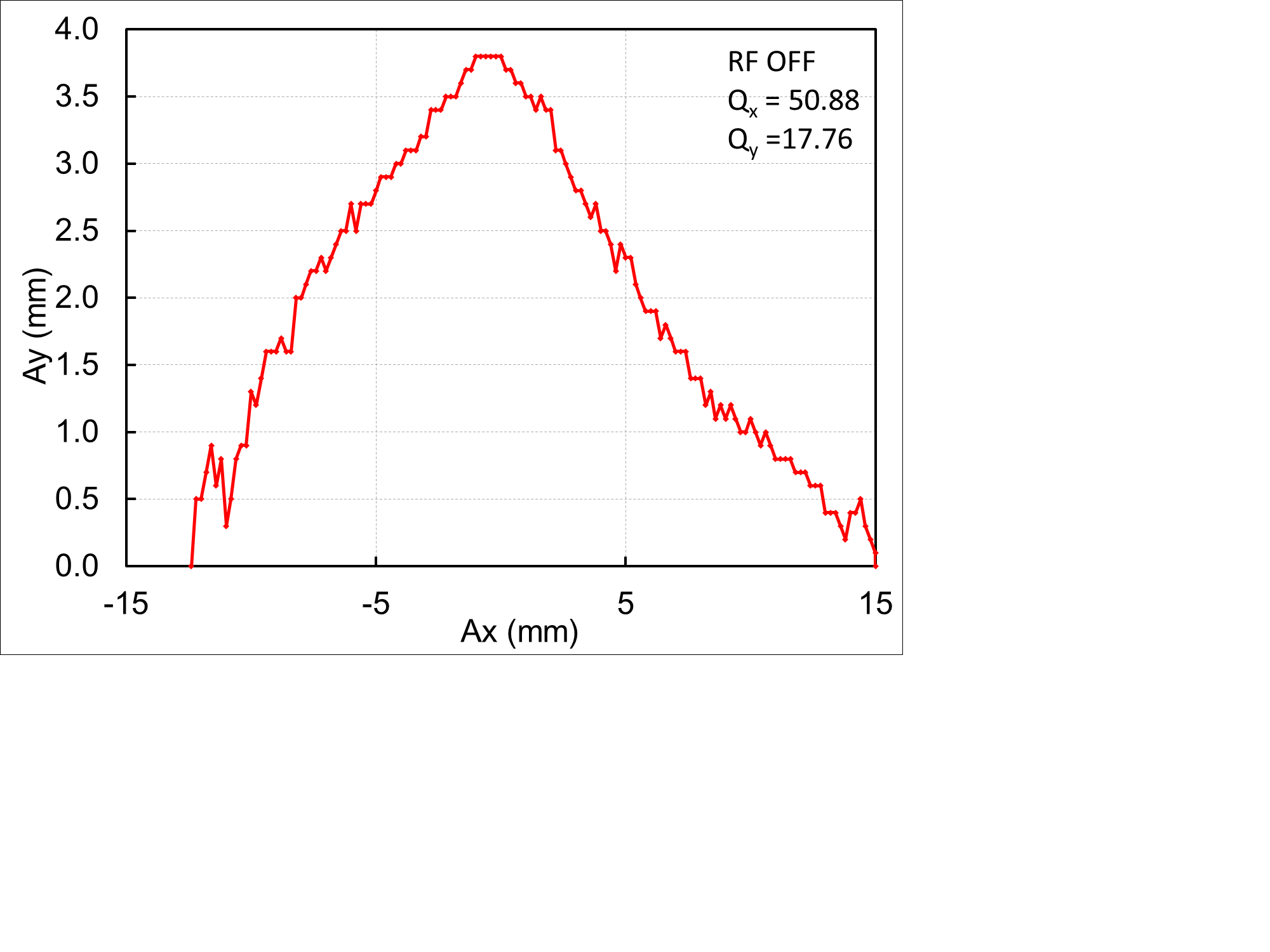}
\includegraphics*[width=.48\columnwidth,angle=0,trim=25 360 450 15,clip]{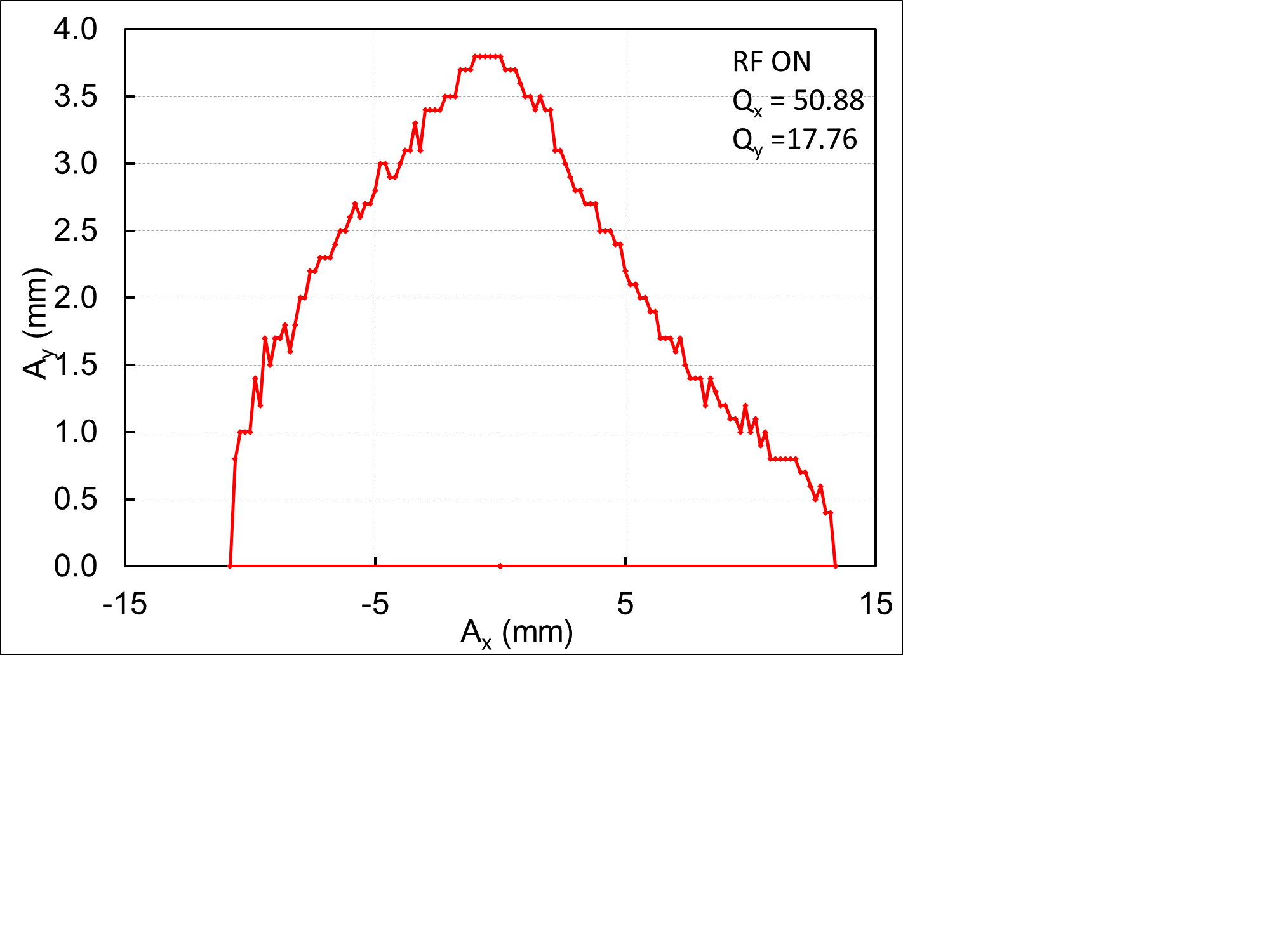}
\caption{Transverse dynamic aperture without (left) and with (right) synchrotron oscillations.}
\label{fig:SKIF-da}
\end{figure}

\begin{figure}[htb]
\includegraphics*[width=.48\columnwidth,angle=0,trim=25 360 450 15,clip]{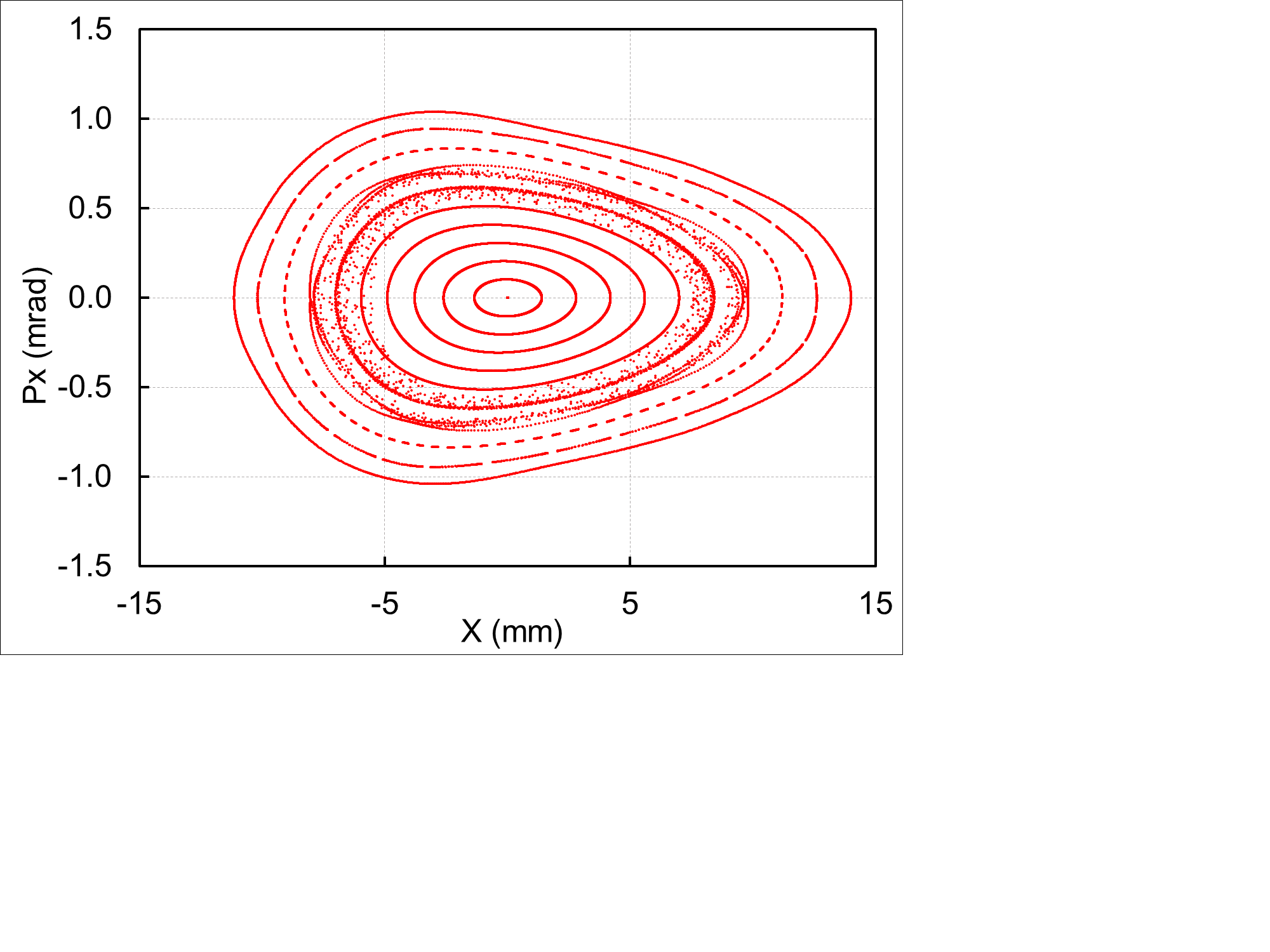}
\includegraphics*[width=.48\columnwidth,angle=0,trim=25 360 450 15,clip]{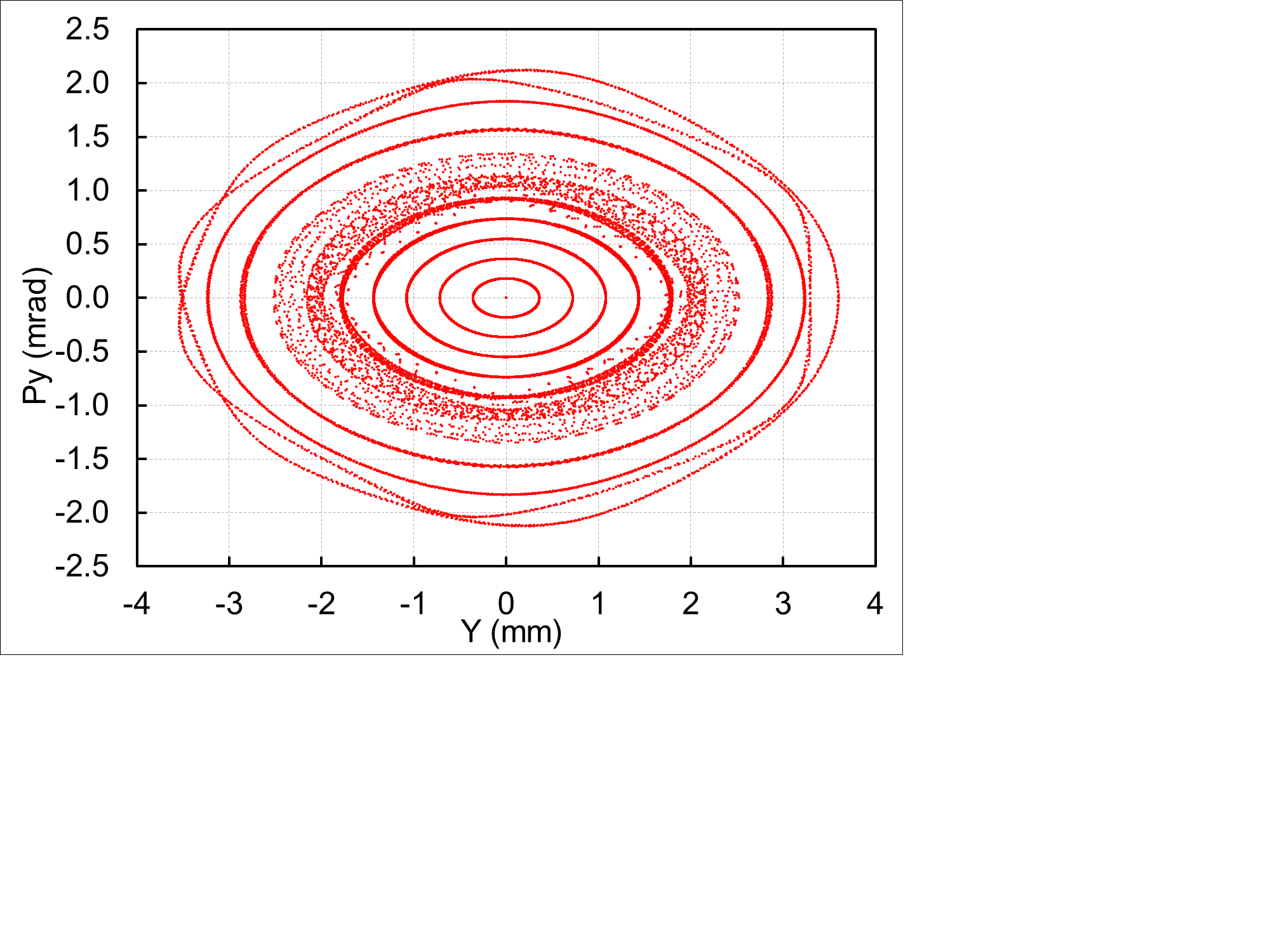}
\caption{Horizontal (left) and vertical (right) phase trajectories up to the last stable particle from FIG.~\ref{fig:SKIF-da}.}
\label{fig:SKIF-phase-traj}
\end{figure}

FIG.~\ref{fig:SKIF-da} presents the transverse dynamic aperture of the whole ring. Each point corresponds to initial conditions $\{x_0,x'_0=0,y_0,y'_0=0\}$ of a particle stable during 4096 turns. The left plot shows the results of the tracking: on the left RF cavities are OFF and on the right are ON.
Even though the initial energy deviation was zero, trajectory lengthening due to betatron motion gave rise to small synchrotron oscillations, which decreased dynamic aperture slightly. Horizontal and vertical phase trajectories up to the maximum aperture are shown in FIG.~\ref{fig:SKIF-phase-traj}.

\begin{figure}[ht!]
\includegraphics*[width=.48\columnwidth,angle=0,trim=25 380 365 15,clip]{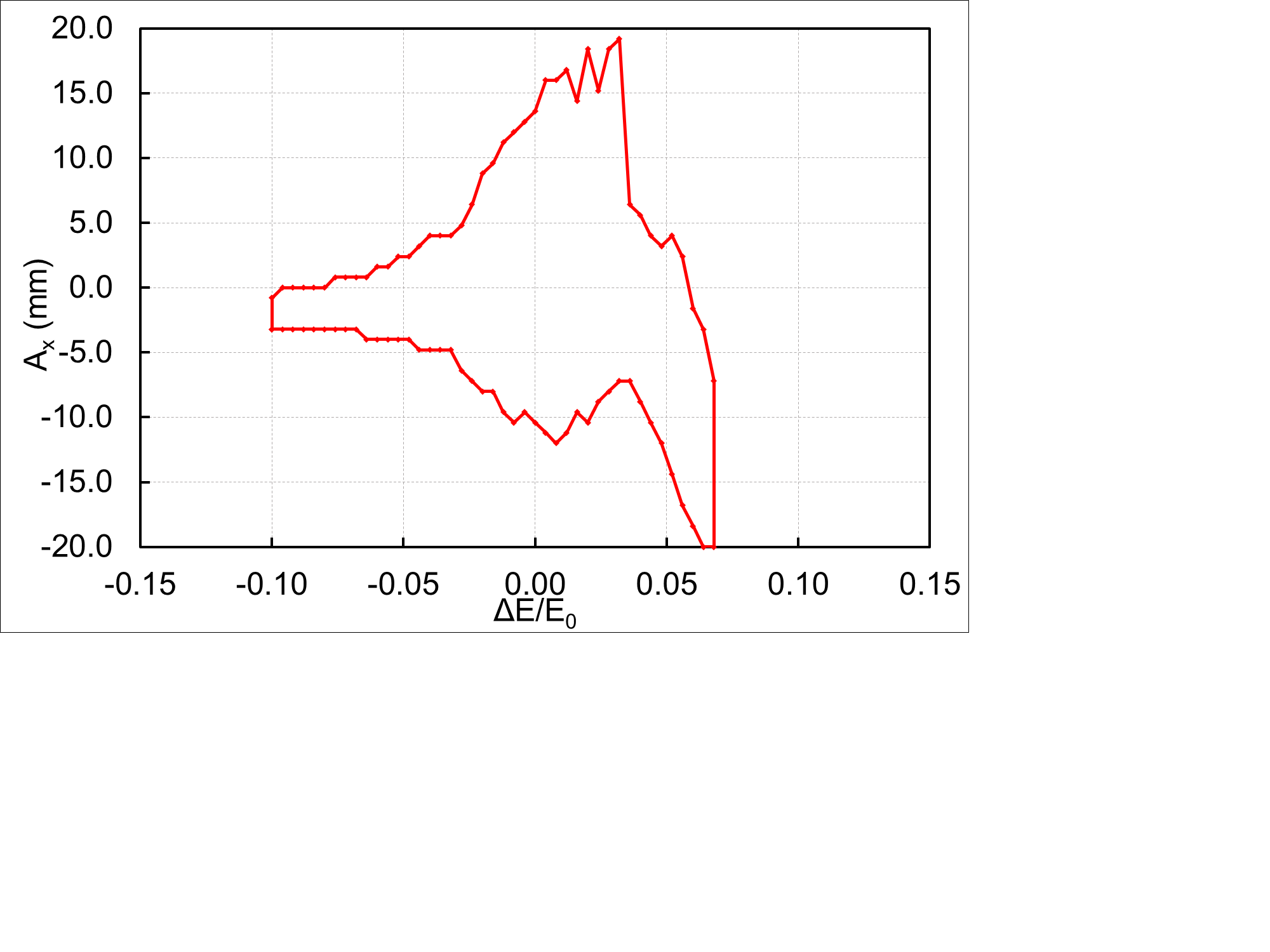}
\includegraphics*[width=.48\columnwidth,angle=0,trim=25 380 365 15,clip]{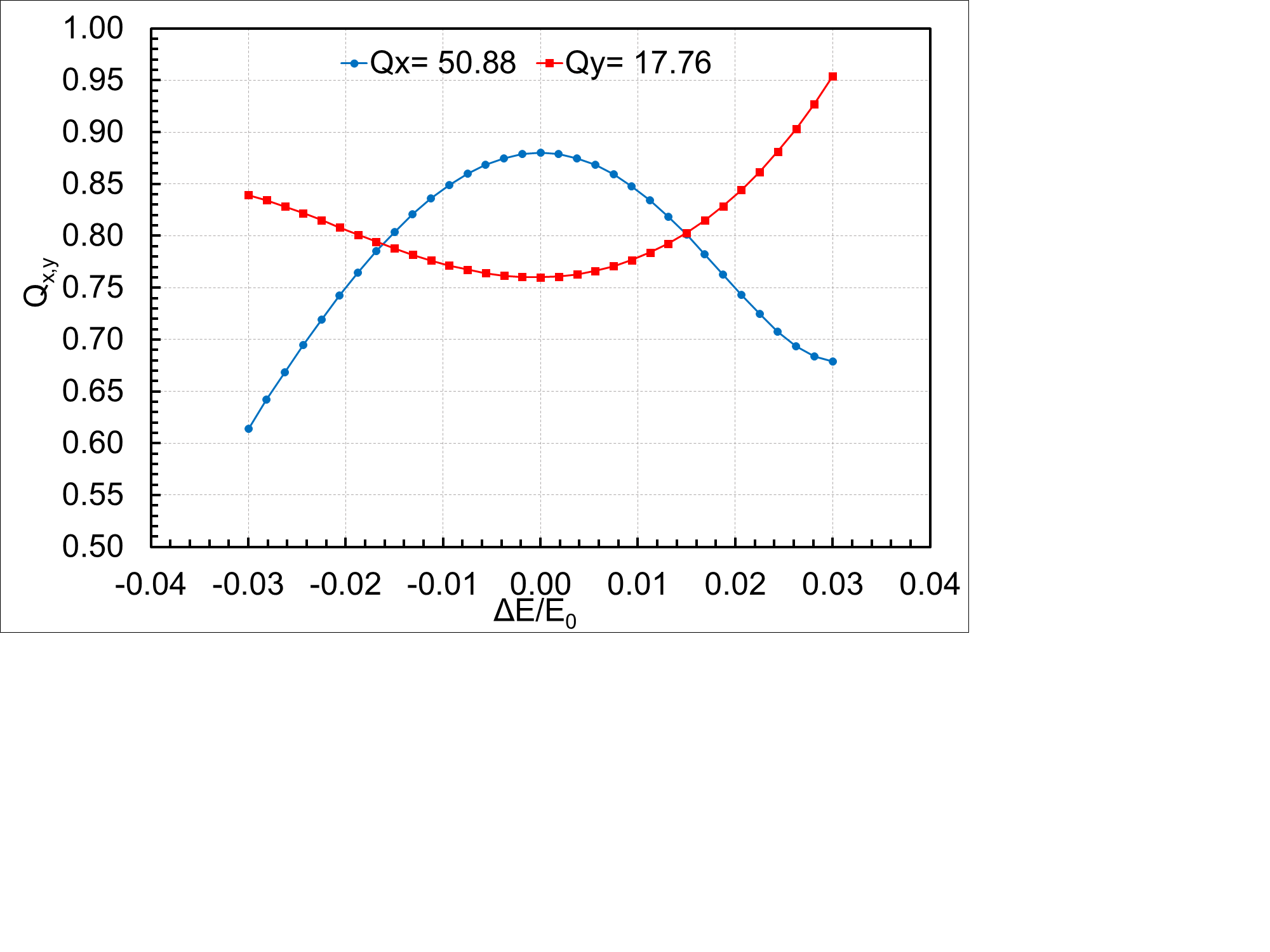}
\caption{Energy acceptance with $Y_0=\sigma_y$ (left) and bandwidth (right) of SKIF with RF OFF.}
\label{fig:SKIF-bandwidth}
\end{figure}

For the light source with small bunch length and transverse dimensions and high current, it is important to have large energy acceptance, which defines beam lifetime through Touschek effect \cite{Bernardini:1997sc,Piwinski:1998qs}. FIG.~\ref{fig:SKIF-bandwidth} shows tunes (betatron tune bandwidth) with respect to momentum deviation.
Betatron tunes cross the integer or half integer value when $\left|\Delta E/E_0\right|>3\%$, which is sufficient for good life time due to Touschek effect.

\section{Optimization of \lowercase{m}TME cell for for 6~GeV light source}
Having notable performance of mTME cell for light source SKIF with energy 3~GeV we reckoned its potential. All contemporary 4th generation light sources with 6~GeV energy -- ESRF-EBS (operating) \cite{ESRF:2018}, APS-U \cite{APS:2019}, Spring-8 II \cite{SPRING-8:2014}, HEPS \cite{Jiao:2018kke}, PETRA IV \cite{Schroer:ig5056} -- make use of HTME (Hybrid TME) magnetic structure, proposed by Pantaleo Raimondi for Italian SuperB factory \cite{Biagini:2008zze}. The logic of designing HTME is the following. When one decreases the bending angle of TME cell to achieve extremely small emittance, dispersion function in the sextupoles drops (if the cell length is preserved), forcing one to increase sextupoles' strengths, resulting in smaller dynamic aperture. 
Raimondi decided to install sextupoles in the two DBA cells with enlarged dispersion (dispersion bump), attached to the ends of the TME sequence, rather than in the TME cells (which makes them more compact)
Since DBA enlarges emittance, one is forced to use magnets with longitudinal field gradient \cite{Nagaoka:2007zza}. Adjusting optical functions and betatron phase advances, the designer creates a -I transformation, which is not ideal because of other nonlinear sources inside the pair of sextupoles. Chromaticity and nonlinear dynamics (dynamic aperture and energy acceptance) are optimized by three sextupoles' families and one octupole family (e.g., the flagship of HMBA -- ESRF-EBS), resulting in horizontal dynamic aperture $A_x\approx\pm13$~mm with $\beta_x=22$~m \cite{ESRF:2018}.

\begin{figure}[htb]
\includegraphics*[width=.48\columnwidth,angle=0,trim=110 55 135 0,clip]{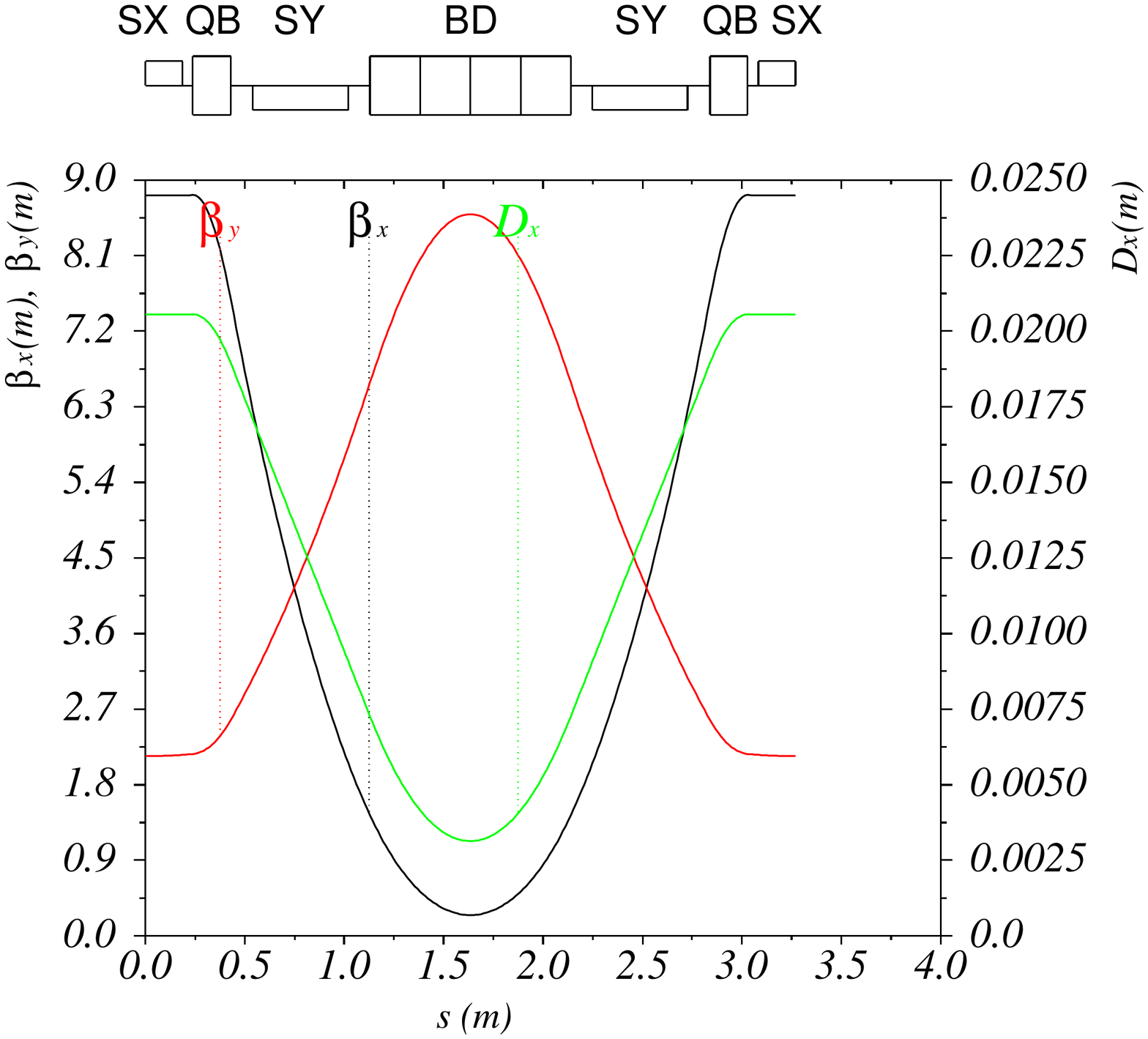}
\includegraphics*[width=.48\columnwidth,angle=0,trim=110 55 135 0,clip]{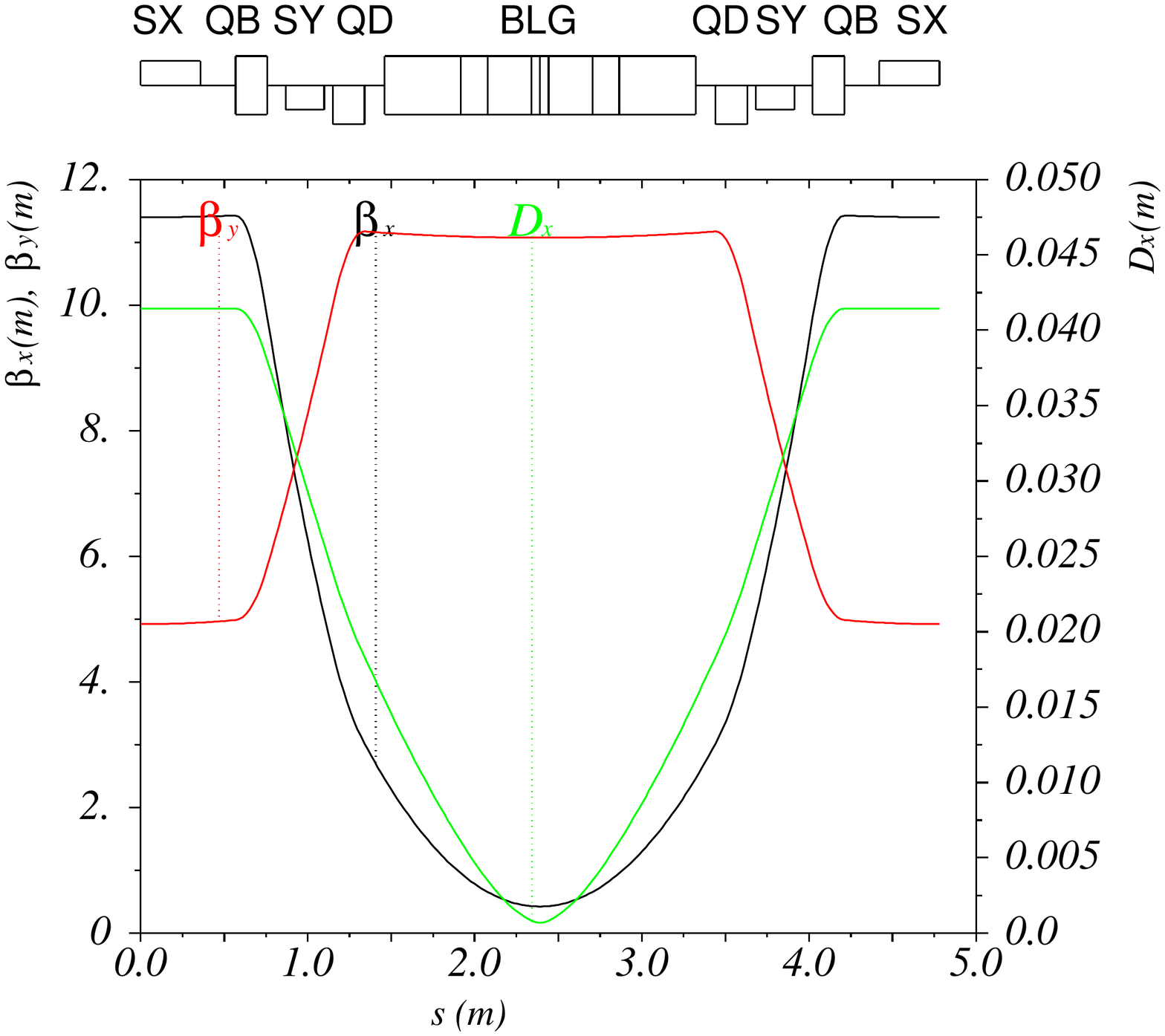}
\caption{Optical functions of scaled mTME SKIF cell on the left ($E=6$~GeV, $\varepsilon_x=30$~pm, $L_c=3.27$~m) and cell with longitudinal field gradient dipole on the right ($E=6$~GeV, $\varepsilon_x=30$~pm, $L_c=4.78$~m).}
\label{fig:6GEV-SKIF-cell}
\end{figure}

Our interest is the question of the existence of fundamental limitations in ``classical'' TME structure (more accurate in mTME) for high energy or the exclusive use of HMBA in 6~GeV light sources is, to some degree, a tradition.
To answer the question, we design mTME lattice of 6~GeV light source with small natural emittance $\varepsilon_x\approx 30$~pm, circumference $\Pi\approx 1100$~m, having 40 straight sections for installation of insertion devices (IDs) and dynamic aperture sufficient for traditional injection. Assuming that straight and matching sections will occupy 30\% of accelerator circumference, the total length of mTME cells is 780~m. Scaling the SKIF lattice to achieve desired parameters, we obtained lattice presented on the left plot of FIG.~\ref{fig:6GEV-SKIF-cell} with details in the first column of Table~\ref{tbl:6GEV-SKIF-cell}. The length of the cell remained almost the same (compared with FIG.~\ref{fig:SKIF-cell}), the cell's bending angle is decreased to $\varphi_c\approx27$~mrad to achieve $\varepsilon_x\approx30$~pm at 6~GeV. Decrease of bending angle and related diminution of dispersion resulted in stronger sextupoles' strengths $(B''l)_D=-1824$~T/m and $(B''l)_F=1235$~T/m (for SKIF $(B''l)_D=-592$~T/m and $(B''l)_F=707$~T/m). Hence, the drop in dynamic aperture by a factor of two. To increase dispersion in sextupoles, to lower their strengths, and to enlarge dynamic aperture, one needs to elongate the cell. If the number of cells is unchanged, then facility circumference grows, which is undesirable because of financial or other reasons. To satisfy the requirements, one has to decrease the number of cells, increasing the bending angle and emittance. The problem is resolved by introducing a longitudinal gradient in the bending magnet, a now-common design practice. The right plot of FIG.~\ref{fig:6GEV-SKIF-cell} shows an example of such a cell with details in the second column of Table~\ref{tbl:6GEV-SKIF-cell}.
\begin{table}[!htb]
\caption{
\label{tbl:6GEV-SKIF-cell}
Comparison of parameters of the two cells for 6~GeV light source with emittance $\varepsilon_x\approx 30$~pm. The column titled ``SKIF-scaled'' is shown on the left of FIG.~\ref{fig:6GEV-SKIF-cell}, ``LGdipole`` -- on the right}
\begin{ruledtabular}
\begin{tabular}{lcc}
							& SKIF-scaled		& LGdipole			\\
\colrule
Length, $L_c$ [m]					& $3.27$			& $4.78$			\\
Cells's bending angle, $\varphi_c$ [mrad]	& $26.8$			& $39.3$			\\
$\varepsilon_x$ [pm]				& $33$			& $30.1$			\\
Momentum compaction				& 				& 				\\
 factor, $\alpha$					& $2.8\cdot 10^{-5}$	& $-2.4\cdot 10^{-5}$	\\
Natural chromaticity, $\xi_x/\xi_y$		& $-1/-0.35$			& $-1/-0.35$			\\
Energy spread, $\sigma_E/E$			& $1.1\cdot 10^{-3}$	& $2.1\cdot 10^{-3}$	\\
Energy loss per turn,				& 				& 				\\
 $U_0$ [keV]						& $15.2$			& $42.1$			\\
Hor. damping time [ms]				& $5.1$			& $2.1$			\\
Long. damping time [ms]				& $6.6$			& $5.2$			\\
$(B''l)_D$ [T/m]					& $-1824$			& $-547$			\\
$(B''l)_F$ [T/m]					& $1235$			& $647$			\\
\end{tabular}
\end{ruledtabular}
\end{table}
Usage of longitudinal gradient dipole enlarges the cell length to $4.78$~m, bending angle to $39.3$~mrad, dispersion, at the end of the cell, from $\eta_c\approx2.45$~cm to $\eta_c\approx4.75$~cm. Hence, the sextupoles' strengths dropped to 
$(B''l)_D=-547$~T/m and $(B''l)_F=647$~T/m, approximately SKIF's values. Therefore, we expect a similar dynamic aperture.

Since it is difficult to superimpose longitudinal and transverse gradients in the same dipole, the latter is transferred from the dipole into two individual defocusing quadrupoles on both sides of the dipole.
\begin{figure}[htb]
\includegraphics*[width=.48\columnwidth,angle=0,trim=100 50 105 0,clip]{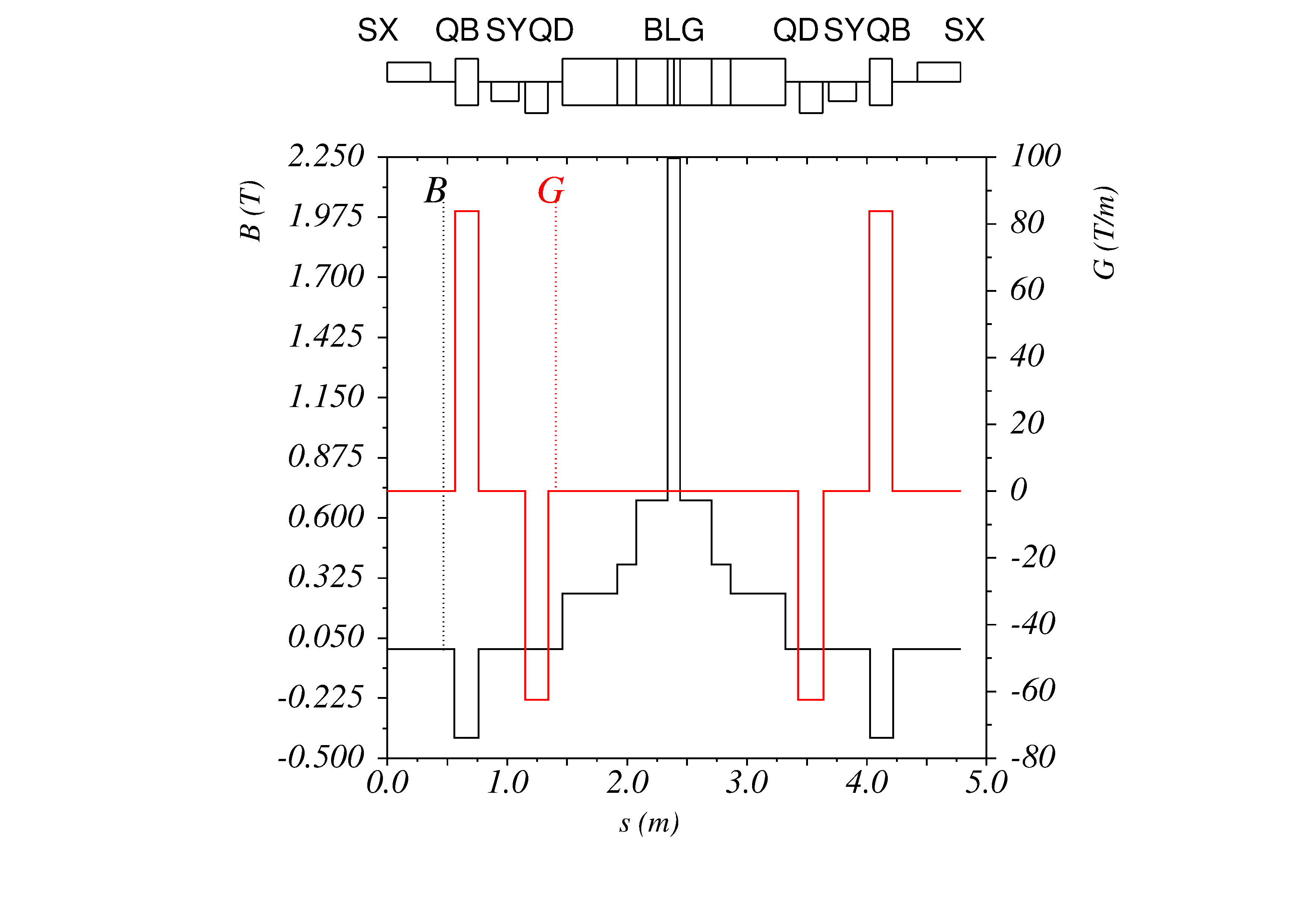}
\includegraphics*[width=.48\columnwidth,angle=0,trim=100 50 105 0,clip]{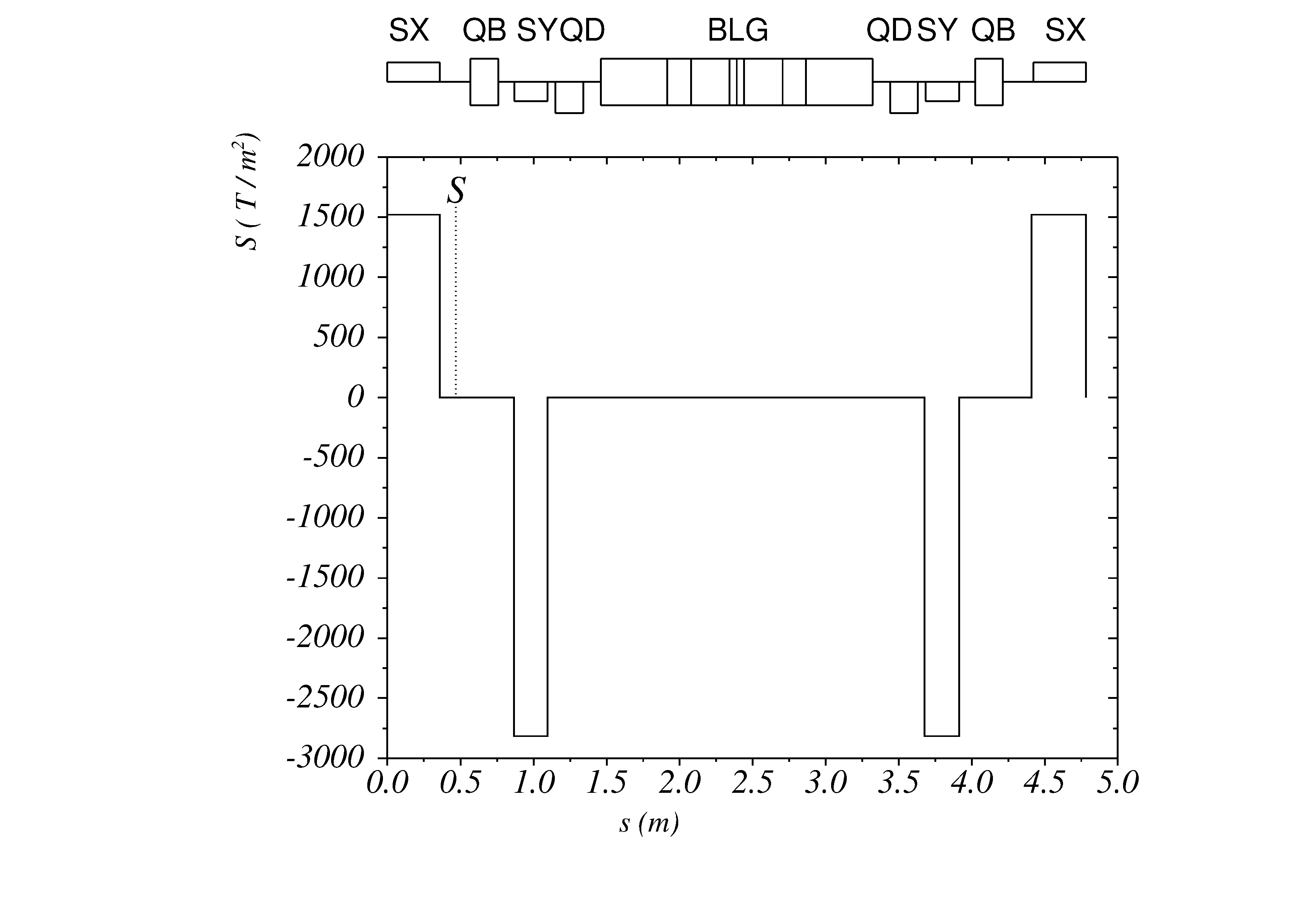}
\caption{Distribution of the dipole field (left, black), the gradient (left, red) and the sextupole gradient (right) in the cell with the longitudinal gradient dipole.}
\label{fig:6GEV-SKIF-cell-field}
\end{figure}
FIG.~\ref{fig:6GEV-SKIF-cell-field} shows the dipole field and the gradient distributions on the left, sextupole gradient --- on the right in the cell with longitudinal gradient dipole. The magnetic field reaches the maximum of 2.24~T in the center of the magnet and gradually decreases in four steps to 0.25~T at the magnet ends. Similar to SKIF, the focusing quadrupoles are shifted in the transverse direction making the reverse bends and helping to reduce emittance. 

The regular TME cell has the minimum emittance $\varepsilon_{TME}=69$~pm with the 39.3~mrad bending angle, whereas longitudinal gradient and transverse bends reduce emittance to 30~pm.

\begin{figure}[htb]
\includegraphics*[width=.48\columnwidth,angle=270,trim=0 0 0 0,clip]{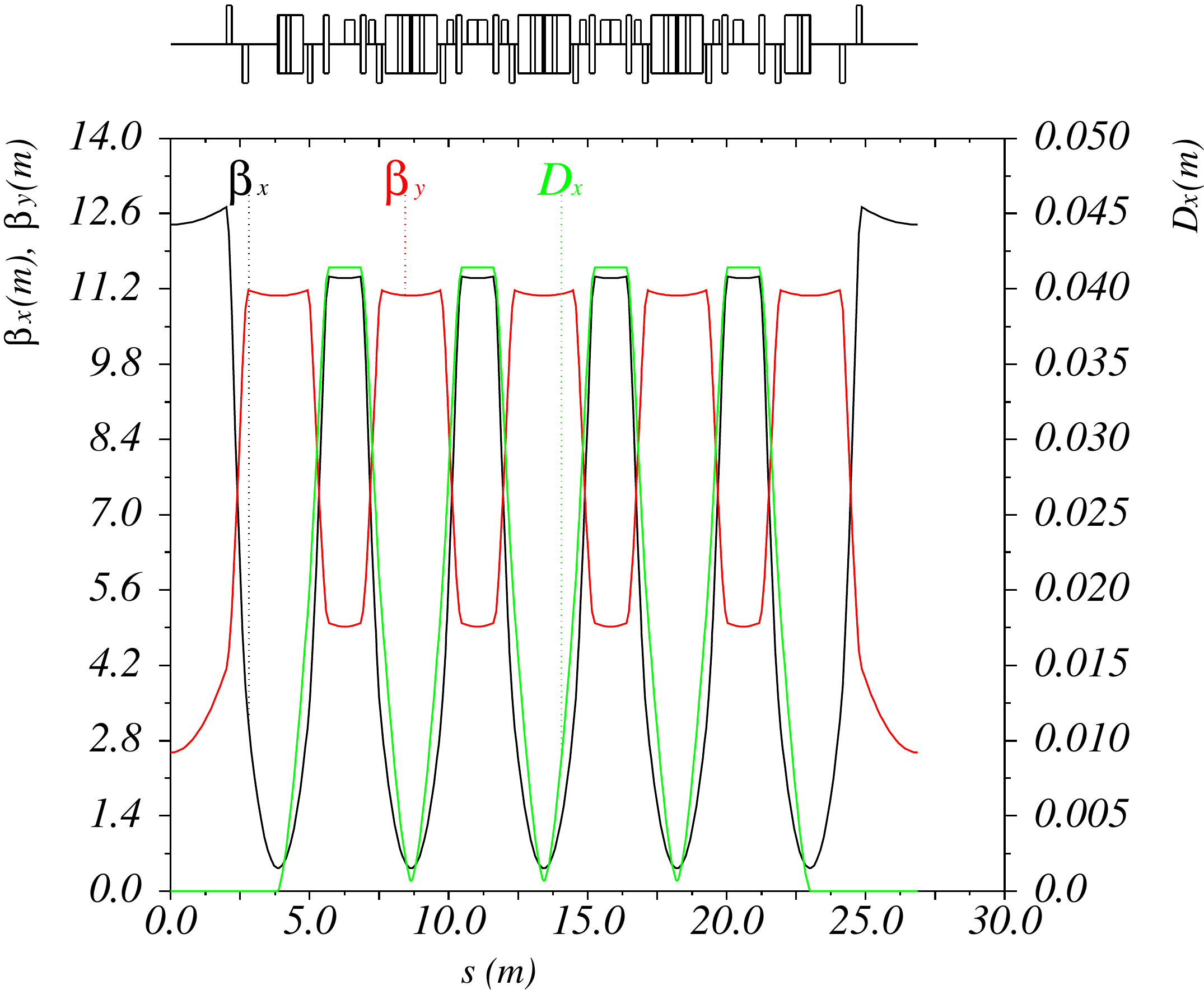}
\caption{Superperiod of the light source with 6~GeV beam energy and 33.5~pm emittance.}
\label{fig:6GEV-SKIF-superperiod}
\end{figure}

Using the cell with longitudinal gradient dipole (right on FIG.~\ref{fig:6GEV-SKIF-cell}) we designed the lattice of the whole ring consisting of 40 superperiods (FIG.~\ref{fig:6GEV-SKIF-superperiod}).
The quadrupoles and sextupoles have feasible lengths and strengths ($B'_{max}\approx80$~T/m, $B''_{max}\approx4000$~T/m$^2$), which are possible with small apertures common for 4th generation light sources.

The parameters of the ring are in TABLE~\ref{tbl:6GEV-SKIF-ring}. The super period is 5BA, the length of the straight sections is 4~m, the circumference is less than 1100~m, the horizontal emittance is 33.5~pm. Dipoles with strong field ensure large synchrotron radiation energy loss (6.74~MeV per turn), the expected outcome for low emittance in a compact ring.

TABLE~\ref{tbl:6GEV-SKIF-ring} presents lattice with negative momentum compaction factor $\alpha=-1.84\times10^{-5}$; however, we also obtained variants with positive momentum compaction factors. Latices with negative momentum compaction factors have been studied theoretically and experimentally \cite{Zobov:2006pp,Ikeda:2004me,Schreiber:2019vor} mainly relating to the topic of collective effects. We chose this negative momentum compaction factor lattice expecting dynamic aperture enhancement with under-compensated chromaticity (e.g. $\xi_{x,y}\approx-5\div-10$). Negative chromaticity is needed for head-tail instability control in case of negative momentum compaction lattice, thus allowing sextupoles diminution.
\begin{table}[!htb]
\caption{
\label{tbl:6GEV-SKIF-ring}
Main parameters of 6~GeV light source based on  mTME cell
}
\begin{ruledtabular}
\begin{tabular}{lc}
Beam energy [GeV]				& $6$				\\
Symmetry					& $40$			\\
Circumference [m]				& $1074.6$			\\
Horizontal emittance [pm]			& $33.46$			\\
Energy spread				& $2.12\cdot 10^{-3}$	\\
Energy loss per turn [MeV]		& $6.74$			\\
Betatron tunes ($x/y$)			& $88.39/27.16$		\\
Momentum compaction factor		& $-1.84\cdot 10^{-5}$	\\
Chromaticity ($x/y$)			& $-212/-79$		\\
Damping numbers ($x/y$)			& $2.13/0.87$		\\
Damping time ($x/y/z$) [ms]		& $3.0/6.4/7.3$		\\
\end{tabular}
\end{ruledtabular}
\end{table}

\begin{figure}[htb!]
\includegraphics*[width=.48\columnwidth,angle=0,trim=20 385 370 20,clip]{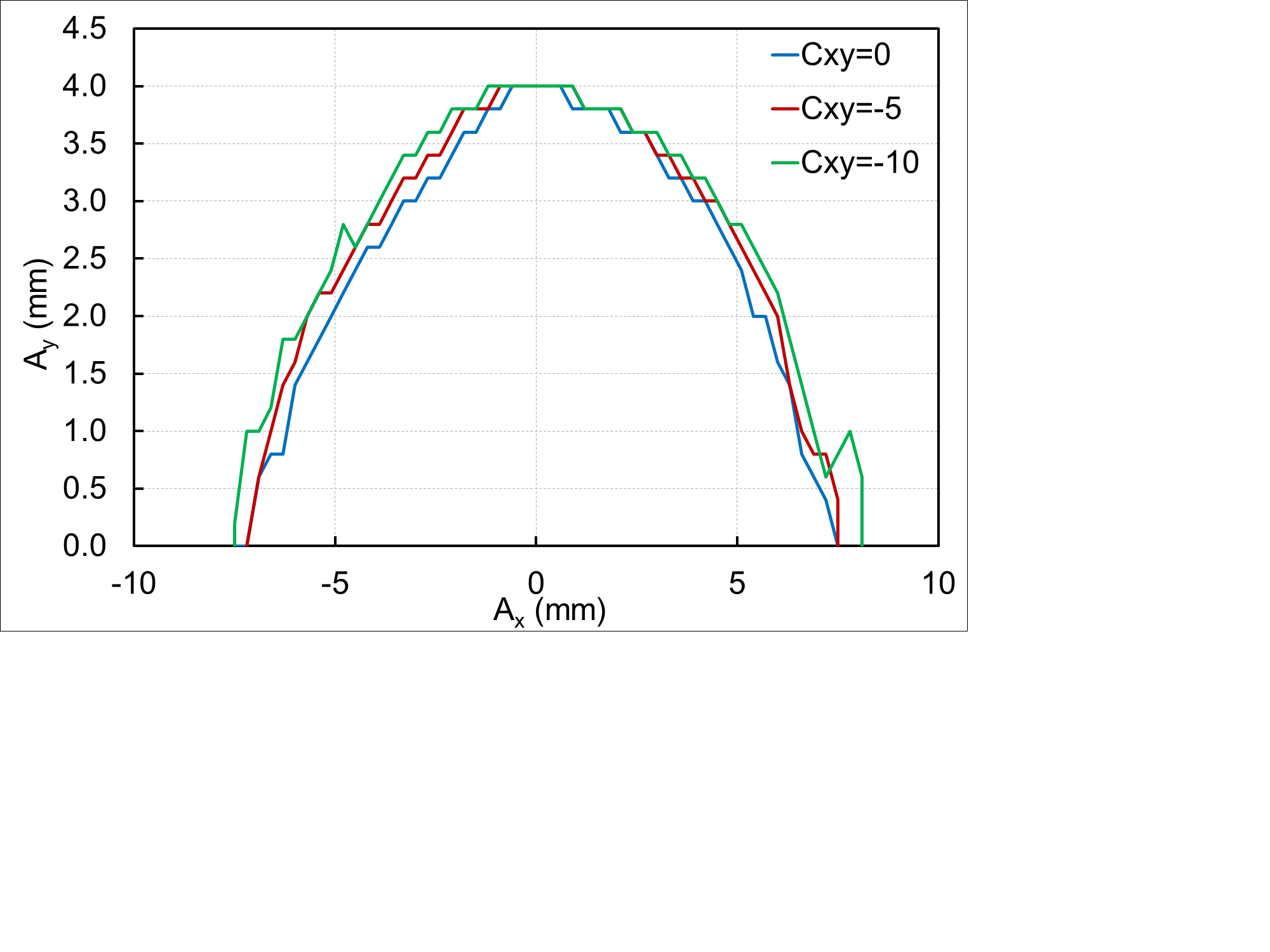}
\caption{Dynamic aperture of 6~GeV light source for three values of under-compensated chromaticity.}
\label{fig:6GEV-ring-da}
\end{figure}

\begin{figure}[ht!]
\includegraphics*[width=.48\columnwidth,angle=0,trim=20 385 370 20,clip]{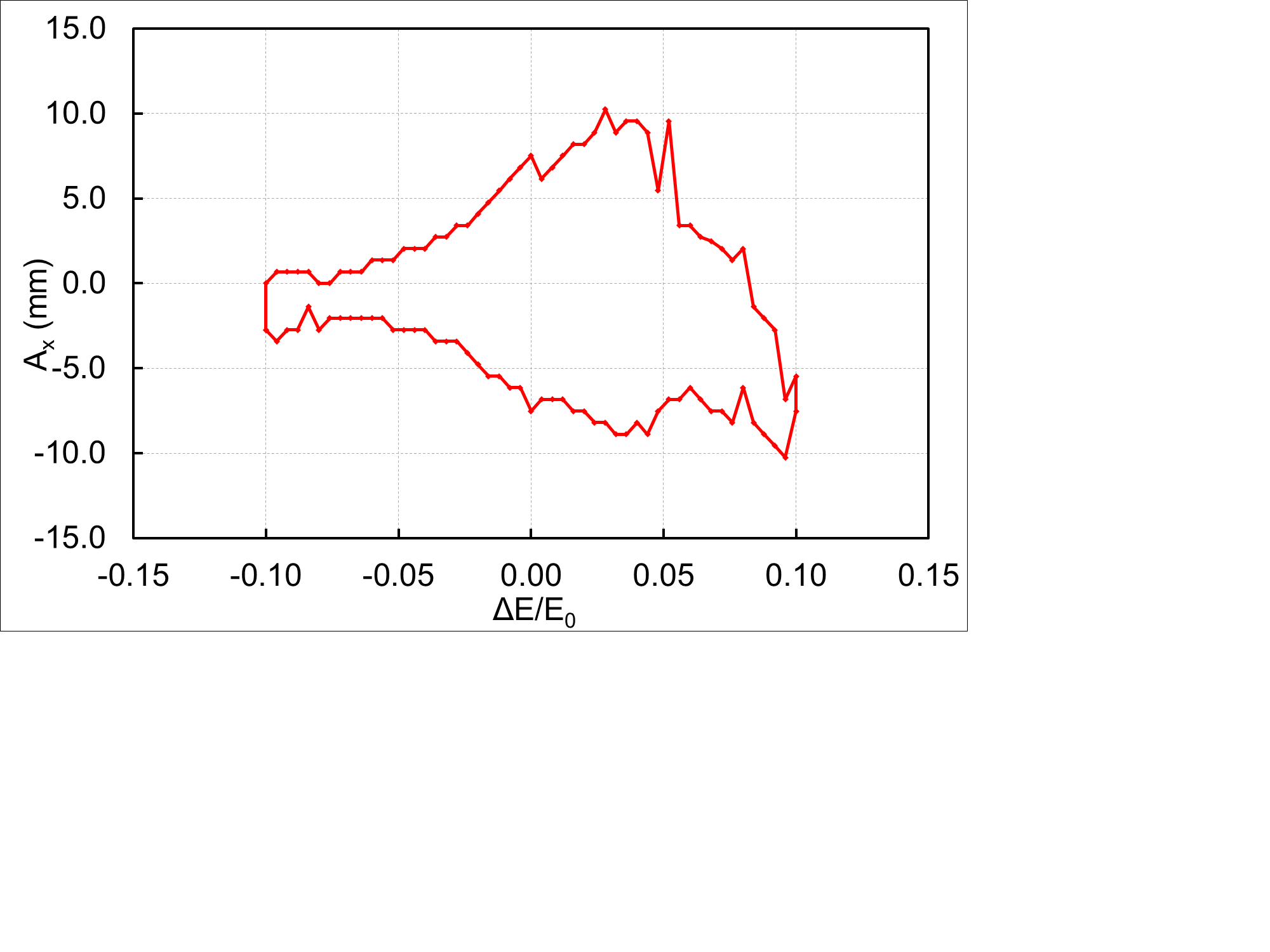}
\includegraphics*[width=.48\columnwidth,angle=0,trim=20 385 370 20,clip]{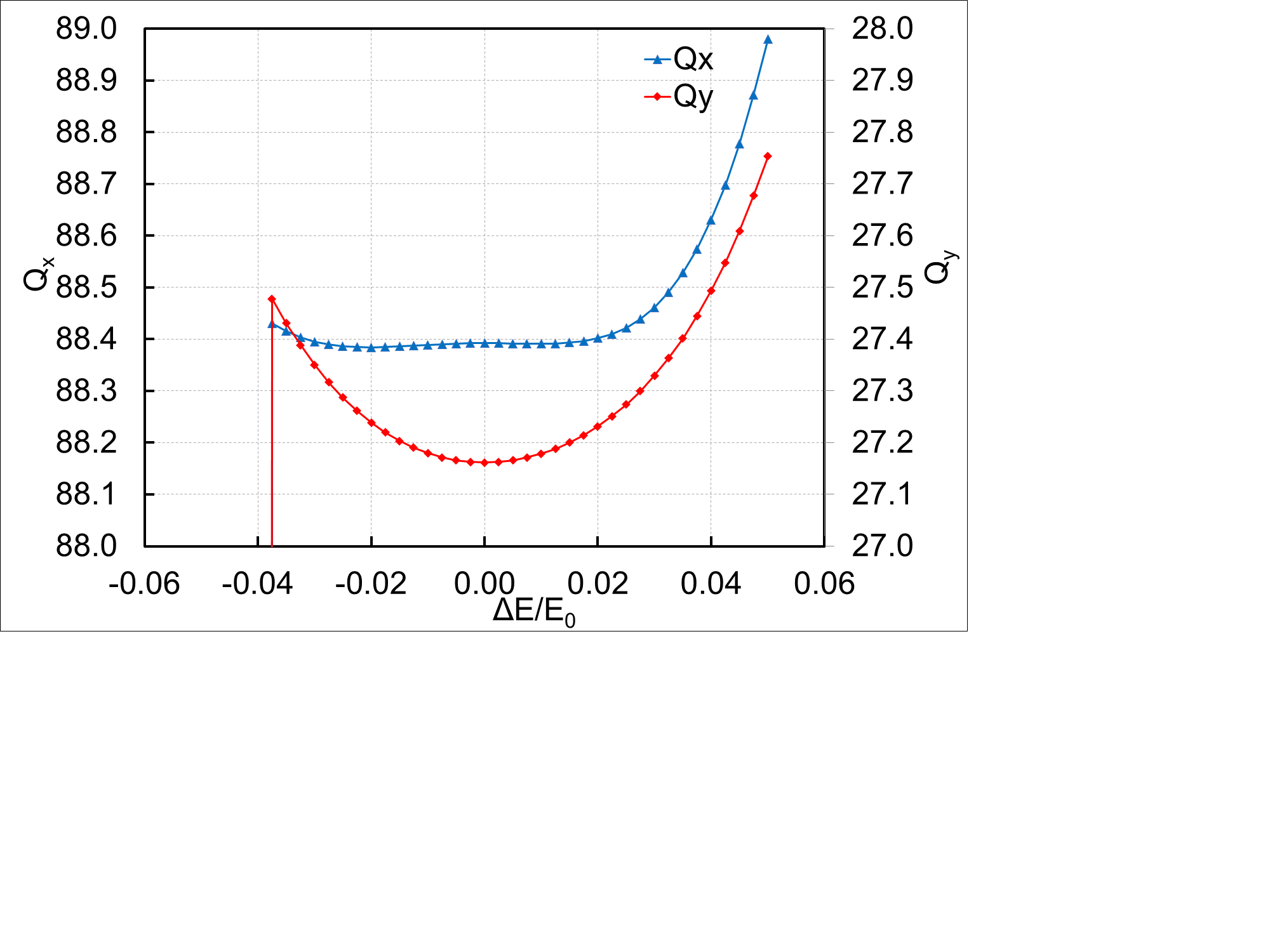}
\caption{Energy acceptance with $Y_0=\sigma_y$ (left) and bandwidth (right) of 6~GeV light source.}
\label{fig:6GEV-SKIF-ring-EA-BW}
\end{figure}

FIG.~\ref{fig:6GEV-ring-da} shows dynamic aperture for three values of chromaticity $\xi_{x,y}=0,-5,-10$ at the observation point corresponding to coordinate frame origin in FIG.~\ref{fig:6GEV-SKIF-cell}. 
Energy acceptance and bandwidth are shown in FIG.~\ref{fig:6GEV-SKIF-ring-EA-BW}.
Dynamic aperture enhancement due to under-compensated chromaticity is insignificant, probably because chromaticity variations are small. The same is true for energy acceptance and bandwidth. Nevertheless, dynamic aperture is feasible for the usual injection in horizontal phase space. Energy acceptance of $\pm10\%$ and bandwidth of $\pm4\%$ ensure sufficient lifetime with intrabeam scattering.

\section{Conclusion}
We studied magnetic lattice cells ensuring low emittance. The study showed that the simplest cell consisting of a central dipole and two symmetrical quadrupole doublets has four solutions. The solution providing minimal emittance (TME) has unacceptably strong sextupoles and a small dynamic aperture. Therefore, for 4th generation light source SKIF, we chose an alternative solution (modified TME, mTME), which allowed sufficient dynamic aperture with feasible sextupole and quadrupole strengths. Based on this cell, the whole ring lattice provides 72~pm horizontal emittance at beam energy 3~GeV, with circumference of 476~m and 16 long (6~m) straight sections for insertion devices, RF cavities, injection equipment etc. Only two families of sextupoles provide horizontal and vertical dynamic apertures of 12~mm and 3.5~mm respectively and energy acceptance (band width) more than 3\%.

We also applied mTME cell with longitudinal gradient dipole to design a 6~GeV machine with 33~pm emittance and showed the validity of the cell for the high energy light sources. The obtained dynamic aperture ($A_x\approx 7$~mm) and energy acceptance ($\Delta E/E_0>$) are feasible for regular injection and could be further optimized.

\bibliography{References.bib}

\begin{thebibliography}{23}%
\makeatletter
\providecommand \@ifxundefined [1]{%
 \@ifx{#1\undefined}
}%
\providecommand \@ifnum [1]{%
 \ifnum #1\expandafter \@firstoftwo
 \else \expandafter \@secondoftwo
 \fi
}%
\providecommand \@ifx [1]{%
 \ifx #1\expandafter \@firstoftwo
 \else \expandafter \@secondoftwo
 \fi
}%
\providecommand \natexlab [1]{#1}%
\providecommand \enquote  [1]{``#1''}%
\providecommand \bibnamefont  [1]{#1}%
\providecommand \bibfnamefont [1]{#1}%
\providecommand \citenamefont [1]{#1}%
\providecommand \href@noop [0]{\@secondoftwo}%
\providecommand \href [0]{\begingroup \@sanitize@url \@href}%
\providecommand \@href[1]{\@@startlink{#1}\@@href}%
\providecommand \@@href[1]{\endgroup#1\@@endlink}%
\providecommand \@sanitize@url [0]{\catcode `\\12\catcode `\$12\catcode
  `\&12\catcode `\#12\catcode `\^12\catcode `\_12\catcode `\%12\relax}%
\providecommand \@@startlink[1]{}%
\providecommand \@@endlink[0]{}%
\providecommand \url  [0]{\begingroup\@sanitize@url \@url }%
\providecommand \@url [1]{\endgroup\@href {#1}{\urlprefix }}%
\providecommand \urlprefix  [0]{URL }%
\providecommand \Eprint [0]{\href }%
\providecommand \doibase [0]{http://dx.doi.org/}%
\providecommand \selectlanguage [0]{\@gobble}%
\providecommand \bibinfo  [0]{\@secondoftwo}%
\providecommand \bibfield  [0]{\@secondoftwo}%
\providecommand \translation [1]{[#1]}%
\providecommand \BibitemOpen [0]{}%
\providecommand \bibitemStop [0]{}%
\providecommand \bibitemNoStop [0]{.\EOS\space}%
\providecommand \EOS [0]{\spacefactor3000\relax}%
\providecommand \BibitemShut  [1]{\csname bibitem#1\endcsname}%
\let\auto@bib@innerbib\@empty
\bibitem [{\citenamefont {lightsources.org}()}]{LightSource}%
  \BibitemOpen
  \bibfield  {author} {\bibinfo {author} {\bibnamefont {lightsources.org}},\
  }\href {https://lightsources.org/} {\emph {\bibinfo {title}
  {https://lightsources.org/}}},\ \bibinfo {type} {Tech. Rep.}\BibitemShut
  {Stop}%
\bibitem [{\citenamefont {Teng}(1984)}]{Teng:1984cz}%
  \BibitemOpen
  \bibfield  {author} {\bibinfo {author} {\bibfnamefont {L.C.}\ \bibnamefont
  {Teng}},\ }\bibfield  {title} {\enquote {\bibinfo {title} {{Minimizing the
  Emittance in Designing the Lattice of an Electron Storage Ring}},}\
  }\href@noop {} {\  (\bibinfo {year} {1984})}\BibitemShut {NoStop}%
\bibitem [{\citenamefont {Teng}(1985)}]{Teng:1985gm}%
  \BibitemOpen
  \bibfield  {author} {\bibinfo {author} {\bibfnamefont {L.C.}\ \bibnamefont
  {Teng}},\ }\bibfield  {title} {\enquote {\bibinfo {title} {{MINIMUM EMITTANCE
  LATTICE FOR SYNCHROTRON RADIATION STORAGE RINGS}},}\ }\href@noop {} {\
  (\bibinfo {year} {1985})}\BibitemShut {NoStop}%
\bibitem [{\citenamefont {Lee}\ and\ \citenamefont {Teng}(1991)}]{Lee:1991st}%
  \BibitemOpen
  \bibfield  {author} {\bibinfo {author} {\bibfnamefont {S.Y.}\ \bibnamefont
  {Lee}}\ and\ \bibinfo {author} {\bibfnamefont {L.}~\bibnamefont {Teng}},\
  }\bibfield  {title} {\enquote {\bibinfo {title} {{Theoretical minimum
  emittance lattice for an electron storage ring}},}\ }\href@noop {} {\bibfield
   {journal} {\bibinfo  {journal} {Conf. Proc. C}\ }\textbf {\bibinfo {volume}
  {910506}},\ \bibinfo {pages} {2679--2681} (\bibinfo {year}
  {1991})}\BibitemShut {NoStop}%
\bibitem [{\citenamefont {Antoniou}\ and\ \citenamefont
  {Papaphilippou}(2014)}]{Antoniou:2013uva}%
  \BibitemOpen
  \bibfield  {author} {\bibinfo {author} {\bibfnamefont {F.}~\bibnamefont
  {Antoniou}}\ and\ \bibinfo {author} {\bibfnamefont {Y.}~\bibnamefont
  {Papaphilippou}},\ }\bibfield  {title} {\enquote {\bibinfo {title}
  {{Analytical considerations for linear and nonlinear optimization of the
  theoretical minimum emittance cells: Application to the Compact Linear
  Collider predamping rings}},}\ }\href {\doibase
  10.1103/PhysRevSTAB.17.064002} {\bibfield  {journal} {\bibinfo  {journal}
  {Phys. Rev. ST Accel. Beams}\ }\textbf {\bibinfo {volume} {17}},\ \bibinfo
  {pages} {064002} (\bibinfo {year} {2014})},\ \Eprint
  {http://arxiv.org/abs/1310.5024} {arXiv:1310.5024 [physics.acc-ph]}
  \BibitemShut {NoStop}%
\bibitem [{\citenamefont {Cai}(2018)}]{Cai:2018bvb}%
  \BibitemOpen
  \bibfield  {author} {\bibinfo {author} {\bibfnamefont {Yunhai}\ \bibnamefont
  {Cai}},\ }\bibfield  {title} {\enquote {\bibinfo {title} {{Single-particle
  dynamics in theoretical minimum emittance cell}},}\ }\href {\doibase
  10.1103/PhysRevAccelBeams.21.114002} {\bibfield  {journal} {\bibinfo
  {journal} {Phys. Rev. Accel. Beams}\ }\textbf {\bibinfo {volume} {21}},\
  \bibinfo {pages} {114002} (\bibinfo {year} {2018})}\BibitemShut {NoStop}%
\bibitem [{\citenamefont {Inc.}()}]{Mathematica}%
  \BibitemOpen
  \bibfield  {author} {\bibinfo {author} {\bibfnamefont {Wolfram~Research{,}}\
  \bibnamefont {Inc.}},\ }\href@noop {} {\enquote {\bibinfo {title}
  {Mathematica, {V}ersion 12.1},}\ }\bibinfo {note} {Champaign, IL,
  2020}\BibitemShut {NoStop}%
\bibitem [{\citenamefont {MADX}()}]{MADX}%
  \BibitemOpen
  \bibfield  {author} {\bibinfo {author} {\bibnamefont {MADX}},\ }\href
  {http://madx.web.cern.ch/madx} {\emph {\bibinfo {title}
  {http://madx.web.cern.ch/madx}}},\ \bibinfo {type} {Tech. Rep.}\BibitemShut
  {Stop}%
\bibitem [{\citenamefont {Jiao}\ \emph {et~al.}(2011)\citenamefont {Jiao},
  \citenamefont {Cai},\ and\ \citenamefont {Chao}}]{Jiao:2011zza}%
  \BibitemOpen
  \bibfield  {author} {\bibinfo {author} {\bibfnamefont {Yi}~\bibnamefont
  {Jiao}}, \bibinfo {author} {\bibfnamefont {Yunhai}\ \bibnamefont {Cai}}, \
  and\ \bibinfo {author} {\bibfnamefont {Alexander~Wu}\ \bibnamefont {Chao}},\
  }\bibfield  {title} {\enquote {\bibinfo {title} {{Modified theoretical
  minimum emittance lattice for an electron storage ring with extreme-low
  emittance}},}\ }\href {\doibase 10.1103/PhysRevSTAB.14.054002} {\bibfield
  {journal} {\bibinfo  {journal} {Phys. Rev. ST Accel. Beams}\ }\textbf
  {\bibinfo {volume} {14}},\ \bibinfo {pages} {054002} (\bibinfo {year}
  {2011})}\BibitemShut {NoStop}%
\bibitem [{\citenamefont {Riemann}\ and\ \citenamefont
  {Streun}(2019)}]{Riemann:2018kga}%
  \BibitemOpen
  \bibfield  {author} {\bibinfo {author} {\bibfnamefont {B.}~\bibnamefont
  {Riemann}}\ and\ \bibinfo {author} {\bibfnamefont {A.}~\bibnamefont
  {Streun}},\ }\bibfield  {title} {\enquote {\bibinfo {title} {{Low emittance
  lattice design from first principles: reverse bending and longitudinal
  gradient bends}},}\ }\href {\doibase 10.1103/PhysRevAccelBeams.22.021601}
  {\bibfield  {journal} {\bibinfo  {journal} {Phys. Rev. Accel. Beams}\
  }\textbf {\bibinfo {volume} {22}},\ \bibinfo {pages} {021601} (\bibinfo
  {year} {2019})},\ \Eprint {http://arxiv.org/abs/1810.11286} {arXiv:1810.11286
  [physics.acc-ph]} \BibitemShut {NoStop}%
\bibitem [{\citenamefont {Streun}(2014)}]{Streun:2014gna}%
  \BibitemOpen
  \bibfield  {author} {\bibinfo {author} {\bibfnamefont {A.}~\bibnamefont
  {Streun}},\ }\bibfield  {title} {\enquote {\bibinfo {title} {{The anti-bend
  cell for ultralow emittance storage ring lattices}},}\ }\href {\doibase
  10.1016/j.nima.2013.11.064} {\bibfield  {journal} {\bibinfo  {journal} {Nucl.
  Instrum. Meth. A}\ }\textbf {\bibinfo {volume} {737}},\ \bibinfo {pages}
  {148--154} (\bibinfo {year} {2014})}\BibitemShut {NoStop}%
\bibitem [{\citenamefont {Bernardini}\ \emph {et~al.}(1963)\citenamefont
  {Bernardini}, \citenamefont {Corazza}, \citenamefont {Di~Giugno},
  \citenamefont {Ghigo}, \citenamefont {Querzoli}, \citenamefont {Haissinski},
  \citenamefont {Marin},\ and\ \citenamefont {Touschek}}]{Bernardini:1997sc}%
  \BibitemOpen
  \bibfield  {author} {\bibinfo {author} {\bibfnamefont {C.}~\bibnamefont
  {Bernardini}}, \bibinfo {author} {\bibfnamefont {G.F.}\ \bibnamefont
  {Corazza}}, \bibinfo {author} {\bibfnamefont {G.}~\bibnamefont {Di~Giugno}},
  \bibinfo {author} {\bibfnamefont {G.}~\bibnamefont {Ghigo}}, \bibinfo
  {author} {\bibfnamefont {R.}~\bibnamefont {Querzoli}}, \bibinfo {author}
  {\bibfnamefont {J.}~\bibnamefont {Haissinski}}, \bibinfo {author}
  {\bibfnamefont {P.}~\bibnamefont {Marin}}, \ and\ \bibinfo {author}
  {\bibfnamefont {B.}~\bibnamefont {Touschek}},\ }\bibfield  {title} {\enquote
  {\bibinfo {title} {{Lifetime and beam size in a storage ring}},}\ }\href
  {\doibase 10.1103/PhysRevLett.10.407} {\bibfield  {journal} {\bibinfo
  {journal} {Phys. Rev. Lett.}\ }\textbf {\bibinfo {volume} {10}},\ \bibinfo
  {pages} {407--409} (\bibinfo {year} {1963})}\BibitemShut {NoStop}%
\bibitem [{\citenamefont {Piwinski}(1998)}]{Piwinski:1998qs}%
  \BibitemOpen
  \bibfield  {author} {\bibinfo {author} {\bibfnamefont {A.}~\bibnamefont
  {Piwinski}},\ }\bibfield  {title} {\enquote {\bibinfo {title} {{The Touschek
  effect in strong focusing storage rings}},}\ }\href@noop {} {\  (\bibinfo
  {year} {1998})},\ \Eprint {http://arxiv.org/abs/physics/9903034}
  {arXiv:physics/9903034} \BibitemShut {NoStop}%
\bibitem [{\citenamefont {ESRF}(2018)}]{ESRF:2018}%
  \BibitemOpen
  \bibfield  {author} {\bibinfo {author} {\bibnamefont {ESRF}},\ }\href
  {https://www.esrf.eu/files/live/sites/www/files/about/upgrade/documentation/Design%20Report-reduced-jan19.pdf}
  {\emph {\bibinfo {title} {EBS Storage Ring Technical Design Report}}},\
  \bibinfo {type} {Tech. Rep.}\ (\bibinfo  {institution} {ESRF, Grenoble},\
  \bibinfo {year} {2018})\BibitemShut {NoStop}%
\bibitem [{\citenamefont {ANL}(2019)}]{APS:2019}%
  \BibitemOpen
  \bibfield  {author} {\bibinfo {author} {\bibnamefont {ANL}},\ }\href
  {https://publications.anl.gov/anlpubs/2019/07/153666.pdf} {\emph {\bibinfo
  {title} {Advanced Photon Source Upgrade Project Final Design Report}}},\
  \bibinfo {type} {Tech. Rep.}\ (\bibinfo  {institution} {Argonne National
  Laboratory},\ \bibinfo {year} {2019})\BibitemShut {NoStop}%
\bibitem [{\citenamefont {Center}(2014)}]{SPRING-8:2014}%
  \BibitemOpen
  \bibfield  {author} {\bibinfo {author} {\bibfnamefont {RIKEN SPring-8}\
  \bibnamefont {Center}},\ }\href {http://rsc.riken.jp/eng/pdf/SPring-8-II.pdf}
  {\emph {\bibinfo {title} {SPring-8-II Conceptual Design Report}}},\ \bibinfo
  {type} {Tech. Rep.}\ (\bibinfo  {institution} {RIKEN SPring-8 Center},\
  \bibinfo {year} {2014})\BibitemShut {NoStop}%
\bibitem [{\citenamefont {Jiao}\ \emph {et~al.}(2018)\citenamefont {Jiao} \emph
  {et~al.}}]{Jiao:2018kke}%
  \BibitemOpen
  \bibfield  {author} {\bibinfo {author} {\bibfnamefont {Yi}~\bibnamefont
  {Jiao}} \emph {et~al.},\ }\bibfield  {title} {\enquote {\bibinfo {title}
  {{The HEPS project}},}\ }\href {\doibase 10.1107/s1600577518012110}
  {\bibfield  {journal} {\bibinfo  {journal} {J. Synchrotron Radiat.}\ }\textbf
  {\bibinfo {volume} {25}},\ \bibinfo {pages} {1611--1618} (\bibinfo {year}
  {2018})}\BibitemShut {NoStop}%
\bibitem [{\citenamefont {Schroer}\ \emph {et~al.}(2018)\citenamefont
  {Schroer}, \citenamefont {Agapov}, \citenamefont {Brefeld}, \citenamefont
  {Brinkmann}, \citenamefont {Chae}, \citenamefont {Chao}, \citenamefont
  {Eriksson}, \citenamefont {Keil}, \citenamefont {Nuel~Gavald{\`{a}}},
  \citenamefont {R{\"{o}}hlsberger}, \citenamefont {Seeck}, \citenamefont
  {Sprung}, \citenamefont {Tischer}, \citenamefont {Wanzenberg},\ and\
  \citenamefont {Weckert}}]{Schroer:ig5056}%
  \BibitemOpen
  \bibfield  {author} {\bibinfo {author} {\bibfnamefont {Christian~G.}\
  \bibnamefont {Schroer}}, \bibinfo {author} {\bibfnamefont {Ilya}\
  \bibnamefont {Agapov}}, \bibinfo {author} {\bibfnamefont {Werner}\
  \bibnamefont {Brefeld}}, \bibinfo {author} {\bibfnamefont {Reinhard}\
  \bibnamefont {Brinkmann}}, \bibinfo {author} {\bibfnamefont {Yong-Chul}\
  \bibnamefont {Chae}}, \bibinfo {author} {\bibfnamefont {Hung-Chun}\
  \bibnamefont {Chao}}, \bibinfo {author} {\bibfnamefont {Mikael}\ \bibnamefont
  {Eriksson}}, \bibinfo {author} {\bibfnamefont {Joachim}\ \bibnamefont
  {Keil}}, \bibinfo {author} {\bibfnamefont {Xavier}\ \bibnamefont
  {Nuel~Gavald{\`{a}}}}, \bibinfo {author} {\bibfnamefont {Ralf}\ \bibnamefont
  {R{\"{o}}hlsberger}}, \bibinfo {author} {\bibfnamefont {Oliver~H.}\
  \bibnamefont {Seeck}}, \bibinfo {author} {\bibfnamefont {Michael}\
  \bibnamefont {Sprung}}, \bibinfo {author} {\bibfnamefont {Markus}\
  \bibnamefont {Tischer}}, \bibinfo {author} {\bibfnamefont {Rainer}\
  \bibnamefont {Wanzenberg}}, \ and\ \bibinfo {author} {\bibfnamefont {Edgar}\
  \bibnamefont {Weckert}},\ }\bibfield  {title} {\enquote {\bibinfo {title}
  {{PETRA IV: the ultralow-emittance source project at DESY}},}\ }\href
  {\doibase 10.1107/S1600577518008858} {\bibfield  {journal} {\bibinfo
  {journal} {Journal of Synchrotron Radiation}\ }\textbf {\bibinfo {volume}
  {25}},\ \bibinfo {pages} {1277--1290} (\bibinfo {year} {2018})}\BibitemShut
  {NoStop}%
\bibitem [{\citenamefont {Biagini}\ \emph {et~al.}(2008)\citenamefont {Biagini}
  \emph {et~al.}}]{Biagini:2008zze}%
  \BibitemOpen
  \bibfield  {author} {\bibinfo {author} {\bibfnamefont {M.E.}\ \bibnamefont
  {Biagini}} \emph {et~al.},\ }\bibfield  {title} {\enquote {\bibinfo {title}
  {{New Low Emittance Lattices for the Super-B Accelerator}},}\ }\href@noop {}
  {\bibfield  {journal} {\bibinfo  {journal} {Conf. Proc. C}\ }\textbf
  {\bibinfo {volume} {0806233}},\ \bibinfo {pages} {WEPP040} (\bibinfo {year}
  {2008})}\BibitemShut {NoStop}%
\bibitem [{\citenamefont {Nagaoka}\ and\ \citenamefont
  {Wrulich}(2007)}]{Nagaoka:2007zza}%
  \BibitemOpen
  \bibfield  {author} {\bibinfo {author} {\bibfnamefont {Ryutaro}\ \bibnamefont
  {Nagaoka}}\ and\ \bibinfo {author} {\bibfnamefont {Albin}\ \bibnamefont
  {Wrulich}},\ }\bibfield  {title} {\enquote {\bibinfo {title} {{Emittance
  minimisation with longitudinal dipole field variation}},}\ }\href {\doibase
  10.1016/j.nima.2007.02.086} {\bibfield  {journal} {\bibinfo  {journal} {Nucl.
  Instrum. Meth. A}\ }\textbf {\bibinfo {volume} {575}},\ \bibinfo {pages}
  {292--304} (\bibinfo {year} {2007})}\BibitemShut {NoStop}%
\bibitem [{\citenamefont {Zobov}\ \emph {et~al.}(2006)\citenamefont {Zobov},
  \citenamefont {Alesini}, \citenamefont {Biagini}, \citenamefont {Drago},
  \citenamefont {Gallo}, \citenamefont {Milardi}, \citenamefont {Raimondi},
  \citenamefont {Spataro},\ and\ \citenamefont {Stella}}]{Zobov:2006pp}%
  \BibitemOpen
  \bibfield  {author} {\bibinfo {author} {\bibfnamefont {M.}~\bibnamefont
  {Zobov}}, \bibinfo {author} {\bibfnamefont {D.}~\bibnamefont {Alesini}},
  \bibinfo {author} {\bibfnamefont {M.E.}\ \bibnamefont {Biagini}}, \bibinfo
  {author} {\bibfnamefont {A.}~\bibnamefont {Drago}}, \bibinfo {author}
  {\bibfnamefont {A.}~\bibnamefont {Gallo}}, \bibinfo {author} {\bibfnamefont
  {C.}~\bibnamefont {Milardi}}, \bibinfo {author} {\bibfnamefont
  {P.}~\bibnamefont {Raimondi}}, \bibinfo {author} {\bibfnamefont
  {B.}~\bibnamefont {Spataro}}, \ and\ \bibinfo {author} {\bibfnamefont
  {A.}~\bibnamefont {Stella}},\ }\bibfield  {title} {\enquote {\bibinfo {title}
  {{DAFNE experience with negative momentum compaction}},}\ }\href@noop {}
  {\bibfield  {journal} {\bibinfo  {journal} {Conf. Proc. C}\ }\textbf
  {\bibinfo {volume} {060626}},\ \bibinfo {pages} {989--991} (\bibinfo {year}
  {2006})},\ \Eprint {http://arxiv.org/abs/physics/0607036}
  {arXiv:physics/0607036} \BibitemShut {NoStop}%
\bibitem [{\citenamefont {Ikeda}\ \emph {et~al.}(2003)\citenamefont {Ikeda},
  \citenamefont {Flanagan}, \citenamefont {Fukuma}, \citenamefont {Hiramatsu},
  \citenamefont {Ieiri}, \citenamefont {Koiso}, \citenamefont {Mimashi},\ and\
  \citenamefont {Mitsuhashi}}]{Ikeda:2004me}%
  \BibitemOpen
  \bibfield  {author} {\bibinfo {author} {\bibfnamefont {H.}~\bibnamefont
  {Ikeda}}, \bibinfo {author} {\bibfnamefont {J.W.}\ \bibnamefont {Flanagan}},
  \bibinfo {author} {\bibfnamefont {H.}~\bibnamefont {Fukuma}}, \bibinfo
  {author} {\bibfnamefont {S.}~\bibnamefont {Hiramatsu}}, \bibinfo {author}
  {\bibfnamefont {T.}~\bibnamefont {Ieiri}}, \bibinfo {author} {\bibfnamefont
  {H.}~\bibnamefont {Koiso}}, \bibinfo {author} {\bibfnamefont
  {T.}~\bibnamefont {Mimashi}}, \ and\ \bibinfo {author} {\bibfnamefont
  {T.}~\bibnamefont {Mitsuhashi}},\ }\bibfield  {title} {\enquote {\bibinfo
  {title} {{Negative momentum compaction at KEKB}},}\ }\href@noop {} {\bibfield
   {journal} {\bibinfo  {journal} {eConf}\ }\textbf {\bibinfo {volume}
  {C0309101}},\ \bibinfo {pages} {THWA002} (\bibinfo {year} {2003})},\ \Eprint
  {http://arxiv.org/abs/physics/0401155} {arXiv:physics/0401155} \BibitemShut
  {NoStop}%
\bibitem [{\citenamefont {Schreiber}\ \emph {et~al.}(2019)\citenamefont
  {Schreiber}, \citenamefont {Boltz}, \citenamefont {Brosi}, \citenamefont
  {H\"arer}, \citenamefont {Mochihashi}, \citenamefont {M\"uller},
  \citenamefont {Papash},\ and\ \citenamefont {Schuh}}]{Schreiber:2019vor}%
  \BibitemOpen
  \bibfield  {author} {\bibinfo {author} {\bibfnamefont {Patrick}\ \bibnamefont
  {Schreiber}}, \bibinfo {author} {\bibfnamefont {Tobias}\ \bibnamefont
  {Boltz}}, \bibinfo {author} {\bibfnamefont {Miriam}\ \bibnamefont {Brosi}},
  \bibinfo {author} {\bibfnamefont {Bastian}\ \bibnamefont {H\"arer}}, \bibinfo
  {author} {\bibfnamefont {Akira}\ \bibnamefont {Mochihashi}}, \bibinfo
  {author} {\bibfnamefont {Anke-Susanne}\ \bibnamefont {M\"uller}}, \bibinfo
  {author} {\bibfnamefont {Alexander}\ \bibnamefont {Papash}}, \ and\ \bibinfo
  {author} {\bibfnamefont {Marcel}\ \bibnamefont {Schuh}},\ }\bibfield  {title}
  {\enquote {\bibinfo {title} {{Status of Operation With Negative Momentum
  Compaction at KARA}},}\ }in\ \href {\doibase
  10.18429/JACoW-IPAC2019-MOPTS017} {\emph {\bibinfo {booktitle} {{10th
  International Particle Accelerator Conference}}}}\ (\bibinfo {year} {2019})\
  p.\ \bibinfo {pages} {MOPTS017}\BibitemShut {NoStop}%
\end{thebibliography}%

\end{document}